# Mapping metabolic aging and disease-associated acceleration using an interpretable NMR-based clock

Alain Ibañez de Opakua[1], Maider Bizkarguenaga[2], Ángela de Diego[2], Ricardo Conde[2], Beatriz González-Valle[2], Rubén Gil-Redondo[2], Thyerre Santana[3], Maria Luisa Seco[4], Ana Cenarro[5,6,7], Virginia Lope[6], Sara Margarita[9], Serena Pelusi[9], Hansjörg Habisch[10], Christoph Haudum[11], Nicolas Verheyen[12], Barbara Obermayer-Pietsch[11], Maria Pilar Sorando-Fernandez[13], Eunate Arana-Arri[14], Susana Meijide[14], Ane Miren Salazar[15], Gonzalo Hernandez[15], Montse Baldán-Martín[16,6], Paula J. Martínez-González[16,6], Fabio Suárez-Trujillo[16,6], Julia Audije-Gil[17], Juan Ignacio Arriaga[18], Milagros Iriberri[19], Miguel Ángel Moran[20,21], Xabier Camino[22], Asun Gracia[23], Ganeko Bernardo-Seisdedos[24], Adriana Sousa[25], Nádia Oliveira[25], Ignacio Verde[25], María José Zaro Bastanzuri[26], María Reyes González Fernández[27], Daniel Fernández-Bergés[26], Laura Navarrete-Arias[28], Mar Llamas-Velasco[28], Umberto Biccari[29], Roberto Morales[29], Tobias Madl[10,30], Miguel Unda-Urzaiz[6,31,32], Arkaitz Carracedo[6,31,33,34,35], Tammo Diercks[35], Adela Castelló[6,8], Hartmut Schäfer[36], Manfred Spraul[36], Claire Cannet[36], Ljubica Tasic[3], Luis Varona[37], Rosario Sanchez-Pernaute[37,33], Elissabetta Buguianessi[38], Quentin Anstee[39,40], Nieves Embade[2], Shelly C. Lu[41], Francisco Mera[42], Luca Valenti[9,43], Claudio Luchinat[44,45], Danay Cibrián[28], Hortensia de la Fuente[28], Fernando Civeira[5,6,46], Dolores Arenas[17], Javier P. Gisbert[16,6], María Chaparro[16,6], Julia Debik[47], Guro Giskeodegard[47], Tone Bathen[47,48], Enrique Zuazua[29,49], José M. Mato[2,6] and Oscar Millet[1,2,6,*]

[1]ATLAS Molecular Pharma, Spain.
[2]CIC bioGUNE, BRTA, Spain.
[3]State University of Campinas, Brazil.
[4]OSARTEN Kooperativa Elkartea, Spain.
[5]IIS Aragón, Hospital Universitario Miguel Servet, Spain.
[6]Instituto de Salud Carlos III, Spain.
[7]IIACS, Zaragoza, Spain.
[8]Cancer and Environmental Epidemiology Unit, ISCIII, Spain.
[9]Fondazione IRCCS Ca' Granda Ospedale Maggiore Policlinico Milano, Italy.
[10]Medical University of Graz, Austria.
[11]Division of Medical University of Graz, Austria.
[12]Department of Internal Medicine, Medical University of Graz, Austria.
[13]Hospital Santa Marina, Spain.
[14]Scientific Coordination, Osakidetza, Spain.
[15]ESAME Biomedical, Spain.
[16]Hospital Universitario de La Princesa, Spain.
[17]Unidad de Investigación de la Fundación Renal Española, Spain.
[18]Pneumology Department, Hospital Universitario de Basurto, Spain.
[19]Pneumology Department, Hospital Universitario de Cruces, Spain.
[20]Bioaraba Health Research Institute, Spain.
[21]Osakidetza, Araba University Hospital, Spain.
[22]Hospital Universitario Donostia, Spain.
[23]Sociedad Española de Médicos Generales y de Familia (SEGM, Spain).
[24]Department of Medicine, Faculty of Health Sciences, University of Deusto, Spain.
[25]University of Beira Interior (UBI), Portugal.
[26]Hospital Don Benito Villanueva, Spain.
[27]Hospital Universitario de Badajoz, Spain.
[28]Instituto de Investigación Sanitaria Princesa (IIS-Princesa), Spain.
[29]DeustoTech, University of Deusto, Spain.
[30]BioTechMed-Graz, Austria.
[31]Biocruces Bizkaia Health Research Institute, Spain.
[32]Department of Urology, Basurto University Hospital, Spain.
[33]Ikerbasque, Basque Foundation for Science, Spain.
[34]University of the Basque Country (UPV/EHU), Spain.
[35]CIC bioGUNE, BRTA, Spain.
[36]Bruker Biospin GmbH, Germany.
[37]BioBizkaia Health Research Institute, Spain.
[38]Gastroenterology Department, University of Turin, Italy.
[39]Newcastle University, UK.
[40]Newcastle Upon Tyne Hospitals NHS Trust, UK.
[41]Cedars-Sinai Medical Center, USA.
[42]Long COVID and Post-Viral Syndromes Unit. Bluehealthcare Clinic, Spain.

[43]Università degli Studi di Milano, Italy.
[44]University of Florence, Florence, Italy.
[45]CIRMMP, Italy.
[46]Universidad de Zaragoza, Spain.
[47]Norwegian University of Science and Technology, Norway.
[48]Trondheim University Hospital, Norway.
[49]Friedrich-Alexander-Universität Erlangen-Nürnberg, Germany.
*Corresponding author. Email: omillet@cicbiogune.es

## Summary

Biological age captures cumulative physiological decline but remains challenging to quantify across the full adult lifespan with mechanistic and clinical interpretability. Here we develop an interpretable metabolic aging clock based on high-throughput serum nuclear magnetic resonance (NMR) profiling in 29,390 individuals aged 7–106 years across multiple cohorts. Using a machine-learning framework that integrates quantified metabolites and NMR-inferred clinical biomarkers, we achieve accurate age prediction ($r = 0.88$; RMSE $\approx 8.7$ years) while resolving the multicollinearity that limits interpretability in spectral models. Feature attribution reveals a consistent biological signature dominated by chronic inflammation, kidney function, and energy metabolism, with albumin and erythrocyte sedimentation rate emerging as principal determinants of metabolic aging. Application to eleven pathogenic cohorts (analyzing serum from 3,882 additional patient donors that represent distinctive and most frequent causes of disease), shows that metabolic age acceleration reflects disease-specific distortions rather than uniform aging shifts, where acute inflammatory states, metabolic dysfunction, and organ-specific pathologies exhibit distinct signatures. In a prospective cardiovascular cohort, individuals who subsequently experienced events showed higher predicted metabolic age at baseline than an age-matched external reference population. Together, these results establish an interpretable, scalable metabolic clock that links systemic biochemical states to aging trajectories and disease-associated metabolic alterationsand provide a framework for precision medicine applications.

**Running title:** Interpretable NMR metabolic aging clock

**Keywords:** metabolic aging, NMR spectroscopy, machine learning, biological age, metabolomics, age acceleration

## INTRODUCTION

Biological age, unlike chronological age, reflects the cumulative physiological decline driven by intrinsic genetic factors and extrinsic influences such as lifestyle, environment, and disease burden (Ferrucci et al. 2020). Metabolic aging clocks are computational models designed to estimate an individual's biological age by analyzing age-dependent patterns in metabolites (Robinson & Lau 2020; Hertel et al. 2016). Because metabolites represent the end products and intermediates of metabolic pathways, they provide an integrated and dynamic snapshot of cellular activity, mitochondrial function, and systemic homeostasis, all of which undergo progressive changes with aging (López-Otín et al. 2023; Panyard et al. 2022).

The development of metabolic clocks builds on the observation that concentrations and interrelationships among circulating metabolites shift predictably with age across populations. Deviations between metabolic and chronological age (commonly termed "age acceleration" or "age deceleration") may signal elevated disease risk, accelerated physiological decline, or conversely, metabolic resilience (Van Den Akker et al. 2020; Mutz et al. 2024). These conceptual foundations parallel those established for epigenetic, transcriptomic, and proteomic aging clocks, while metabolomic data provide a dynamic readout reflecting both long-term biological processes and short-term physiological adaptations (Johnson et al. 2020; Panyard et al. 2022).

Current metabolic aging clocks employ a variety of analytical platforms, most prominently mass-spectrometry-based metabolomics and large-scale serum nuclear magnetic resonance (NMR) profiling. NMR spectroscopy has emerged as a particularly attractive platform for constructing aging clocks due to its high reproducibility, minimal sample preparation, direct quantifiability, and suitability for large-scale clinical or epidemiological studies. Furthermore, serum and plasma NMR provide unique access to clinically relevant features, such as detailed lipoprotein subclass distributions and inflammatory glycoprotein markers (e.g., GlycA, GlycB), that are tightly linked to aging biology and cardiometabolic risk (Gil-Redondo et al. 2024; Soininen et al. 2015; Fung et al. 2023).

Several NMR-based metabolic clocks have already been developed, providing an expanding methodological foundation. Early efforts used NMR urine spectra to model metabolic age, demonstrating feasibility but limited precision, in the context of the substantial short-term variability characteristic of urinary metabolites (Tynkkynen et al. 2019; Hertel et al. 2016). Subsequent large-scale studies leveraging serum or plasma NMR profiling have substantially improved performance: analyses of tens of thousands of individuals across Estonian, Finnish, and Dutch cohorts have shown that NMR-derived biomarker patterns can predict chronological age and identify metabolic signatures associated with all-cause mortality (Van Den Akker et al. 2020; Fischer et al. 2014). However, validation across independently acquired datasets and distinct laboratory environments remains an important additional test of generalizability. The UK Biobank cohort has also been extensively used to develop metabolic clocks using conventional algorithms (Zhang, Wang, Wang, et al. 2024) and machine learning techniques (Mutz et al. 2024), but over a limited age range (approximately 40-70 y.o.). In turn, prospective analyses have been conducted on the same cohort to evaluate the sensitivity of metabolomic profiles in predicting the risk of up to 50 chronic diseases (Shang et al. 2024). Despite these advances, published NMR-based aging models suffer from regression to the mean, systematically overestimating the age of younger individuals and underestimating it in older subjects, which severely limits their performance at the extremes of the lifespan (Mutz et al. 2024). Moreover, epidemiological studies suggest that NMR spectra may also capture information related to clinical parameters (Soininen et al. 2015; Bragg et al. 2022; Klén et al. 2020; Bizzarri et al. 2022) but this possibility has not been explored in the development of metabolic clocks so far.

To address the problem of lifespan coverage in metabolic clocks and expand the clinical interpretability of the models, we assembled one large cohort of 29,390 individuals, covering a broad age range (7-106 years) with focus on adults, by combining multiple independent cohorts. In addition to the general population cohort, this resource incorporates patient groups totaling 3,882 donors and covering 11 diseases that include major non-communicable conditions as well as several infectious diseases. Building on the methodological perspective and initial proof-of-concept established in earlier NMR-based metabolomic aging research (Ibáñez de Opakua et al. 2025), we here developed a machine-learning model for age prediction that prioritizes interpretability while considering both quantified metabolite concentrations and additional clinical parameters inferred from the same NMR spectra. The objective was twofold: i) to characterize metabolic and clinical patterns that define aging trajectories within a large cohort covering the entire adult lifespan; and ii) to examine how different disease cohorts of similar geographic origin and ethnicity localize within this metabolic aging space.

# RESULTS

### Cohort Assembly and NMR Spectral Acquisition

To develop metabolic aging clocks, we assembled a large cohort of 29,390 individuals by integrating six independent studies. The cohort was designed to represent the general population, with balanced sex distribution and a wide age range (7-106 years) but mainly targeting adult population. The constituent studies differed substantially in sample size and age distribution, with individual cohorts contributing differently across the lifespan. This heterogeneity was subsequently considered through leave-one-cohort-out (LOCO) validation (Figure S1) which showed lower performance than the random-split evaluation, with a mean R of 0.49 (range, 0.25–0.72) and a mean RMSE of 14.2 years (range, 9.0–20.4). To independently validate the model, we also incorporated an external cohort obtained and measured in a different country and laboratory (Graz, Austria, n = 121). In addition, we included several disease-specific cohorts covering major non-communicable diseases (chronic kidney disease (CKD, n = 828), prostate cancer (PRC, n = 717), metabolic syndrome (MetS, n = 660), MASLD (n = 160), inflammatory bowel disease (IBD, n = 310), psoriasis (PSO, n = 226), cardiovascular disease with associated posterior event (CVD, n = 59) and amyotrophic lateral sclerosis (ALS, n = 63)) as well as infectious diseases and related syndromes, including influenza (FLU, n = 158),

COVID-19 (COV, n = 470), and long COVID (LCOV, n = 231). Comprehensive descriptions of the study cohorts can be found in the supporting information. Additionally, the general characteristics for each individual cohort are detailed in Tables S1-S20.

For each cohort (model training and validation, n = 29,390; external validation, n = 121; disease groups, n = 3,882), serum samples were analyzed using NMR spectroscopy. Multiple spectra were acquired (see Experimental procedures), each optimized for sensitivity to specific molecular features, including inflammatory markers, lipoproteins, and circulating metabolites (Figure 1).

**NMR signal–based models for metabolic age prediction**

We first developed an age-prediction model using the binned 1D NOESY dataset, which captures signals from all molecular components detectable by NMR. To that end, we refined the full set of 29,390 samples to derive an *age-balanced* subset of 8,640 individuals (Figure 1). The NOESY-based model achieved high predictive accuracy, with a Pearson correlation of R = 0.93 and a Root Mean Square Error (RMSE) of 6.5 years on the independent test set (Figure S2A).

To assess whether age-associated information was specific to the NOESY acquisition, we evaluated two complementary models based on the CPMG spectrum and the NOESY information in the time domain (free-induction decay, FID). The CPMG-based model achieved a predictive performance comparable to NOESY (r = 0.92; RMSE = 6.8 years; Figure S2B), consistent with recent analyses demonstrating that NOESY and CPMG spectra encode highly overlapping structural variations, to the extent that CPMG profiles can be accurately derived directly from NOESY acquisitions (Vignoli et al. 2025). Interestingly, the model trained on the unstructured FID data also remained highly predictive (r = 0.89; RMSE = 8.6 years; Figure S2C). This indicates that the dominant signals shaping the initial FID profile provide sufficient independent features for accurate age prediction without the need for a Fourier transform or accounting for the data's intrinsic time-domain sequentiality.

SHapley Additive exPlanations (SHAP) were used to interpret the model. SHAP values represent the fair, weighted average contribution of a feature across all possible combinations. When applied to the NOESY dataset, the feature contributions exhibited a characteristically broad and non-specific distribution of SHAP values (Figure S3A). Direct interpretation of the NOESY model in the metabolic space proved unreliable due to the extreme multicollinearity inherent to binned NMR spectra: adjacent and partially overlapping resonances produced large blocks of highly correlated bins. To robustly quantify this redundancy, we computed the condition number (κ) of the feature matrix, a metric that reflects how multicollinearity affects the overall numerical stability of the dataset. Binned NOESY data yielded $\kappa \approx 1.5 \times 10^4$, far exceeding the commonly accepted threshold of κ = 30, which indicates problematic multicollinearity (Belsley et al. 1980).

**Interpretable aging clocks based on NMR-derived biomarkers**

The conventional approach to extract clinical value from serum NMR spectra involves assigning spectral signals to specific metabolites and quantifying them (Würtz et al. 2017; Jiménez et al. 2018). However, epidemiological studies based on NMR data suggest that these spectra may also capture information related to other well-established clinical parameters (directly or indirectly) (Soininen et al. 2015; Bragg et al. 2022). Building on this premise, we trained supervised regression models to quantify 49 metabolites and infer 25 routinely measured clinical biomarkers, directly from NOESY spectra (see Experimental procedures and Table S21). The selected parameters span key physiological domains, including carbohydrate and amino acid metabolism, ketone bodies, TCA cycle intermediates, and one-carbon metabolism, as well as lipid profiles, inflammatory markers, renal function indices, hematological variables, and protein-related measures. For metabolite quantification, a dedicated pipeline was applied to J-res spectra, using targeted spiking experiments as the primary validation reference. The predicted biomarkers exhibited moderate to strong correlations with their corresponding laboratory measurements (Figure S4), supporting their use as semiquantitative proxies rather than precise clinical measurements.

This combined feature set (74 variables in total) served as the foundation for constructing an initial interpretable metabolic age model. Although this model achieved good predictive performance (r = 0.89; RMSE = 8.6 years; Figure S5A), analysis of the feature matrix still revealed clear signs of multicollinearity, yielding a condition number of κ ≈ 38 (well above the conventional threshold of 30). This redundancy structurally dispersed the resulting SHAP values, thereby compromising a clear and robust biological interpretation (Figure S5B)[26]. For example, GlycA and GlycB both exhibited high SHAP importance but with opposite directional effects on predicted age, a pattern that is physiologically implausible. For comparison, models trained exclusively on IVDr automated metabolite and lipoprotein reports (143 parameters) also demonstrated robust predictive capacity (r = 0.86; RMSE = 9.3 years), but showing high collinearity ($\kappa \approx 180$) due to the lipoprotein subfraction intrinsic cross-correlation, which also introduced additional complexity in feature attribution.

To address this issue and obtain a fully interpretable biochemical representation, we intentionally bypassed automated dimensionality reduction techniques in favor of a manual, expert-curated preprocessing strategy specifically aimed at reducing multicollinearity by targeting the strongest inter-feature correlations. Starting from the original 74 features, redundant variables were removed, and certain biochemical parameters were manually reformulated or aggregated to reduce redundancy while preserving physiologically distinct information, resulting in a final set of 75 variables. These decisions were based exclusively on the correlation structure among predictors and were not informed by their association with chronological age or by model performance. Accordingly, hemoglobin, fructosamine, and total cholesterol were discarded due to redundancy with other biomarkers. The GlycB/GlycA ratio was computed and Glyc B removed because of its strong correlation with Glyc A. Additional metabolically meaningful ratios (including urea/creatinine, BCAA/AAA, lactate/pyruvate, and glutamine/glutamate) were introduced as potentially informative composite indicators of metabolic balance. Total protein was also decomposed into a non-albumin fraction to isolate a more specific signal. With this representation, we substantially reduced multicollinearity (from $\kappa \approx 38$ to $\kappa \approx 16$), while preserving the original biochemical meaning, thereby yielding a well-conditioned structure suitable for stable interpretation.

The interpretable model achieved high predictive performance (r = 0.88; RMSE = 8.7 years; Figure 2A), with a regression slope of 0.74 compared with 0.86 for the NOESY-based model. Although this represents a modest trade-off in predictive accuracy and regression-to-the-mean bias, the model retained performance close to that of the NOESY-based model while relying on substantially fewer and more biologically constrained features. This trade-off was accompanied by a substantial gain in interpretability: the reduction in multicollinearity resulted in stable and coherent SHAP value distributions (Figure 2B), in contrast to the diffuse or contradictory attribution patterns observed in the unprocessed biomarker model (Figure S5C).

To rigorously validate that the calculated metabolic distortion is not a statistical artifact driven by differences in chronological age distributions, we implemented an iterative random sub-sampling approach by using the pool of initially discarded samples (n = 20,750) (Figure S6, see Experimental procedures). Two-sample Kolmogorov-Smirnov tests comparing the metabolic distortion (differences between metabolic and chronological ages) revealed an exceptional structural alignment between the discarded samples and their matched reference counterparts.

To independently assess the generalizability of our interpretable model developed using all six source cohorts, we evaluated its predictive performance on an independent external control cohort from general population in Austria (n = 121, age range 18-89), a geographically unrelated area and whose samples were acquired and measured independently in a different laboratory. As shown in Figure 2C, the regression of predicted metabolic age against chronological age in this external cohort (R = 0.81; slope = 0.66) closely mirrored that of an age-matched subset from our primary reference dataset (R = 0.89; slope = 0.69). Furthermore, Kolmogorov-Smirnov test distributions of both groups yielded a *p*-value of 0.1, statistically confirming that the model's behavior does not differ between populations.

Finally, to further assess the generalizability of the unified metabolic age model, we evaluated its performance separately in males and females and compared it with sex-specific models trained exclusively on each subgroup (Figure S7). The unified model achieved similarly strong accuracy in females (r = 0.909, RMSE = 8.3 years) and males (r = 0.850, RMSE = 9.0 years). Training sex-specific models did not yield substantial improvements: the female-only model performed nearly identically to the unified model (r = 0.908, RMSE = 8.2 years), and the male-only model showed just a modest increase in accuracy (r = 0.877, RMSE = 8.5 years). Importantly, this agreement was maintained at the individual level after removing the common effect of chronological age: age-adjusted predictions from the unified model remained strongly correlated with those from the female-specific (R = 0.947) and male-specific (R = 0.919) models (Figure S8A-B). The slightly superior performance in females is likely explained by the larger number of older female participants in the dataset, reflecting their longer life expectancy.

**Biological architecture of metabolic aging as revealed by SHAP analysis**

To obtain a higher-level view of the biological processes underlying metabolic aging, SHAP values were grouped into predefined metabolic and clinical categories representing major physiological domains, including energy metabolism, kidney function, nitrogen homeostasis, inflammation, and cardiovascular risk markers (Table S22). This SHAP-group analysis revealed a clear and reproducible structure in the model's decision-making (Figure 3).

The grouped SHAP analysis identifies chronic inflammation as the dominant axis of metabolic aging in this model, followed by nitrogen homeostasis, energy metabolism and structural integrity and systemic frailty. At the individual-feature level, albumin and erythrocyte sedimentation rate (ESR) displayed the largest mean absolute SHAP values (3.1 and 2.4 years, respectively), almost doubling the contribution of the next most influential features, alanine (Ala, 1.6 years) and 1,5-anhydrosorbitol (AnHsorb, 1.4 years).

It is well-known that sex is a key determinant for metabolism. Consequently, SHAP analyses also revealed clear sex-specific differences in the biochemical determinants of metabolic age. Importantly, these differences were reproducible both when male and female samples were analyzed within the unified model and when SHAP values were derived from the sex-specific models themselves (Figure S7). At the global feature-importance level, mean absolute SHAP values were strongly correlated between the unified and female-specific ($R = 0.898$) and unified and male-specific ($R = 0.891$) models, while concordance between the two sex-specific models was lower but remained substantial ($R = 0.705$; Figure S8C–E). In females, urea and estimated glomerular filtration rate (eGFR) contributed more strongly to age prediction, whereas in males, Ala and AnHsorb ranked among the most influential features. Despite these sex-dependent differences, the dominant contributors to metabolic aging remained consistent across analyses: albumin and ESR emerged as the top predictors in both sexes and in both modeling strategies.

**Application of the metabolic age model to disease cohorts**

We next investigated how the general population model behaves across a range of disease cohorts. Diseases included acute infections as well as non-communicable diseases and covering the main axis of pathology: cardiometabolic disease (CVD, MetS, CKD, MASLD), cancer (PRC), immunological disorders (IBD, PSO), neurodegenerative disease (ALS) and infection or infection-derived syndromes (COV, LCOV and FLU). For each condition, metabolic age was compared against chronological age, and the deviation from the reference regression line (termed metabolic distortion) was quantified. Positive distortion reflects an older-than-expected metabolic profile (accelerated aging), whereas negative distortion indicates relatively preserved or "younger" metabolic signatures.

To benchmark disease-associated deviations, each cohort was compared against a reference group of individuals from the general population matched by age. For both cases and references, we computed the distribution of metabolic distortion and represented it as histograms (Figure 4A-K). As expected from the calibration of the model in the balanced population, the matched reference groups showed distortion distributions centered around zero. In contrast, all disease cohorts exhibited right-shifted distributions, indicative of accelerated metabolic aging, and increased dispersion, reflecting enhanced inter-individual heterogeneity. All case–reference comparisons were statistically significant according to the Kolmogorov–Smirnov test, with p-values far below $10^{-5}$ in most cohorts and $p = 0.001$ for LCOV. To further validate these findings, the same procedure was applied to two control groups: an internal control consisting of the model's independent test set (ITCL, Figure 4L), and the external control comprising the general population cohort from Austria (ETCL, Figure 4M). Both groups yielded distributions that were not statistically distinguishable from the reference baseline, with Kolmogorov–Smirnov p-values of 0.2 and 0.1 for the internal and external controls, respectively. Consistently, complementary cohort-specific regression models with predicted metabolic age as the outcome and chronological age, sex, and disease status as predictors showed a significant positive effect of disease status in all disease cohorts, whereas no significant effect was observed in either control cohort (Figure S9). These results confirm that the observed metabolic shifts are disease-specific and robust to inter-laboratory variability.

To explore the reversibility of disease-driven metabolic aging, we investigated the IBD cohort by comparing patients in clinical remission against those with active disease (Figure S10). Patients in remission showed a metabolic distortion approximately 3 years lower than those in the active group, accompanied by a 3-year reduction in the standard deviation. This decreased variability was primarily driven by a lower prevalence of individuals with extreme metabolic age acceleration. These findings suggest that the metabolic distortion captured by our model is not a fixed state but can partially shift back toward the reference aging trajectory following clinical recovery (allostasis).

The prospective cardiovascular cohort is of particular interest in our analysis. Considering that all patients in this group experienced a CVD episode, examining their individualized metabolic age profiles is highly informative. The corresponding SHAP waterfall plots illustrating these individual trajectories are presented in Figure S11. Overall, the patterns are highly heterogeneous, underscoring the intrinsic complexity of individuals at the metabolic level and highlighting the multifactorial nature of CVD events. Most patients exhibited an increased metabolic age compared with their chronological age, although with diverse combinations of SHAP values, often consistent with pronounced lipid dysregulation, elevated glucose levels, and other CVD-associated markers. Interestingly, a common feature was elevated eGFR (Figure S12), which contributed to a lower predicted age (negative SHAP delta), likely reflecting a state of glomerular hyperfiltration; an early maladaptive hemodynamic response that often precedes cardiovascular events (Kim et al. 2025).

**Representation of diseases along the metabolic aging trajectory**

To investigate how the different diseases under consideration differ from the normal aging we have performed a PCA analysis of all the groups, specifically centering the analysis on the general population (Figure 5A). As expected, the closest group to the age model is the Austrian external validation cohort (Figure 5A, inset). In the opposite end falls the CKD cohort, consistent with the severity of the disease of patients bound to dialysis for survival. In the kidney disease cohort, while uremic toxins (urea, myo-inositol, urate) strongly pushed the metabolic age upward (Figure 5B), creatinine exhibited a large negative SHAP difference (note that this refers to the feature's SHAP value impact on the model, not its raw clinical value), effectively acting as a "rejuvenating" factor. This occurs because, in the healthy training population, higher creatinine is a marker of preserved muscle mass and robustness. The model, unaware of the renal failure context, interprets the pathological retention of creatinine as a sign of muscular health.

Other diseases that are well discriminated by the aging model include MASLD and infectious diseases (FLU and COV). MASLD profoundly disrupts liver metabolism (the liver being the central metabolic organ) likely affecting serum homeostatic concentrations. In contrast, infections such as COV are known to induce a significant, albeit temporary, metabolically accelerated aging (Bruzzone et al. 2023). Consistent with these observations, the category-level analysis of SHAP changes revealed distinct disease-specific functional profiles, with inflammatory and immune-related contributions predominating in FLU and COV, kidney-related alterations dominating CKD, and metabolic pathways such as nitrogen homeostasis contributing prominently to MASLD (Figure 5C). The remaining diseases show moderate acceleration, consistent with the idea that their main contributing factors are already embedded within the aging process. For example, IBD and PSO are largely driven by inflammatory markers similar to those observed in age-related inflammation (inflammaging), while aging and MetS are often difficult to distinguish, likely reflecting the influence of Western dietary and lifestyle patterns in the cohort used to develop the metabolic clock. However, direct comparison with the physiological aging trajectory showed that these disease-associated alterations vary substantially in their degree of aging-like behavior, with age-orthogonal components predominating in all examined diseases and the age-aligned component reaching at most 36.1% (Figure S13 and Figure 5D).

# DISCUSSION

The present study demonstrates that high-throughput serum NMR dataset analysis, combined with machine-learning modeling, yields a robust and biologically interpretable metabolic aging clock that captures disease-related deviations from healthy aging trajectories. Unlike previous studies, which often adjust for sex as a covariate or necessitate separate sex-specific clocks due to known dimorphisms in metabolic aging (Peng et al. 2026), we found that a single, unified model trained jointly on men and women performed as well as stratified versions, although predictive performance was slightly lower in men. Interestingly, a similar direction of sex difference was reported by Hertel et al., whose metabolic age score correlated slightly less strongly with chronological age in men than in women (r = 0.74 vs. 0.78). This difference may partly reflect greater underlying biological and metabolic heterogeneity in men, potentially related to sex-linked genetic architecture. Importantly, the use of a unified model did not obscure sex-specific biological nuances: markers of kidney function (urea, eGFR) contributed more strongly to the age estimate in women, whereas glycemic markers (Ala, AnHsor) were relatively more important in men, consistent with established sex differences in renal aging and glucose handling (Lodge et al. 2025).

It is well-known that high-dimensional omics data often obscure single-feature explanations (Huang et al. 2025). To overcome this effect, we implemented a framework based on quantified metabolites and clinical parameters obtained from the NMR data. This strategy yielded numerically stable and physiologically consistent SHAP patterns with only a modest trade-off in predictive performance, in line with previous observations where pathway-level features facilitated mechanistic insight without sacrificing robustness (Macdonald-Dunlop et al. 2022).

This platform-specific feature representation limits direct benchmarking against previously published metabolomic aging clocks, as their required inputs cannot be fully reconstructed from our NMR dataset. Nevertheless, published metrics provide useful descriptive context. Our NOESY and interpretable models showed R = 0.93 and 0.88, with slopes of 0.86 and 0.74, respectively, compared with lower reported values for Hertel et al. (R = 0.74–0.78, slope ≈ 0.60), van den Akker et al. (R = 0.65, slope ≈ 0.43), Lau et al. (R = 0.64, slope ≈ 0.42), and Mutz et al. (R up to 0.59, slope ≈ 0.40). Importantly, the values reported here for Hertel et al. correspond to their development cohort; correlations were lower in their independent replication cohort (R = 0.53–0.61), whereas an equivalent regression slope was not reported. These cross-study comparisons should be interpreted cautiously because of differences in cohort composition and validation strategies.

Within this interpretable framework, the principal marker related to aging is the inflammation and immune activity axis with the specific selection of albumin and ESR by the model over more transient markers like C-reactive protein. While cytokines and acute-phase proteins fluctuate rapidly, albumin (a negative acute-phase reactant) and ESR (driven by fibrinogen and immunoglobulins) provide a time-integrated readout of the systemic inflammatory state (Watson et al. 2020). Glyc and SPC signals (uniquely measured by NMR spectroscopy) are well-known to be involved in acute inflammation (Bizkarguenaga, Bruzzone, et al. 2022) and they did moderately contribute to the age-related inflammation marker. All this fits neatly within the concept of "inflammaging," the chronic, low-grade, sterile inflammation that accumulates with age. Inflammaging is now recognized as a central mechanistic axis linking many of the hallmarks of aging to age-related diseases (Franceschi et al. 2018). Moreover, the inflammatory burden seems to disrupt systemic energy homeostasis, generating a self-reinforcing cycle of immunometabolic dysregulation. The prominence of nitrogen homeostasis alongside energy metabolism indicates that alterations in systemic amino acid handling may be as critical as direct glycemic dysregulation in the aging process (Zhang, Wang, Yu, et al. 2024). Closely linked to these shifts, structural integrity and systemic frailty emerged as a major determinant of the clock, largely driven by albumin levels. This reflects the dual nature of albumin in the context of aging: acting both as a negative acute-phase reactant linked to systemic inflammaging and as a fundamental indicator of nutritional status and protein reserve (López-Otín et al. 2013).

The metabolic clock also identifies kidney function and nitrogen waste as critical determinants of metabolic age. This relationship is consistent with recent evidence positioning the kidney as a paradigmatic organ of age-related decline, where the senescence-associated secretory phenotype and chronic inflammaging accelerate tissue dysfunction and impair regenerative capacity (Luca et al. 2025). The combined SHAP contribution of urea, urate, and eGFR suggests that the model captures the progressive accumulation of uremic retention solutes and the gradual loss of filtration capacity associated with these processes. Furthermore, the interplay between one-carbon and methylation metabolism and redox homeostasis underscores the systemic impact of aging on cellular maintenance. The contribution of methylation-related metabolites and glutathione-precursor amino acids likely reflects the convergence of epigenetic drift and the gradual exhaustion of antioxidant defenses under persistent oxidative stress.

Finally, the persistence of an atherogenic profile within lipoprotein and cardiovascular risk markers, alongside alterations in muscle and BCAA metabolism, probably acts as a downstream integrator of this systemic deterioration. Heavily influenced by the dominant overall contribution of inflammation and immune activity, this signature suggests that cytokine-mediated disruption of proteostasis and systemic insulin resistance synergistically promote both atherogenic dyslipidemia and the metabolic signatures of muscle frailty.

The application of the model to disease cohorts revealed that "metabolic age acceleration" is not a uniform phenomenon but rather a disease-specific distortion of the healthy aging trajectory. We quantified this as the difference in SHAP values between patients and age-matched controls (Figure 4, S14 and S15). The most profound distortions were observed in acute infections and active inflammatory states. Patients with FLU, COV, and IBD exhibited large positive shifts in metabolic age driven by a massive intensification of the inflammatory axis. In these cohorts, the contributions of ESR, C-reactive protein, and albumin deviated dramatically from the reference (e.g., ESR SHAP delta > +5 years in FLU). This surge highlights that the "inflammaging" clock, when exposed to acute-phase responses, captures a biochemical mimicry of advanced catabolic aging (Henry et al. 2020). In contrast to the acute infection cohorts, LCOV showed a more modest metabolic deviation from the healthy reference population, consistent with the subsidence of the overwhelming inflammatory signal. Instead, the main drivers shifted toward subtler metabolic inefficiencies, particularly glutamic acid and glucose, suggesting that persistent post-acute sequelae may be driven less by systemic inflammation and more by lingering dysregulation in bioenergetics and neurotransmitter metabolism (Holmes et al. 2021).

MASLD exhibited a unique hierarchical reordering of metabolic drivers. Although inflammatory markers remained relevant, the primary driver of age acceleration was glutamic acid, which exhibited a massive positive SHAP contribution, clearly distinct from other cohorts. Glutamic acid occupies a central metabolic node, linking amino acid metabolism, the TCA cycle, nitrogen handling and the urea cycle, and glutathione synthesis. Although not specific to liver disease and reflective of systemic metabolic dysfunction, elevated glutamic acid in MASLD likely arises from increased protein turnover and amino flux, enhanced transamination reactions (e.g., ALT and AST), mitochondrial dysfunction, oxidative stress, and insulin resistance, which shifts central carbon metabolism toward amino acid-derived gluconeogenesis. Together, these processes lead to altered glutamic acid handling and its accumulation in circulation (Masoodi et al. 2021). Similarly, in the CKD cohort, the model captured a pronounced organ-specific metabolic rewiring. The overall metabolic distortion was heavily dominated by the systemic accumulation of uremic toxins (including urea, myo-inositol, and urate) and major alterations in creatinine levels, which collectively emerged as the principal drivers of metabolic age prediction, reflecting the profound consequences of declining renal filtration capacity.

In line with these observations, PCA analysis across the cohorts delineated three clear axes of metabolic distortion (Figure 5A). One major axis reflects renal impairment, heavily weighted by eGFR and diverse uremic toxins. A second axis maps the inflammatory burden, driven by albumin, ESR, and C-reactive protein, which clearly isolates the infectious disease cohorts (COV and FLU). The third axis captures a strictly metabolic signature, primarily governed by glutamic acid and sarcosine, which distinctly characterizes the MASLD cohort. This spatial separation further illustrates how different diseases influence the metabolic clock through distinct, biologically coherent pathways. Other conditions displayed broader contributions distributed across several categories.

This interpretation was further supported by the functional categorization of disease-associated SHAP changes (Figure 5C). Inflammatory and immune-related contributions predominated in the infectious cohorts, kidney function and nitrogen-waste markers overwhelmingly characterized CKD, while MASLD showed a distinct metabolic profile dominated by nitrogen homeostasis and one-carbon and methylation metabolism.

To determine whether these disease-associated metabolic alterations followed the physiological aging trajectory, we decomposed each condition-specific effect vector into components aligned with and orthogonal to healthy metabolic aging (Figure 5D), and independently quantified directional similarity using cosine similarity with bootstrap confidence intervals (Figure S13). The degree of alignment varied markedly across conditions, with infectious diseases showing the strongest similarity to physiological aging, although age-orthogonal alterations predominated in all examined diseases. These results indicate that increases in metabolic age can arise through different combinations of aging-like and disease-specific metabolic changes rather than through a uniform acceleration of physiological aging. The consistently low age-aligned component therefore emphasizes that disease-associated increases in predicted metabolic age should not be interpreted simply as an acceleration of physiological aging, highlighting the importance of examining the underlying metabolic signature rather than the predicted age alone.

A clinically relevant finding emerged from the prospective CVD cohort. Although clinically healthy at baseline, individuals who subsequently developed cardiovascular events exhibited a distinct metabolic acceleration of approximately 8 years relative to an age-matched external reference population.

The study has several limitations. Although the cohort is large and age-balanced, it predominantly represents a Caucasian population. The heterogeneous LOCO performance may reflect reduced age–sex representation after cohort exclusion and other cohort-specific differences; correlations also require caution within narrow age ranges. Together with the relatively small Austrian validation cohort (n = 121), these findings highlight the need for further validation in larger and more diverse independent populations. Moreover, demographic, clinical, and preanalytical characteristics across disease cohorts could not be fully harmonized beyond age and sex, and residual cohort-related confounding cannot therefore be excluded. At the individual feature level, the paradox of "false rejuvenators" represents a further conceptual challenge. This phenomenon occurs because certain parameters reflect different physiological states in normal aging compared to specific diseases. For example, while elevated creatinine is a sign of preserved muscle mass and robustness in the healthy general population, it is a marker of severe renal damage in patients with CKD. Such context-dependent reversals can theoretically mislead models into assigning a lower biological age to sick individuals, as has been previously observed in predictive clocks applied to colorectal cancer patients (Zhang, Mo, et al. 2024). Therefore, the model is not intended for individual clinical interpretation without disease-specific contextualization, and the specific variables driving metabolic deviations must always be interpreted within a clinical framework, as a "younger" molecular signal is not inherently beneficial if it stems from a disruption of normal homeostasis. That said, although individual features could exhibit isolated "rejuvenating" SHAP values in our study, this effect was consistently outweighed by the broad coverage of our biomarker panel across diverse physiological processes, and our interpretable model never produced a net deviation toward a younger overall metabolic age in any of the analyzed disease cohorts.

In conclusion, this work establishes an interpretable and scalable NMR-based metabolic aging clock that captures both the shared and disease-specific biochemical processes underlying human aging. By integrating metabolite quantification with clinically relevant biomarkers and addressing multicollinearity through a biologically informed framework, the model achieves a balance between predictive performance and mechanistic insight. Beyond accurate age estimation, the concept of metabolic distortion provides a useful tool to understand how different pathologies diverge from normative aging trajectories, highlighting the heterogeneity of disease-associated metabolic remodeling. Future studies should focus on extending validation across more diverse populations, incorporating longitudinal and prospective outcome data to assess temporal dynamics and future health trajectories, and integrating additional omics layers to further refine the biological interpretation of metabolic aging.

# EXPERIMENTAL PROCEDURES

**Cohorts and study population.** This study incorporated data from a multi-national consortium. The reference population was recruited across Spain and Portugal, whereas independent disease cohorts were collected from Spain, Italy and the United Kingdom. Additionally, a general population control cohort from Austria was included for external validation of the reference population model. All samples were double coded prior to analysis to ensure full anonymization. The cohorts included in the foundational dataset are described in the supporting information, with detailed metadata and descriptive statistics for each cohort provided in Tables S1-S20.

**Blood collection and serum preparation.** Venous blood was obtained from fasting individuals (unless otherwise specified) following standardized operating procedures, as previously described (Bizkarguenaga, Gil-Redondo, et al. 2022). After allowing coagulation at room temperature, samples were centrifuged, and the serum was aliquoted into cryogenic vials. All aliquots were stored at –80 °C until analysis, to ensure sample stability and comparability across cohorts. The protocol was implemented uniformly across all participating centers to minimize pre-analytical variability.

**NMR spectra acquisition.** High-throughput $^{1}$H-NMR spectra were acquired on Bruker Avance III HD and Neo 600 MHz instruments equipped with BBI or TXI probes and SampleJet™ automation systems. Measurements were performed at 310K following previously validated calibration and QC procedures to secure reproducibility (Loo et al. 2020). For each serum sample, three spectral datasets were generated: (i) a quantitative 1D $^{1}$H NOESY experiment with water suppression (Bruker pulse program *noesygppr1d*), (ii) a 1D CPMG (Carr–Purcell–Meiboom–Gill) sequence (*cpmgpr1d*), used to attenuate signals from macromolecules such as proteins and lipoproteins and single out low-molecular-weight metabolites, and (iii) a fast 2D J-resolved experiment (J-Res, *jresépprqf*), to resolve peak overlap, assist in metabolite assignment and improve quantification. The automated Bruker metabolite and lipoprotein profiles used for comparison were generated automatically from these spectra using B.I.Quant-PS™ and B.I.LISA™ software respectively. Additionally, 1D-PGPE (Pulsed Gradient Perfect Echo, *zggppezf.ivdr*) spectra were acquired to quantify specific inflammation markers (GlycA, GlycB, and SPC) using Bruker algorithms (PhenoRisk PACS™).

**Data analysis.** Software description and data analysis procedures, including data preprocessing, metabolite quantification, prediction of clinical parameters, balancing strategy, machine learning models for age prediction and model interpretability are detailed in the Supplementary Information.

(Olson & Moore 2019)(Hoerl & Kennard 1970)(Geurts et al. 2006)(Lundberg et al. n.d.)

## Acknowledgements

We thank all cohort members and researchers who participated in this study, and especially the patients and the Biobank of the Aragon Health System (PT23/00146), integrated in the ISCIII Platform for Biobanks and Biomodels, for their collaboration. This work was supported by the Department of Economic Development and Infrastructures of the Basque Government (Elkartek bg2021, bg2023, bg2025) and the Plan Complementario en Biotecnología Aplicada en Salud. Funding was also provided by the European Research Council (ERC) under the European Union's Horizon Europe programme (grant agreement No. 101096251-CoDeFeL). We also acknowledge support from the Alexander von Humboldt Professorship, the Horizon Europe MSCA project ModConFlex (101073558), DFG Transregio 154, AFOSR (24IOE027), SURE-AI (Research Council of Norway, 357482), and the Madrid Government-UAM Excellence Programme (V PRICIT). We were also supported by MICIU (PID2024-160696OB-I00, PDC2025-165310-I00, PID2023-146872OB-I00-DyCMaMod), and by the Basque Government through the consolidated research group "CMRG-Computational Mathematics Research Group" under Grant IT1856-26. We thank the Center for Medical Research, Medical University of Graz, Graz, Austria for laboratory access. We are grateful to the Austrian Science Fund (FWF) for excellence cluster 10.55776/COE14, Grants DOI 10.55776/P28854, 10.55776/I3792, 10.55776/DOC130, and 10.55776/W1226, the Austrian Research Promotion Agency (FFG) grants 864690 and 870454; the Integrative Metabolism Research Center Graz; the Austrian Infrastructure Program 2016/2017; the Styrian Government (MetAGE, Zukunftsfonds, doc.fund program); the City of Graz (MetAGE, doc.fund); and BioTechMed-Graz (flagship project). This project was funded in part by the FFG and the European Union (EFRE) under grant 912192. We would like to express our sincere gratitude to all the representatives of the Research Unit of the Spanish Kidney Foundation and the Asociación Long COVID de Euskal Herria, who have participated in this work.

## Conflict of interest

The authors declare no conflict of interest.

## Author contributions

Conceptualization: AIdeO, UB, RM, LV, CL, EZ, JMM, and OM. Data curation: AIdeO, BG-V, RG-R, HH, CH, NV, BO-P, EA-A, SuM, AMS, GH, MB-M, JA-G, AS, NO, IV, MJZ-B, MRG-F, DF-B, TM, MU-U, ArC, AdC, RS-P, EB, QA, FM, LV, DC, HdlF, FC, DA, JP-G, MC, GG, and TB. Formal analysis: AIdeO, RG-R, HH, CH, NV, BO-P, and TM. Funding acquisition: AnC, UB, TM, NE, EZ, JMM, and OM. Investigation: AIdeO, MB, AdD, RC, BG-V, TS, MLS, AnC, VL, SaM, SP, HH, CH, NV, BO-P, MB-M, PJM-G, FS-T, JA-G, JIA, MI, MAM, XC, AG, GB-S, AS, NO, IV, MJZ-B, MRG-F, DF-B, LN-A, ML-V, UB, RM, MU-U, ArC, AdC, LT, RS-P, EB, QA, FM, LV, DC, HdlF, FC, DA, JP-G, MC, GG, and TB. Methodology: AIdeO, UB, RM, TD, HS, MS, CC, EZ, and OM. Project administration: JMM and OM. Resources: TM, HS, MS, CC, NE, and OM. Writing – original draft: AIdeO. Writing – review & editing: AIdeO, MB, AdD, RC, BG-V, RG-R, TS, MLS, AnC, VL, SaM, SP, HH, CH, NV, BO-P, MPS-F, EA-A, SuM, AMS, GH, MB-M, PJM-G, FS-T, JA-G, JIA, MI, MAM, XC, AG, GB-S, AS, NO, IV, MJZ-B, MRG-F, DF-B, LN-A, ML-V, UB, RM, TM, MU-U, ArC, TD, AdC, HS, MS, CC, LT, RS-P, EB, QA, NE, SCL, FM, LV, CL, DC, HdlF, FC, DA, JP-G, MC, GG, TB, EZ, JMM, and OM.

## Data availability statement

The complete source code, featuring a Jupyter notebook with the analytical functions and the Docker environment configurations, is currently accessible at https://github.com/AlainIOLA/MetAge-NMR. Upon publication, the repository will be permanently archived on Zenodo. Underlying data are available upon reasonable request to the corresponding author, subject to cohort-specific data-sharing agreements and ethical restrictions.

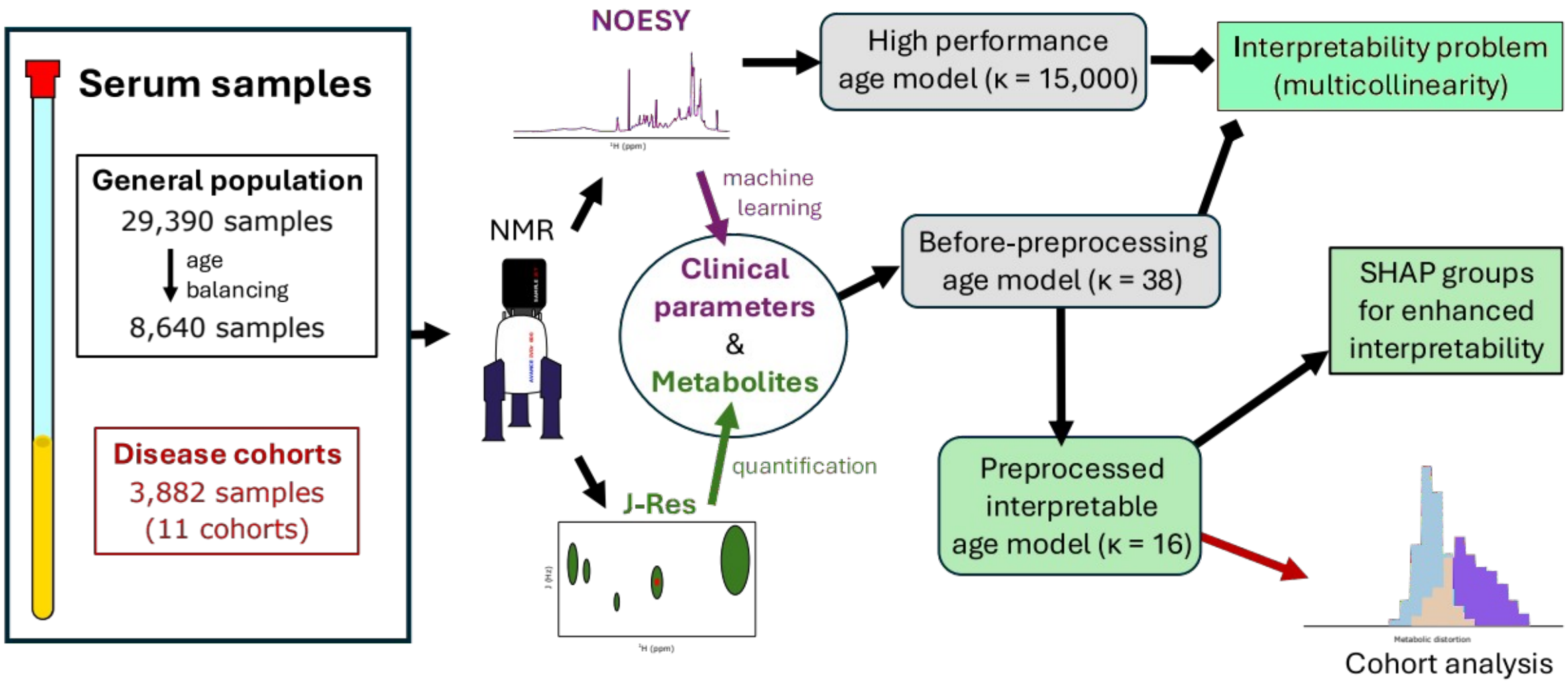

**Figure 1 | Methodological workflow for the development of the interpretable metabolic aging clock.** The flowchart illustrates the study design, starting with serum samples from a general population (balanced by age) and eleven disease cohorts. NMR spectroscopy is performed to acquire NOESY and J-Res spectra. The NOESY data is used to predict clinical parameters and to build a high-performance age model; however, this age spectral model suffers from an interpretability problem due to high multicollinearity (κ=15,000). J-Res spectra are utilized for metabolite quantification. These metabolites and clinical parameters are combined into a preliminary age model, which is also limited by interpretability issues linked to multicollinearity (κ=38). To resolve this, a preprocessing step refines the features into a final interpretable age model (κ=16). Finally, SHAP groups are applied for enhanced interpretability, enabling metabolic distortion analysis in disease cohorts.

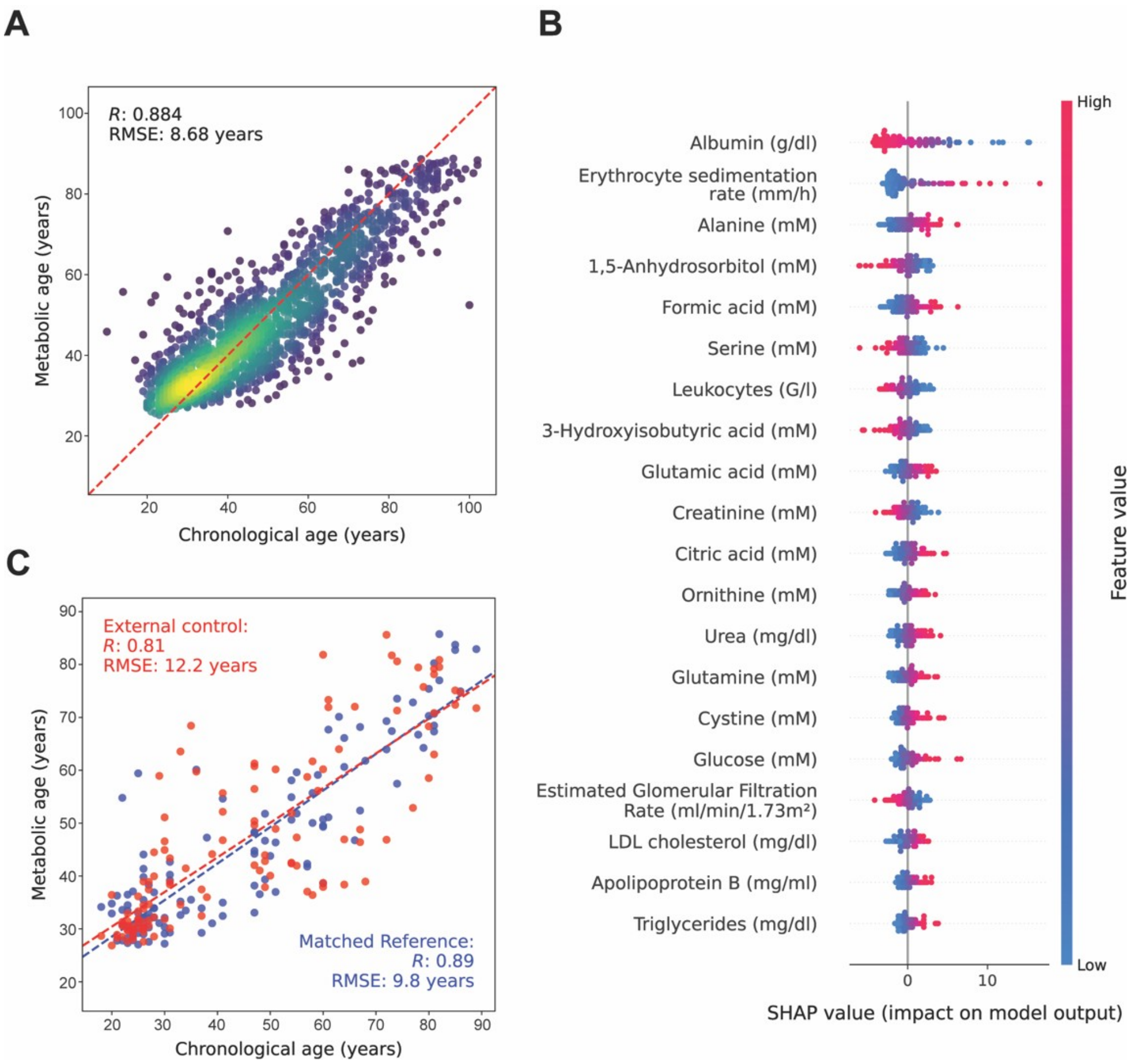

**Figure 2 | Biomarker-based interpretable metabolic age prediction model.** (**A**)**,** Pearson correlation plot for the preprocessed biomarker feature set. The diagonal line represents the identity line, illustrating the minimal regression to the mean (slope = 0.74). (**B**)**,** SHAP summary plot for the interpretable model. Each point corresponds to one sample, positioned along the x-axis by its SHAP value for a given feature; features are ordered by mean absolute SHAP value. Point color indicates the feature value (blue = low, red = high). (**C**)**,** External validation of the model showing chronological versus metabolic age. The external control cohort (Austria) is shown in red, plotted alongside an age-matched reference dataset in blue.

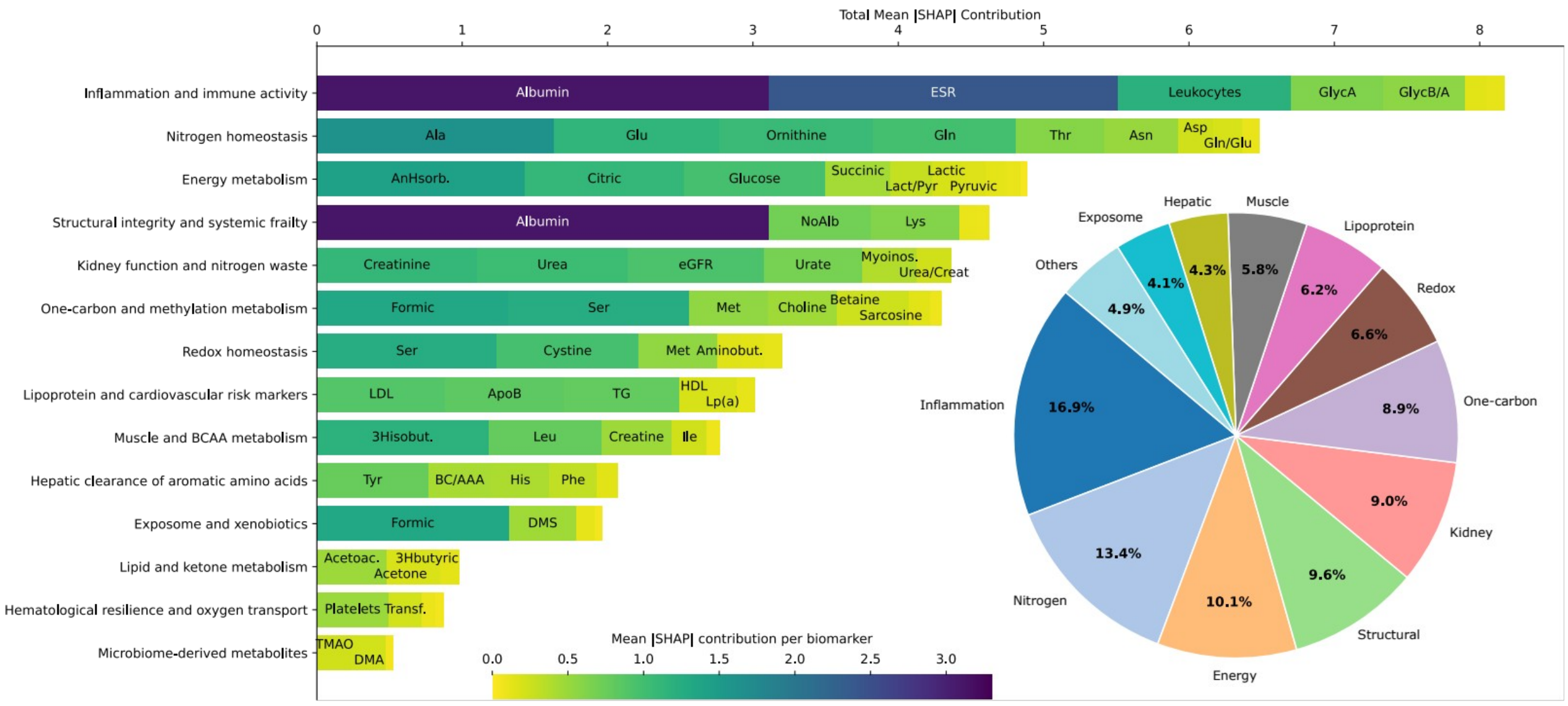

**Figure 3 | Contribution of SHAP groups.** The main plot displays the total SHAP magnitude aggregated by predefined biological or clinical groups. An inset pie chart below illustrates the relative percentage contribution of each group to the overall metabolic age prediction. A complete list of the abbreviations used in this figure is available in the Supporting Information.

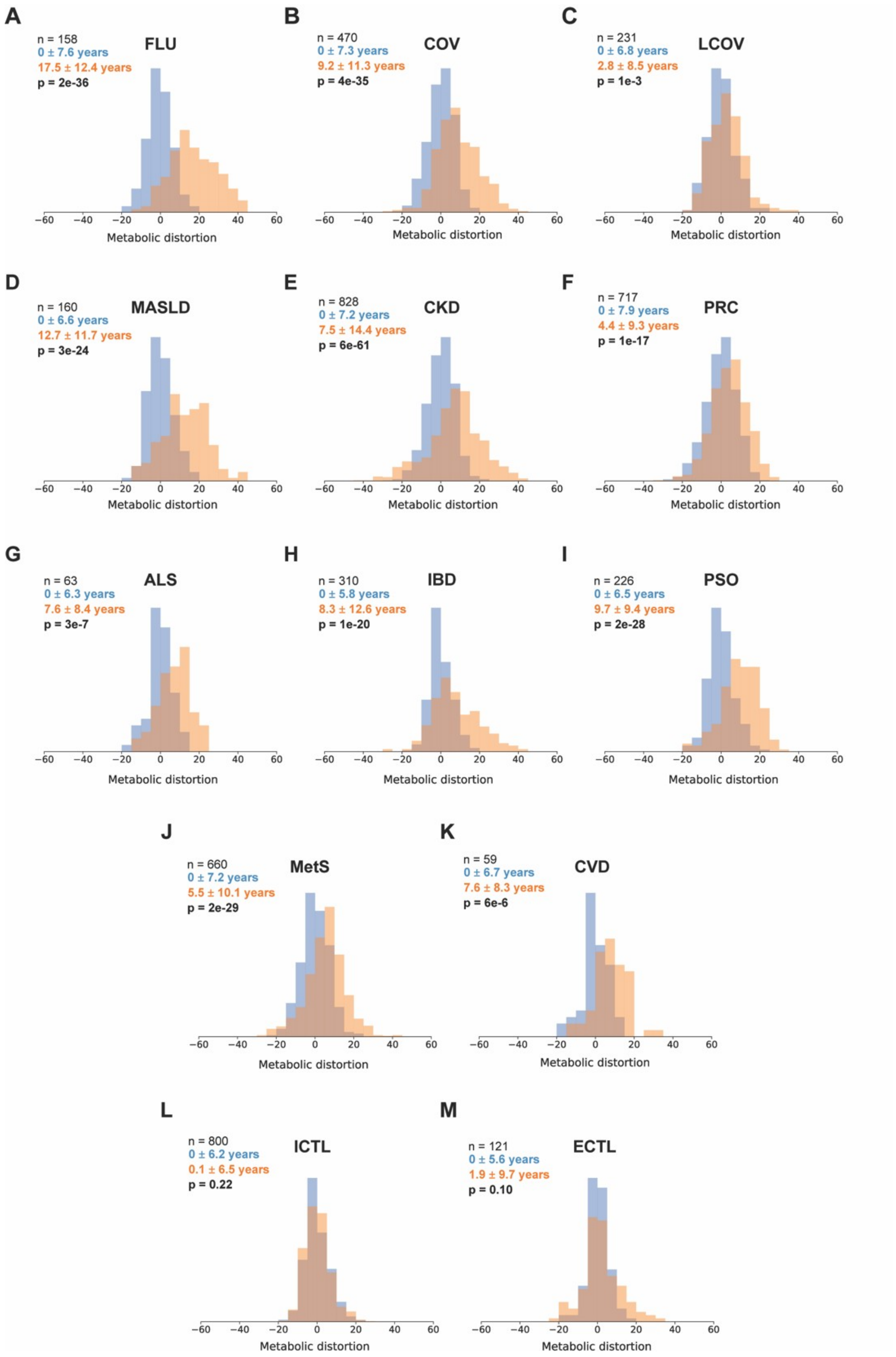

**Figure 4 | Metabolic distortion and SHAP differences across disease cohorts. (A-K),** Histograms of metabolic distortion (deviation from the reference regression line of metabolic age vs. chronological age) for each disease cohort (orange) and its age-matched reference group (blue), which was additionally sex-matched for the PRC cohort. Reference values correspond to out-of-fold predictions obtained during cross-validation, such that each reference individual was predicted by a model that had not been trained on that individual. For each panel, the sample size (n), the mean ± standard deviation of metabolic distortion for both groups, and the corresponding p-values from the Kolmogorov-Smirnov (K-S) test are indicated. **(L-M),** Distribution of metabolic distortion for the internal control (independent model test set) and the external validation cohort (Austria), respectively, showing their alignment with the reference baseline.

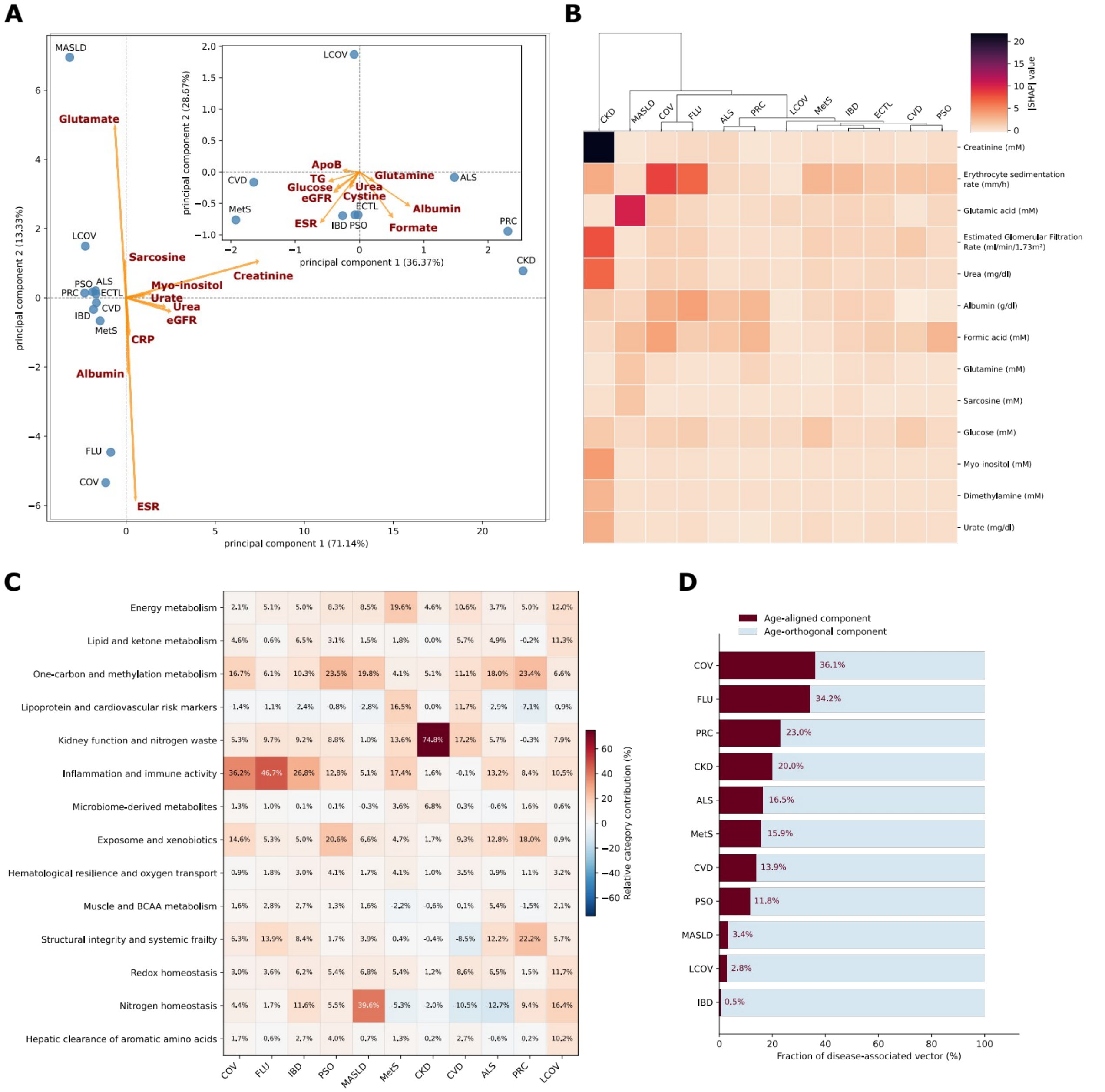

**Figure 5 | Principal component analysis and functional characterization of disease-associated metabolic age signatures. (A),** Principal component analysis (PCA) biplot displaying the metabolic aging signatures of all evaluated disease cohorts relative to the baseline healthy control state (centered at the origin). The analysis is entirely driven by the absolute SHAP (|SHAP|) values extracted from the general population metabolic age model. The loading vectors (arrows) indicate the top 10 most influential metabolites driving the variance across diseases. **Inset:** A focused PCA excluding the highly dispersed cohorts (CKD, MASLD, COV and FLU) to resolve the spatial distribution and metabolic similarities of the remaining tightly clustered diseases. **(B),** Heatmap and hierarchical clustering of the disease cohorts based on their feature importance profiles. Only metabolites or clinical variables with a substantial impact (|SHAP| value > 1.5) in at least one cohort are represented. The dendrogram groups the diseases using Ward's minimum variance method, revealing shared patterns of metabolic deviation from the healthy baseline. Color intensity corresponds to the magnitude of the SHAP values. (**C**) Relative contribution of predefined functional categories to disease-associated changes in SHAP values. For each disease cohort, feature-level changes in mean absolute SHAP values relative to the corresponding age-matched healthy reference group (Δ|SHAP|) were aggregated by functional category. (**D**) Decomposition of disease-associated metabolic alterations into age-aligned and age-orthogonal components relative to physiological metabolic aging. Disease-associated effect vectors were estimated in the standardized space by comparison with age-matched healthy controls, while healthy-aging vectors were derived from nonlinear age- and sex-dependent trajectories estimated in the healthy reference population. The age-aligned component represents the fraction of the disease-associated vector following the healthy-aging direction, whereas the remaining fraction corresponds to alterations orthogonal to physiological aging.

# Supporting Information for

## Mapping metabolic aging and disease-associated acceleration using an interpretable NMR-based clock

Alain Ibañez de Opakua *et al.*

*Corresponding author. Email: omillet@cicbiogune.es

**This PDF file includes:**

Supporting text (extended experimental procedures, cohorts descriptions and metabolites abbreviations)
Figures S1 to S17
Tables S1 to S23

## Supporting text

Extended Experimental Procedures

**Data preprocessing.** Intensity normalization was performed using the electronic reference signal for in-tube calibration (ERETIC) method, which introduces an artificial reference signal in each spectrum with an integral corresponding to a 10 mM $^1$H concentration. Subsequently, chemical-shift alignment was refined by calibrating each spectrum to the alanine $CH_3$ resonance, ensuring that peak positions were harmonized across the dataset. To guarantee that all spectra shared an identical chemical-shift grid, each spectrum was interpolated to a common ppm axis prior to feature extraction. After alignment, a spectral binning optimization procedure was carried out to determine the best trade-off between dimensionality reduction and information retention. For the age model, a final bin width of 0.0115 ppm (n = 1,039 points) was selected, corresponding to a 50-fold reduction of the processed spectral resolution (Figure S16). All NMR data for the clinical-parameter prediction models was preprocessed using an analogous bin optimization strategy.

For alternative modelling based on raw time-domain NMR data, the FID was used directly as input instead of the processed spectrum. As an analogue to spectral binning, each FID was truncated after 512 complex points (with 512 real and 512 imaginary components), representing an approximately 96-fold dimensionality reduction, comparable to the 50-fold reduction achieved by the binning used in the spectral model. This truncation preserved the early, high–signal-to-noise region of the acquisition, which contains most of the informative time-domain dynamics, while discarding late-time points dominated by noise and fine structural information. No additional apodization or zero-filling was applied, thus minimizing any bias and allowing the model to learn directly from the intrinsic FID structure.

**Metabolite quantification.** To resolve overlapping resonances and obtain clear signals for complex metabolites, we utilized two-dimensional J-resolved (J-res) spectroscopy. However, because J-res is inherently non-quantitative, targeted spiking experiments were required to extract specific calibration coefficients. These coefficients were subsequently incorporated into a quantitative pipeline applied to the J-res spectra to calculate the final concentrations for up to 49 metabolites. Conversely, for metabolites exhibiting well-resolved, non-overlapping resonances directly in the 1D NOESY spectra, concentrations were derived from the standard Bruker quantitative profiling (B.I.Quant-PS™) (*22*). This combined strategy allowed us to optimize precision and robustness across metabolites with different spectral characteristics. As metabolite levels were used as model features, the procedure prioritized internal precision and cross-sample consistency over absolute quantification. The quantified metabolites are listed in Table S21.

In addition, special care was taken to resolve specific peak overlaps for trimethylamine-N-oxide (TMAO) and sarcosine. TMAO and betaine share a major resonance at approximately 3.28 ppm, which is the only detectable signal for TMAO, but one of several for betaine. To disentangle their contributions, betaine was first quantified from its well resolved resonance at ~3.91 ppm; its expected contribution to the signal integral at 3.28 ppm was then back-calculated and subtracted to derive the TMAO concentration from the residual signal integral. Sarcosine shows an intense singlet at 2.76 ppm from its $CH_3$ group that overlaps with the triplet signal from an aliphatic fatty acid. Considering the expected 1:2:1 intensity distribution within a triplet then allowed to extrapolate its overlapping central from one well resolved lateral signal integral and subtracting the former to derive the sarcosine concentration from the residual integral.

**Prediction of clinical parameters.** Concomitant with metabolite quantification, we derived a complementary set of 25 clinical biomarkers directly from the quantitative 1D $^1$H spectra using supervised regression models. These were trained to reproduce standard laboratory measurements alongside Bruker-derived quantifications for GlycA, GlycB, and SPC. Pipeline selection and hyperparameter optimization were carried out using the TPOT AutoML framework (*1*). Each model was evaluated by computing the Pearson correlation between NMR estimated values and their corresponding clinical measurements on an independent 20% test subset excluded from the model training. The full list of inferred biomarkers is given in Table S21.

**Balancing strategy.** From a total initial dataset of 29,390 samples from the general population, and to avoid age related sampling biases, the dataset was stratified into five age intervals (7–30, 31–40, 41–50, 51–65, and >65 years). To ensure balanced representation across both age and sex (total $n$ = 8,640), an equal number of participants was randomly selected from the first three intervals. For the oldest category, where women are naturally more represented due to higher life expectancy, the sampling was adjusted to ensure the overall sex balance across the entire dataset. This led to a compensatory imbalance within the older groups, where the smaller number of male participants selected in the >65 age group (735) was offset by selecting a correspondingly larger number of male participants in the 51-65 age group (993), as summarized in Table S23. Figure S17 compares the age distribution of the full dataset with that of the balanced subset.

**Machine learning models for age prediction.** All models were trained using an identical machine learning architecture derived from the TPOT AutoML framework. After feature standardization, predictions were generated through a stacking ensemble combining a Ridge linear regression (*2*) model with an ExtraTrees regressor (*3*). For each modeling strategy, the dataset was partitioned such that 20% of the samples were excluded from the training to lateron serve as an independent test set for final evaluation. The remaining 80% of the data was used for training and internal validation through five-fold cross-validation.

For the iterative negative control approach (Figure S6), a pool of initially discarded samples ($n$ = 20,750) was used in 1,000 iterations of 1,000 healthy discarded samples each. For each iteration, a new specific sub-cohort from the reference data was dynamically drawn to precisely match the age distribution of the discarded subset, and their metabolic distortions were computed. Two-sample Kolmogorov-Smirnov tests revealed an exceptional structural alignment between the discarded samples and their matched reference counterparts, as characterized by an insignificantly small mean D-statistic (Figure S6B) and a vast majority of non-significant iterations (Figure 5C).

**Model interpretability.** Feature relevance and model behavior were analyzed using SHAP explanations (*4*). To obtain a stable and biologically meaningful interpretation focused on the magnitude of influence, individual SHAP values were aggregated into predefined SHAP groups corresponding to specific metabolic pathways and clinical domains (Table S22). The SHAP value for each group was calculated as the sum of the absolute SHAP values of all its constituent features. Although group size can influence the accumulated SHAP magnitude, this effect is limited when individual features contribute little to prediction, while the more problematic inflation caused by large opposing SHAP values from correlated predictors was minimized through the preceding multicollinearity-reduction step.

**Software and computational environment.** All data processing, automated machine learning, and statistical analyses were implemented in Python (v. 3.9.17). The core computational environment utilized NumPy (v. 1.23.5) and Pandas (v. 1.5.3) for data manipulation, nmrglue (v. 0.9) for NMR data extraction and processing, scikit-learn (v. 1.4.1.post1) and TPOT (v. 0.12.2) for machine learning workflows, SHAP (v. 0.42.1) for model interpretation, and Matplotlib (v. 3.7.2) for data visualization. To guarantee strict cross-platform reproducibility, the complete analytical pipelines for both metabolite quantification and the prediction of clinical parameters were encapsulated within a custom Docker container based on a lightweight Debian environment.

## Description of the cohorts

The **Mondragón Corporation** cohort (Osarten and Akribea) consists of 23,909 individuals employed within the Mondragón Corporation, a federation of worker cooperatives in the Basque Country (Spain). Participants were recruited during routine annual occupational health examinations. The cohort covers a broad spectrum of working-age adults across diverse industrial, financial, retail, and knowledge-based sectors, offering a heterogeneous but population-representative sample. The study was approved by the Ethics Committee of Research with Medication of Euskadi (approval numbers: CEIC-E 16-114 and CEIC-E 19-13).

The **DDM-Madrid** cohort comprises 933 women aged 39–50 years who underwent gynecological evaluations at the Madrid Salud Diagnostic Center between 2013 and 2015. Recruitment was conducted via telephone invitations. This cohort offers a focused snapshot of mid-life female metabolic profiles. The study was approved by the Ethics and Animal Welfare Committee of this Institute and all participants signed an informed consent (approval number: CEI PI 02_2012-v2).

The **Biosilver** cohort includes 686 older adults recruited from geriatric centers, hospitals, and community health facilities across Spain. All participants provided fasting serum samples collected between 2023 and 2025. This cohort enriches the dataset at advanced ages, a critical segment for modeling late-life metabolic changes. The study was approved by the Ethics Committee of Research with Medication of Euskadi (approval number: CEI: PI2023063).

The **AGEPORTUGAL** cohort consists of 246 elderly participants recruited from geriatric institutions in Portugal. This cohort contributes additional geographic diversity within the Iberian Peninsula and strengthens representation of advanced age ranges. The study was approved by the Ethics Committee of University of Beira Interior (approval number: CE-UBI-Pj-2017–012).

The **SPBB** cohort contains 3,480 samples sourced from several national biobanks belonging to the ISCIII Biobanks and Biomodels Platform. Samples were selected specifically to reinforce representation of older age groups that were underrepresented in occupational cohorts, thereby improving the age distribution of the overall dataset. Additionally, a subset of 458 samples meeting the criteria for metabolic syndrome was specifically selected from this cohort to form part of a new, distinct disease group (**MetS**). Each biobank followed standard operating procedures and obtained appropriate approval from their respective Ethics and Scientific Committees. The contributing biobanks are as follows: 1. HCB-IDIBAPS Biobank (Registry No. B00575) - 2. Valdecilla Biobank (Registry No. B00162) from the Cantabria Cohort collection - 3. BioBank Hospital Ramón y Cajal-IRYCIS (Registry No. B00678, integrated in PT23/00098) - 4. CIBERES Pulmonary Biobank Consortium (Registry No. B00140, integrated in PT23/00102), a network of twelve tertiary Spanish hospitals - 5. Biobank of the Aragon Health System (Registry No. B00873, integrated in PT20/00112) - 6. Biobank IBSP-CV (integrated in PT23/00034 and the Valencian Biobanking Network). All samples and data were processed in compliance with ethical and scientific standards and approved by the Ethics Committee of Research with Medication of Euskadi (approval number: CEIC-E 21-199).

The **external control (ECTL)** cohort consists of 121 individuals serving as controls in various studies conducted at the Medical University of Graz (Group of Tobias Madl). All participants were included as healthy or negative controls according to the protocols of their respective original studies(*5*). These studies were conducted with the approval of the relevant institutional review boards and ethics committees, specifically: the Ethics Committee of North-West and Central Switzerland (EKNZ-2015-351) and the Ethics Committee of the Medical University of Graz (EK 32-475 ex 19/20). All subjects provided informed consent in accordance with the Declaration of Helsinki. Notably, all samples in this cohort were processed and measured at the Medical University of Graz (Group of Tobias Madl). These samples were specifically included to evaluate and account for potential cohort effects, as the analytical procedures were conducted in a different laboratory setting from the rest of the study.

The **LA2** cohort contains individuals from a prospective frailty study in the Extremadura region, Spain, specifically including patients with cardiac insufficiency aged over 65 years. From this overall cohort, 89 samples meeting the criteria for metabolic syndrome were selected to be included in the disease group (**MetS**). All samples and data were processed in compliance with ethical standards, and the study was approved by the Ethics Committee of Research of Badajoz on February 14th, 2022.

The **LiverBible** cohort contains apparently healthy blood donors from Italy aged 40–65 years who were enrolled up to July 2021 for comprehensive liver disease, metabolic, and cardiovascular screening. Participants were selected based on the presence of at least three criteria of altered metabolic regulation: being overweight (BMI >25 kg/m$^2$), arterial hypertension, hyperglycaemia, low HDL cholesterol, and increased triglycerides. All individuals were negative for HBV and HCV markers and reported no excessive alcohol use. For the present study, a subset of 113 samples fulfilling the diagnostic criteria for metabolic syndrome was specifically selected to be included in the disease group (**MetS**). The study protocol was approved by the Ethical Committee of the Fondazione IRCCS Ca' Granda (approval number: CE 125_2018bis).

The **Covid-19** cohort (**COV**) consists of 470 adults hospitalized during the acute phase of COVID-19 infection. Patients were recruited from two hospitals in the Basque Country, Hospital Universitario Cruces and Hospital Universitario Basurto, during the 2020 COVID-19 pandemic. Diagnostic and inclusion criteria were based on the presence of symptoms compatible with COVID-19 and confirmation of SARS-CoV-2 infection by RT-PCR on nasal swab samples. Serum samples were collected during the acute phase of SARS-CoV-2 infection. All samples were provided by the Basque Biobank for Research (BIOEF). The study was approved by the corresponding ethics committee (CEIC-E 20-26).

The **Long Covid-19** cohort (**LCOV**) consists of 231 adults presenting persistent symptoms associated with post-acute COVID-19 syndrome. Patients were recruited from hospitals in the Basque Country (Hospital Universitario Cruces, Hospital Universitario Basurto, and Hospital Universitario Txagorritxu), and serum samples were provided by the Basque Biobank for Research (BIOEF), as well as from Madrid (BlueHealthCare Clinic) and Zaragoza. Diagnostic and inclusion criteria were based on the presence of symptoms compatible with Long COVID and a previous confirmed diagnosis of SARS-CoV-2 infection. The study was evaluated and approved by the corresponding ethics committee (PI2020215).

The **Influenza** cohort (**FLU**) consists of 158 adults presenting symptoms compatible with influenza infection. Diagnostic and inclusion criteria were based on the presence of clinical symptoms of influenza and a confirmed diagnosis of influenza A (H1N1 or H3N2) or influenza B infection. All serum samples were provided by the Basque Biobank for Research (BIOEF). The study was approved by the corresponding ethics committee (PI+CES-BIOEF 2023-13).

The **Inflammatory Bowel Disease** cohort (**IBD**) consists of 310 adults, diagnosed with inflammatory bowel disease (IBD), encompassing both Crohn's disease and ulcerative colitis. 189 of that patients were recruited across 16 Spanish hospitals between 2017 and 2019 following ECCO diagnostic criteria. Only treatment-naïve individuals at the time of diagnosis were included, while those with prior IBD therapy, other immune-mediated conditions, active infection, neoplasms, unrelated immunosuppressive treatment, or pregnancy/lactation were excluded. Fasting serum samples were collected at diagnosis, with written informed consent obtained from all participants. The other 121 patients were included from the EXIT trial, a prospective, multicentre, randomised (1:1), controlled study with a quadruple-blind design (patients, physicians, data managers, and statisticians), conducted across 33 inflammatory bowel disease (IBD) units in Spain. Adult patients with IBD (Crohn's disease or ulcerative colitis) in sustained clinical remission for at least 6 months under standard-dose anti-TNF therapy (infliximab or adalimumab), in combination with stable immunomodulator treatment, were eligible. The study protocol was approved by the Clinical Research Ethics Committee of the Hospital Universitario de La Princesa (Register number: 2728).

The **Psoriasis** cohort (**PSO**) includes 226 patients with moderate-to-severe plaque psoriasis. Patients were consecutively recruited from the Dermatology Department at Hospital Universitario de La Princesa. Blood samples were collected, following written informed consent approved by the institutional ethics committee (CEIm, Hospital Universitario de La Princesa; approval numbers 3519, 4585, and 5103). Samples are from the private collection C.0005113, registered in the Spanish National Biobank Registry.

The **Amyotrophic Lateral Sclerosis** cohort (**ALS**) comprise a total of 63 patients diagnosed with amyotrophic lateral sclerosis who were evaluated at the ALS Unit of Basurto University Hospital. Serum samples were obtained at the time of lumbar puncture, performed at disease diagnosis. Patients with either rapid progression or slow progression were included. Samples were collected in strict compliance with ethical and biosafety regulations, and their use for research purposes was approved by the Basque Country Research Ethics Committee for Medicinal Products (CEIm-E), under approval number PI2025129.

The **Chronic Kidney Disease** (FRIAT) cohort (**CKD**) consists of 828 adult patients enrolled in a hemodialysis program for a minimum of three months. Inclusion criteria comprised patients aged over 18 years with a diagnosis of chronic kidney disease undergoing maintenance hemodialysis. Crucially, all blood draws were collected post-prandially (non-fasting) and prior to the hemodialysis session. All serum samples were provided by the Fundación Renal Íñigo Álvarez de Toledo. The study was approved by the corresponding ethics committee (PIC126-24_FRIAT and EO145-25_FRIAT).

The **Metabolic Dysfunction-Associated Steatotic Liver Disease** (LITMUS, Liver Investigation: Testing Marker Utility in Steatohepatitis) cohort (**MASLD**) comprises 160 individuals with biopsy-confirmed metabolic dysfunction-associated steatotic liver disease recruited from tertiary liver centres in Italy (including Milan, Turin, and Palermo) and the United Kingdom (Newcastle upon Tyne). The study was approved by the UK Health Research Authority (NRES Committee North East – Newcastle & North Tyneside 1; Reference: 17/NE/0007).

The **Prostate Cancer** cohort (**PRC**) comprises a total of 1,032 serum samples collected at Basurto University Hospital (Spain) and managed by the Basque Biobank for Research (BIOEF). Among these individuals, 717 were diagnosed with prostate cancer, while the remaining 315 presented benign prostatic hyperplasia (BPH). Given the benign nature of BPH, samples from these individuals were incorporated into the general population dataset, whereas the prostate cancer cases were retained as a disease-specific subgroup. This structure enables the model to capture metabolic aging deviations associated with oncological processes while maintaining adequate representation of non-malignant urological conditions. The study was approved by the Ethics Committee of Research with Medication of Euskadi (approval numbers: CEIC-E 11-12, 14-14 and 19-20).

The **Cardiovascular Disease** cohort (**CVD**) is a prospective study designed to identify early risk factors for cardiovascular disease. The cohort specifically comprises 59 individuals who had no clinical diagnosis of any cardiovascular disease at the time of biological sample collection but who subsequently developed CVD during follow-up (weeks or years later). The main inclusion criterion was the absence of CVD at recruitment with a later diagnosis of the disease. Exclusion criteria included the presence of any autoimmune or oncological disease at the time of inclusion or a diagnosis of such conditions within the three months prior to recruitment. Samples and data from patients included in this study were provided by the Biobank of the Aragon Health System. All procedures were conducted in accordance with ethical standards and received approval from the Ethics Committee of Research with Medication of Euskadi (approval number: CEIC-E 21-199).

<u>List of abbreviations for metabolites and clinical markers</u>

- **2Hbutyric:** 2-Hydroxybutyric acid
- **3Hbutyric:** 3-Hydroxybutyric acid
- **3Hisobut.:** 3-Hydroxyisobutyric acid
- **Acetoac.:** Acetoacetic acid
- **Ala:** Alanine
- **Aminobutyric:** 2-Aminobutyric acid
- **AnHsorb.:** 1,5-Anhydrosorbitol
- **ApoB:** Apolipoprotein B
- **Arg:** Arginine
- **Asn:** Asparagine
- **Asp:** Aspartate
- **BC/AAA:** BCAA to AAA (Phe+Tyr) ratio
- **Ca:** Calcium
- **CRP:** C-reactive protein
- **D-Gal:** D-Galactose
- **DMA:** Dimethylamine

- **DMG:** N,N-Dimethylglycine
- **DMS:** Dimethylsulfone
- **eGFR:** Estimated Glomerular Filtration Rate
- **ESR:** Erythrocyte sedimentation rate
- **Gln:** Glutamine
- **Glu:** Glutamic acid
- **Gln/Glu:** Glutamine to glutamate ratio
- **Gly:** Glycine
- **GlycB/A:** GlycB to GlycA ratio
- **HDL:** HDL cholesterol
- **His:** Histidine
- **Ile:** Isoleucine
- **Lact/Pyr:** Lactate to pyruvate ratio
- **LDL:** LDL cholesterol
- **Leu:** Leucine
- **Lp(a):** Lipoprotein(a)
- **Lys:** Lysine
- **Met:** Methionine
- **Myoinos.:** Myo-inositol
- **NoAlb:** Non-albumin protein
- **Oxoglutaric:** 2-Oxoglutaric acid
- **Phe:** Phenylalanine
- **Pro:** Proline
- **Ser:** Serine

- **Succinic:** Succinic acid
- **TG:** Triglycerides
- **Thr:** Threonine
- **TMAO:** Trimethylamine-N-oxide
- **Transf.:** Transferrin
- **Tyr:** Tyrosine
- **Val:** Valine

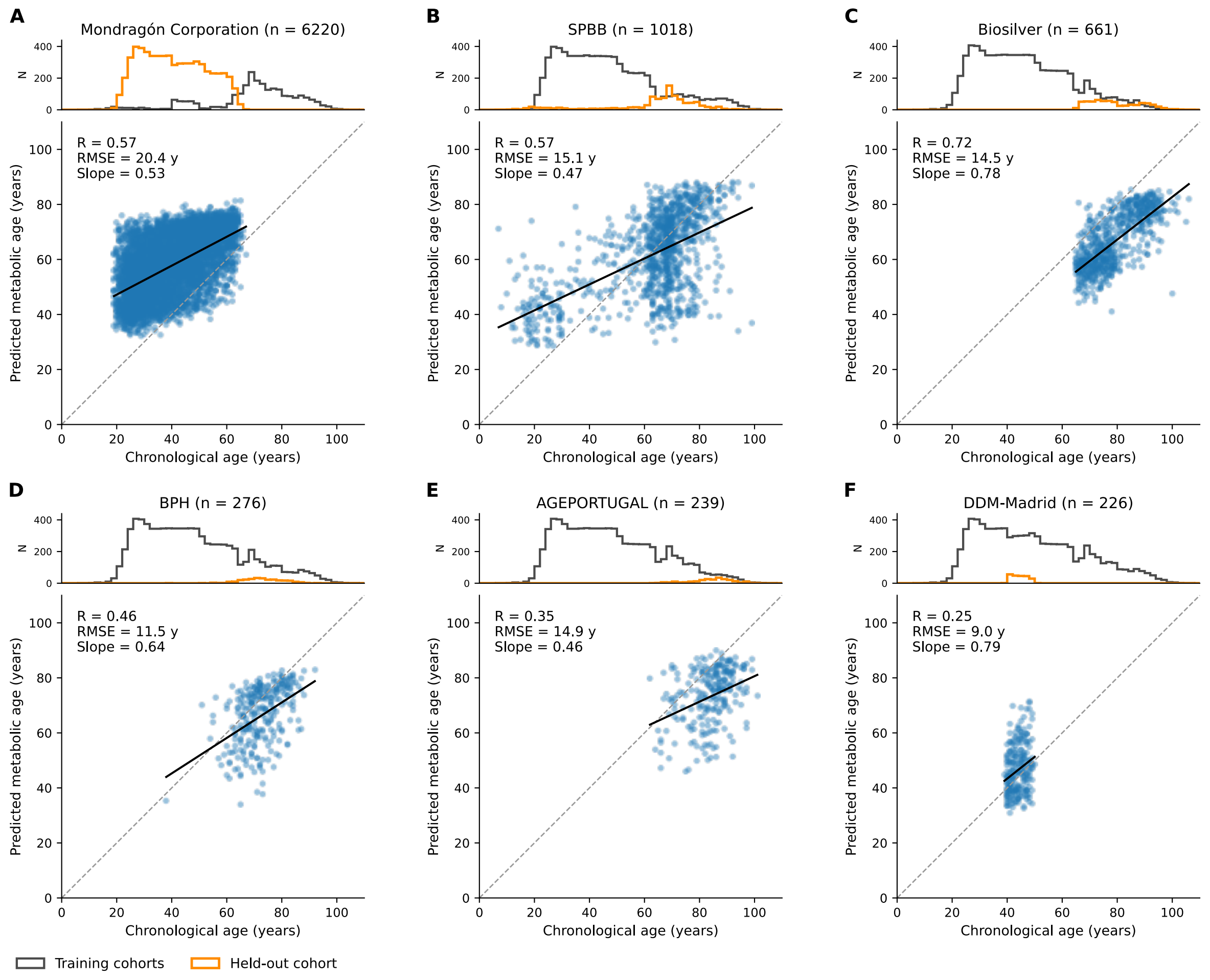

**Figure S1 | Leave-one-cohort-out validation of the interpretable metabolic aging model.** The model was sequentially trained on five of the six constituent cohorts and evaluated exclusively on the remaining held-out cohort. Cohorts are ordered by decreasing size of the held-out set. For each cohort, the upper panel shows the chronological-age distributions of the training population (black) and the held-out cohort (orange) as raw sample counts, illustrating the heterogeneous contribution of individual cohorts to different regions of the lifespan. The lower panel shows predicted metabolic age versus chronological age in the corresponding held-out cohort; dashed lines indicate identity and solid lines indicate linear regression fits. Sample size, Pearson correlation coefficient (R), root mean squared error (RMSE), and regression slope are reported within each panel.

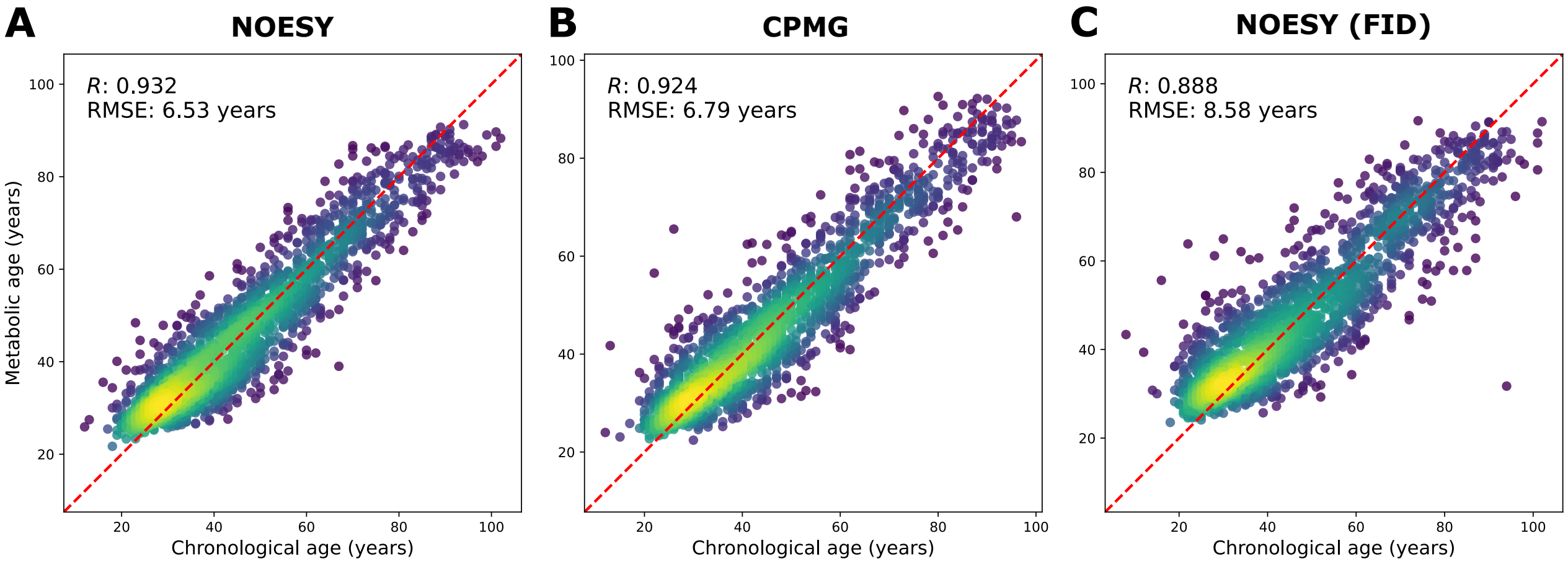

**Figure S2 | NMR signal-based metabolic age prediction models. (A),** Predicted metabolic age versus chronological age for the NOESY-based model, the CPMG-based model **(B)** and the NOESY FID model **(C)**. The solid red dashed line represents the line of identity (y=x). The performance metrics, Pearson's correlation coefficient (r) and Root Mean Square Error (RMSE) in years, are shown in each subplot. The color gradient of the data points, from dark blue (low density) to bright yellow (high density), represents the local concentration of samples.

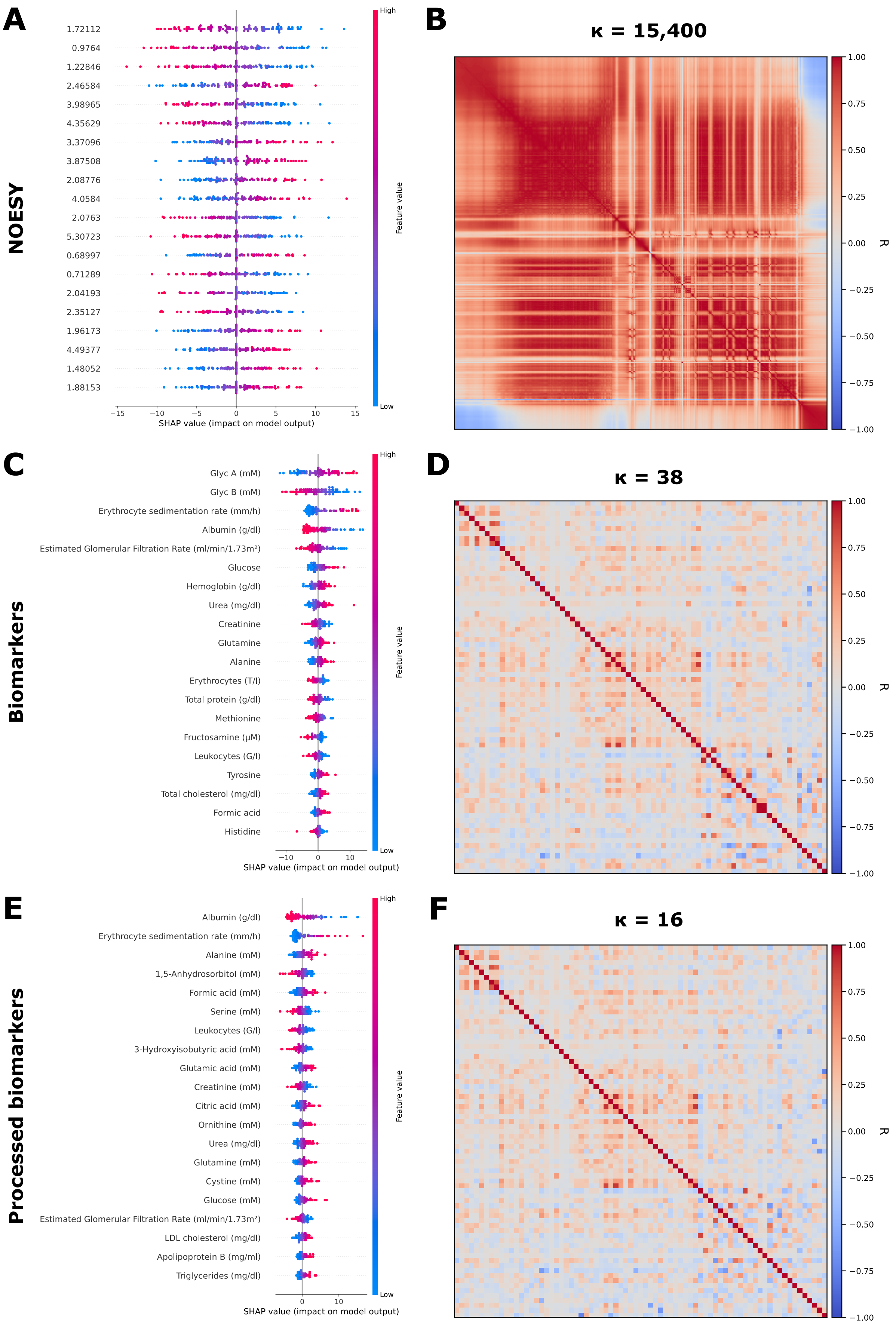

**Figure S3 | See next page for caption.**

**Figure S3 | Multicollinearity and SHAP value structure across modeling frameworks.** The panels display the evaluation of three feature sets: binned 1D NOESY spectra (**A**, **B**), quantified biomarkers (**C**, **D**), and biomarkers after preprocessing to reduce multicollinearity and enrich informational independence (**E**, **F**). SHAP summary plots (**A**, **C**, **E**) illustrate the distribution and magnitude of the top 20 feature contributions to age prediction (color scale from blue to red). The corresponding full Pearson correlation matrices (**B**, **D**, **F**), computed using all numeric features, reveal the extent and structure of inter-feature correlations. The NOESY dataset exhibits large contiguous blocks of highly correlated bins, whereas biomarker-based representations show progressively reduced collinearity following preprocessing.

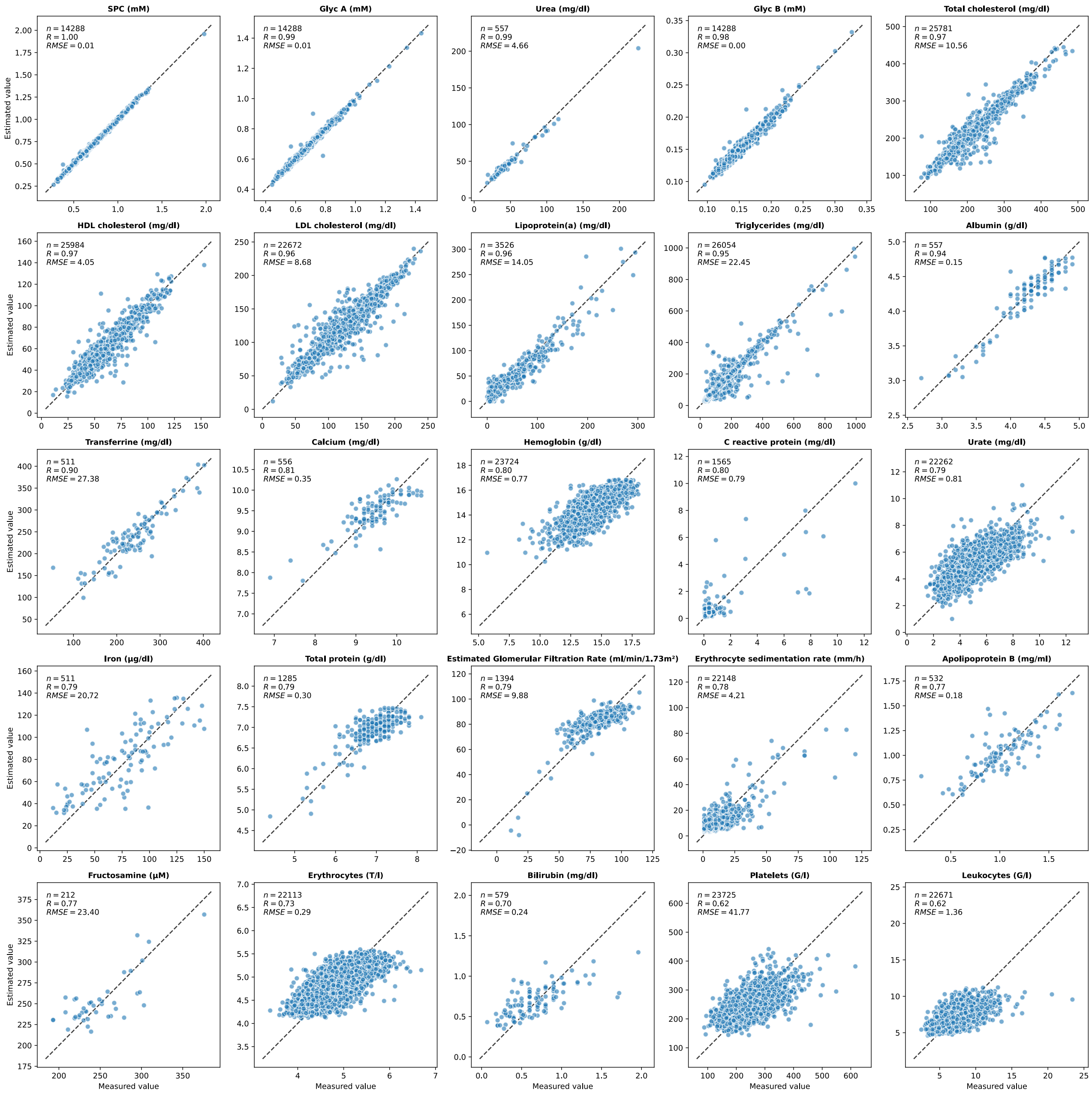

**Figure S4 | Performance of NMR-based clinical biomarker prediction models.** Each scatter plot compares the NMR-estimated value (y-axis) with the matched laboratory measurement (x-axis) in the independent 20% test set. The Pearson correlation coefficient (R), the root-mean-square error (RMSE), and the sample size (n) are shown within each panel.

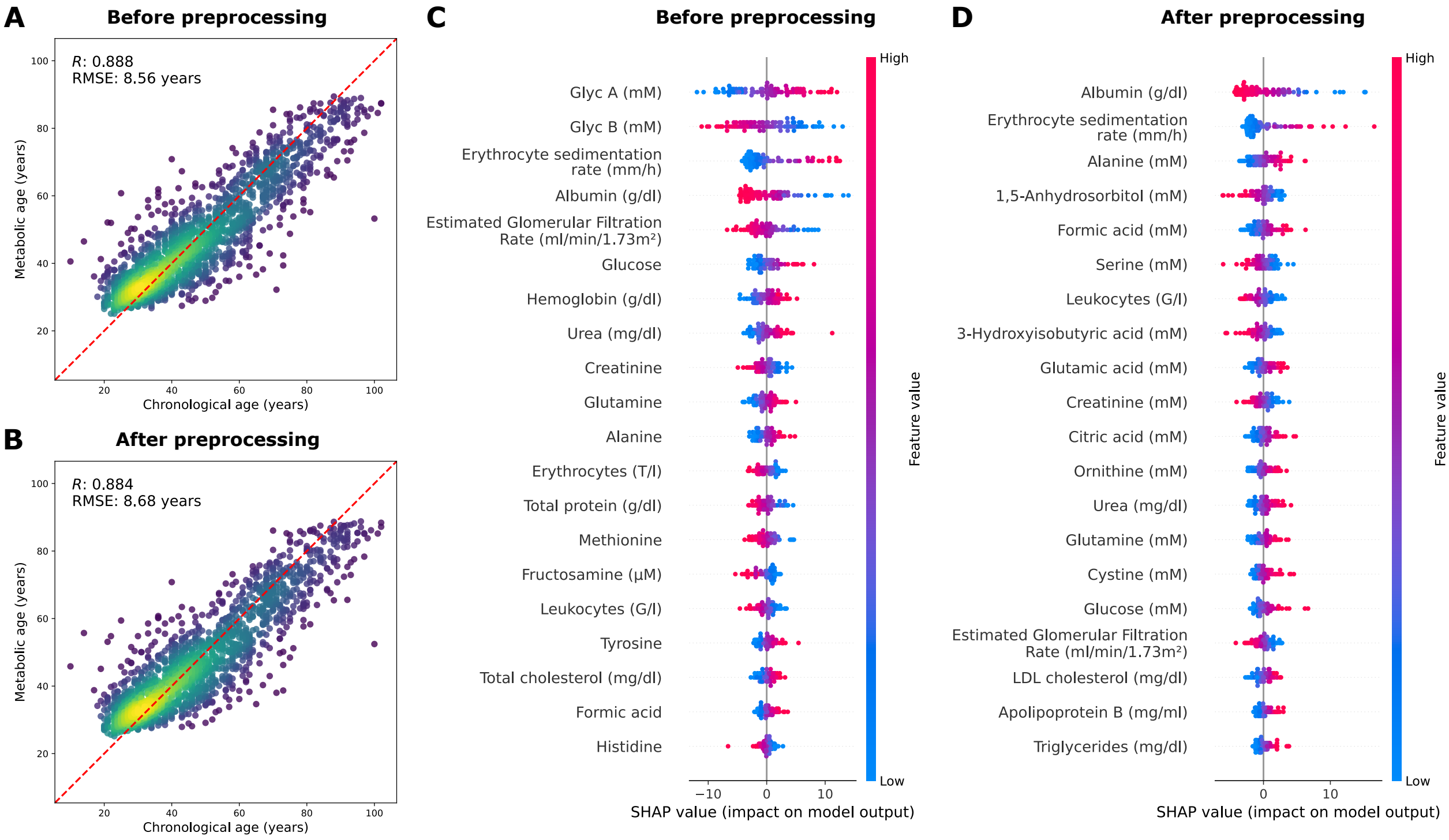

**Figure S5 | Biomarker-based metabolic age prediction models.** Pearson correlation matrices for the biomarker-based model before preprocessing (**A**) and after preprocessing (**B**). At the right, SHAP summary plots for the same two models: unprocessed (**C**) and preprocessed (**D**). Each point corresponds to one sample, positioned along the x-axis by its SHAP value for a given feature; features are ordered by mean absolute SHAP value. Point color indicates the feature value (blue = low, red = high).

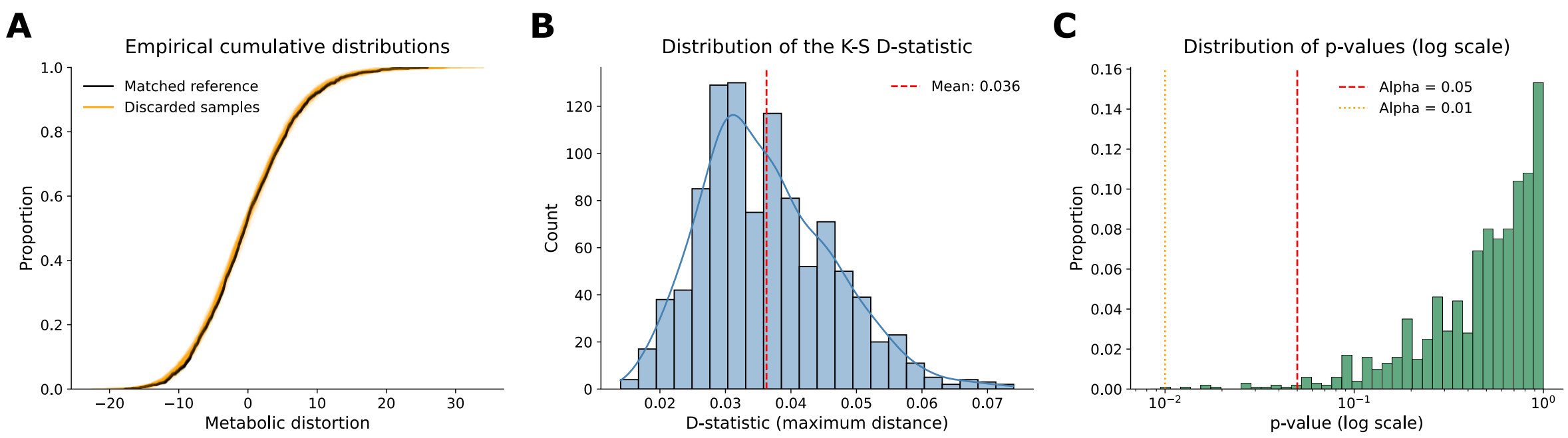

**Figure S6 | Validation of metabolic distortion using an iterative negative control. (A),** Empirical cumulative distribution functions (ECDF) comparing the metabolic distortion of 1,000 random subsets of previously discarded healthy samples (orange, N = 1,000 per subset) against their respectively age-matched reference sub-cohorts (black). (**B),** Distribution of the Kolmogorov-Smirnov D-statistic across the 1,000 iterations, representing the maximum distance between the ECDF curves. **(C),** High-resolution distribution of p-values plotted on a logarithmic scale.

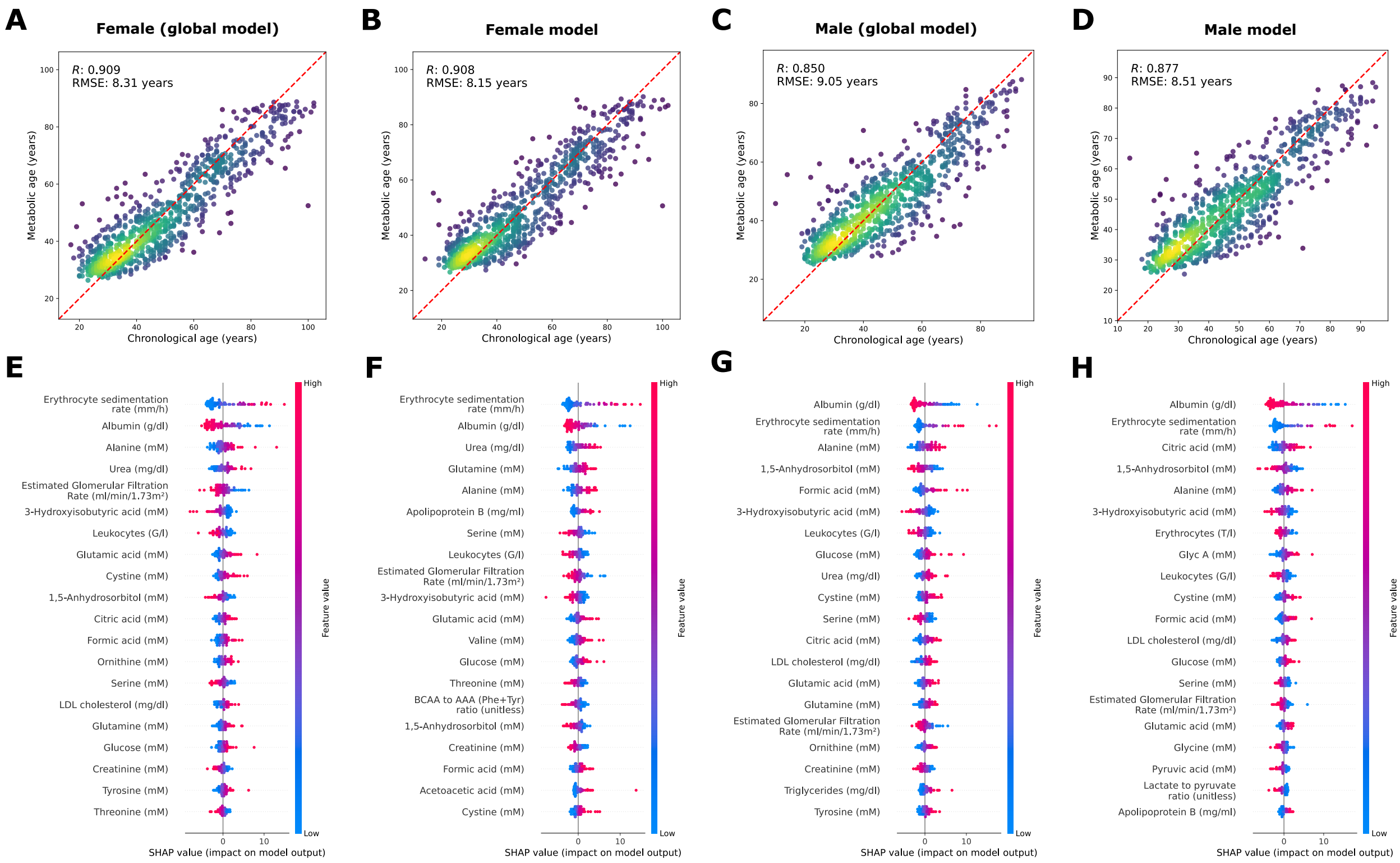

**Figure S7 | Comparison of predictive performance and feature contributions between unified and sex-specific models for metabolic age prediction.** Correlation plots (**A-D**) display predicted versus chronological age, along with the corresponding Pearson correlation coefficient (r) and root-mean-square error (RMSE), for the unified model applied to females (**A**), the female-only model (**B**), the unified model applied to males (**C**), and the male-only model (**D**). SHAP summary plots (**E-H**) illustrate the contribution of the top 20 features to the predicted metabolic age for the same four models, respectively. In the SHAP plots, features are ordered by mean absolute SHAP value, and point color represents relative feature magnitude (low to high).

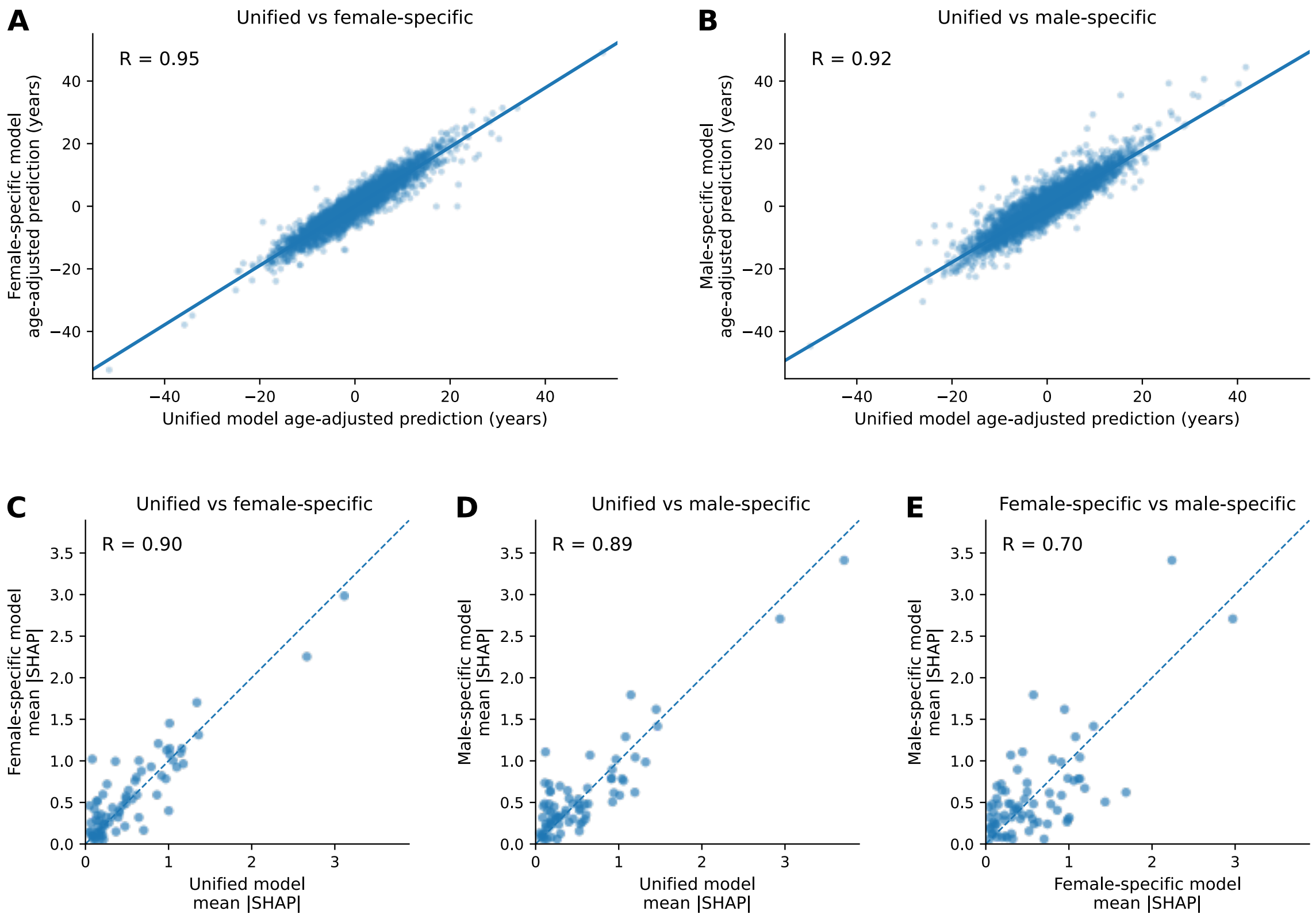

**Figure S8 | Agreement between unified and sex-specific metabolic aging models at the individual-prediction and feature-importance levels.** (**A–B**) Agreement between out-of-sample metabolic age predictions from the unified and sex-specific models after adjustment for chronological age. Predictions from each model were independently residualized against chronological age before calculating the correlation. (A) Unified versus female-specific model (R = 0.95). (B) Unified versus male-specific model (R = 0.92). (**C–E**) Feature-level comparison of mean absolute SHAP values across models, with each point representing one of the 75 model features. (C) Unified versus female-specific model (R = 0.90). (D) Unified versus male-specific model (R = 0.89). € Male-specific versus female-specific models (R = 0.70). For panels C and D, SHAP values from the unified and sex-specific models were calculated using identical subsets of 100 male or female participants, respectively. Dashed lines indicate equality (y = x).

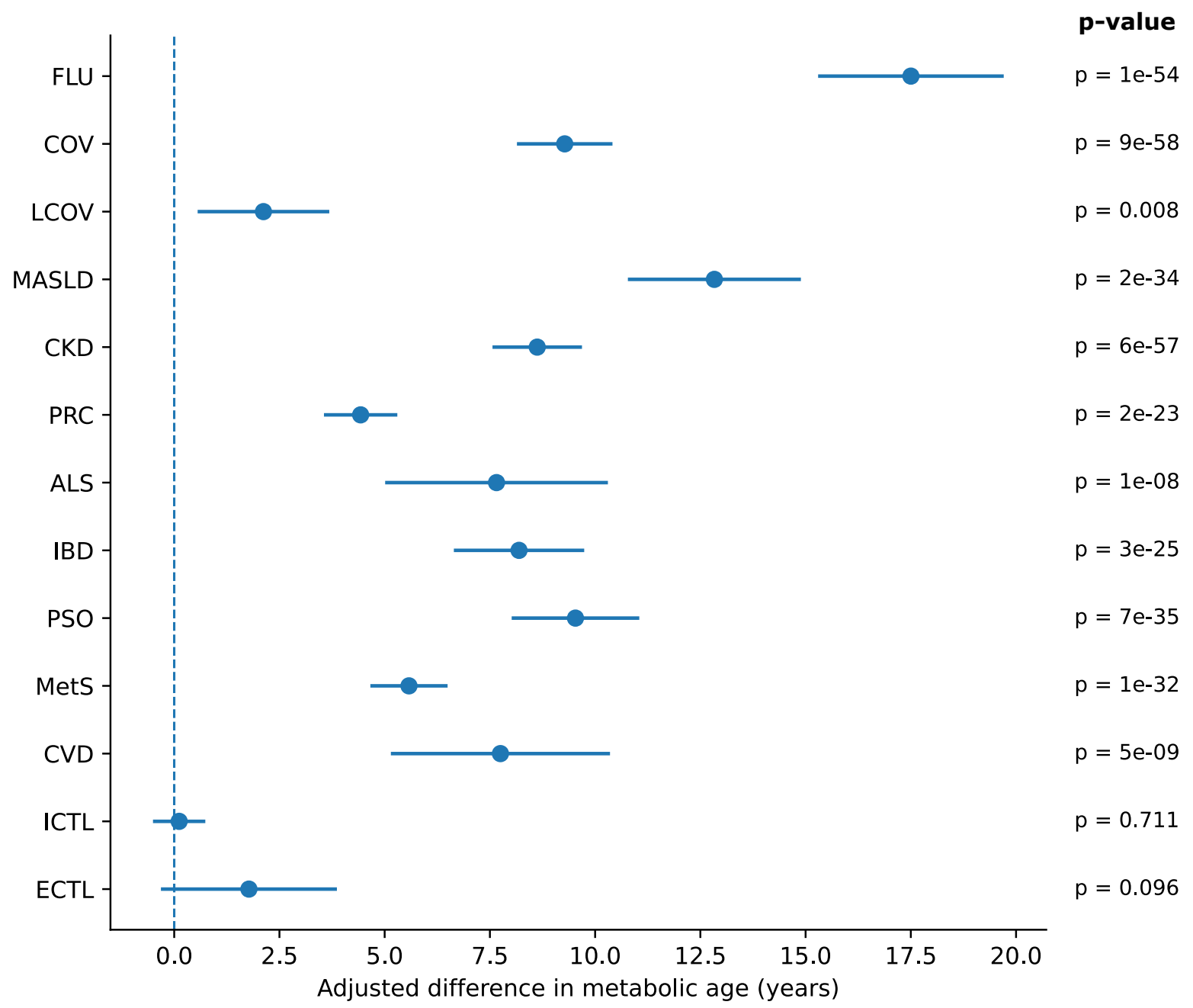

**Figure S9 | Adjusted disease-associated differences in predicted metabolic age**. For each disease cohort, a linear regression model was fitted using predicted metabolic age as the outcome and chronological age, sex, and disease status as predictors, using the same age-matched out-of-sample reference individuals employed in the distributional analyses. Points indicate the regression coefficient for disease status, expressed as the adjusted difference in metabolic age in years relative to the corresponding reference population; horizontal lines indicate 95% confidence intervals. Heteroskedasticity-robust HC3 standard errors were used. Sex was omitted only for the prostate cancer cohort because all individuals were male. Internal (ICTL) and external (ECTL) control cohorts are shown as negative controls.

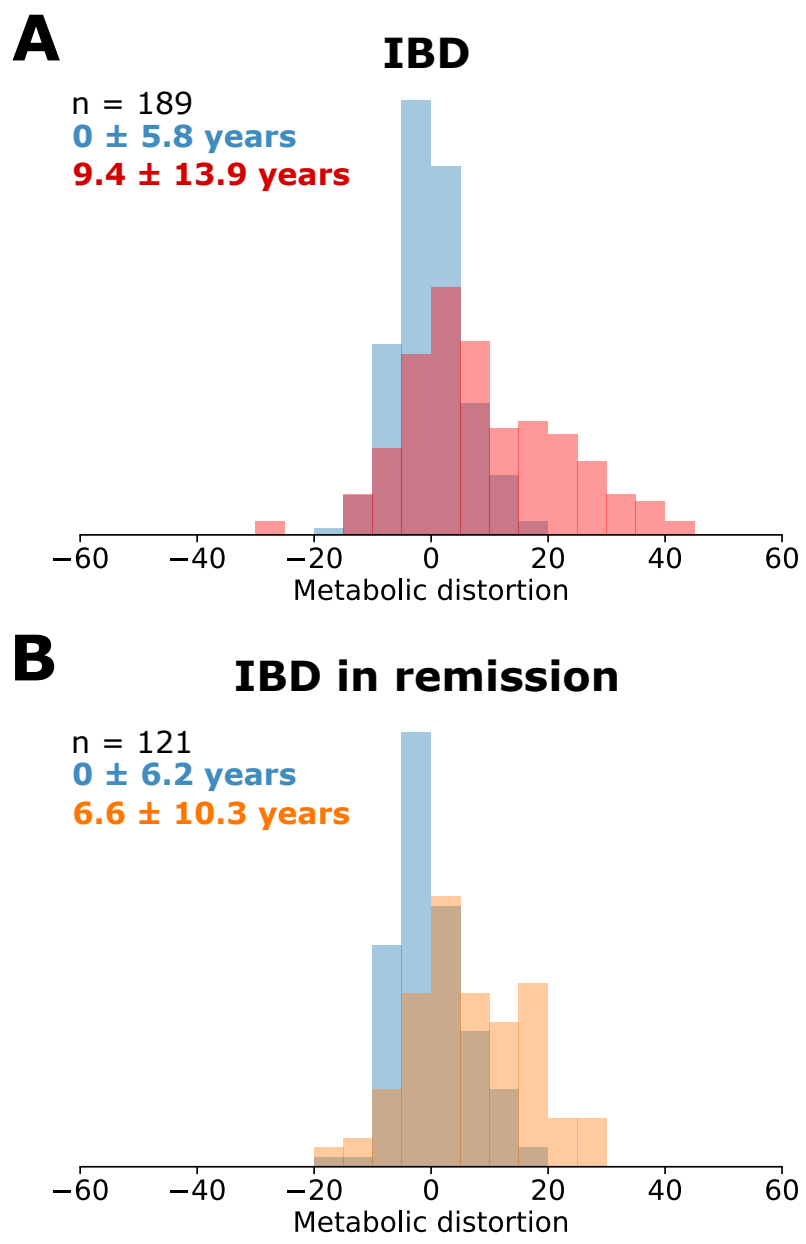

**Figure S10 | Metabolic distortion in the inflammatory bowel disease cohort based on clinical remission. (A),** Histogram of metabolic distortion (deviation from the reference regression line of metabolic age vs. chronological age) for the IBD cohort patients with active disease (red) and its age-matched reference group (blue). (**B),** Histogram of metabolic distortion for the subgroup of IBD patients in clinical remission (orange) and their age-matched reference group (blue). For each panel, the sample size (n) and the mean ± standard deviation of metabolic distortion for both groups are indicated.

<<See FigureS11.pdf file>>

**Figure S11 | Individual SHAP waterfall plots for all 59 participants from the prospective cardiovascular cohort.** Each panel corresponds to a single participant who subsequently experienced a cardiovascular event. The plots illustrate the decomposition of the model's prediction for each individual, showing how specific features push the metabolic age upward (red bars, positive contribution) or pull it downward (blue bars, negative contribution). The final predicted metabolic age represents the cumulative sum of the model's baseline value and these individual feature contributions.

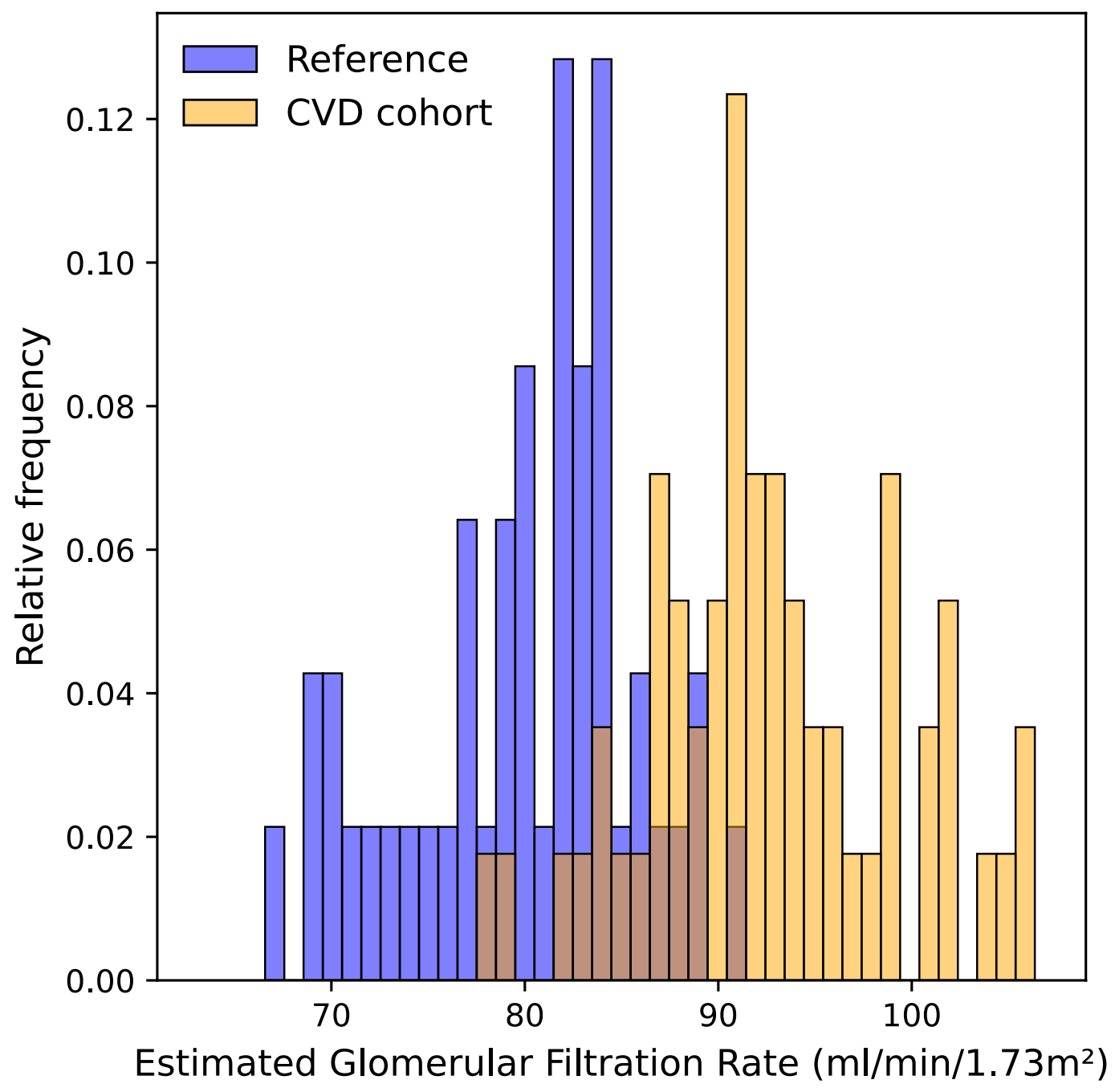

**Figure S12 | Distribution of estimated glomerular filtration rate (eGFR) in the reference cohort and the cardiovascular disease (CVD) cohort.**

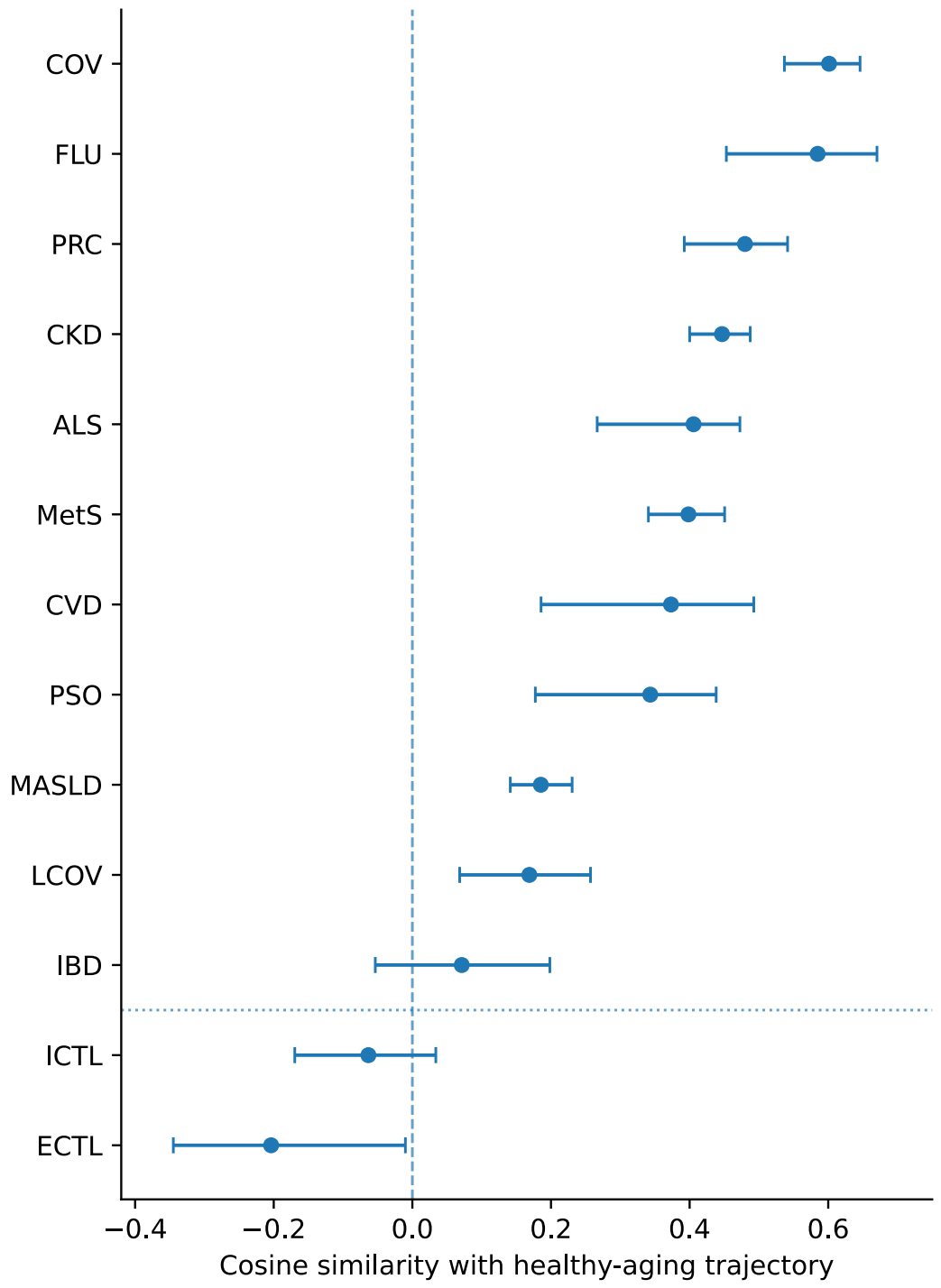

**Figure S13 | Alignment of disease-associated metabolic alterations with physiological metabolic aging.** Cosine similarity between each condition-associated metabolic effect vector and the corresponding healthy-aging vector. Disease effects were estimated in the standardized 75-feature space by comparison with age-matched healthy controls, with age and sex included as covariates. Healthy-aging vectors were derived from nonlinear age- and sex-dependent trajectories estimated in the healthy reference population and averaged according to the age and sex distribution of each condition. Points indicate cosine similarity and error bars bootstrap 95% confidence intervals. ICTL and ECTL denote internal and external healthy-control comparisons, respectively.

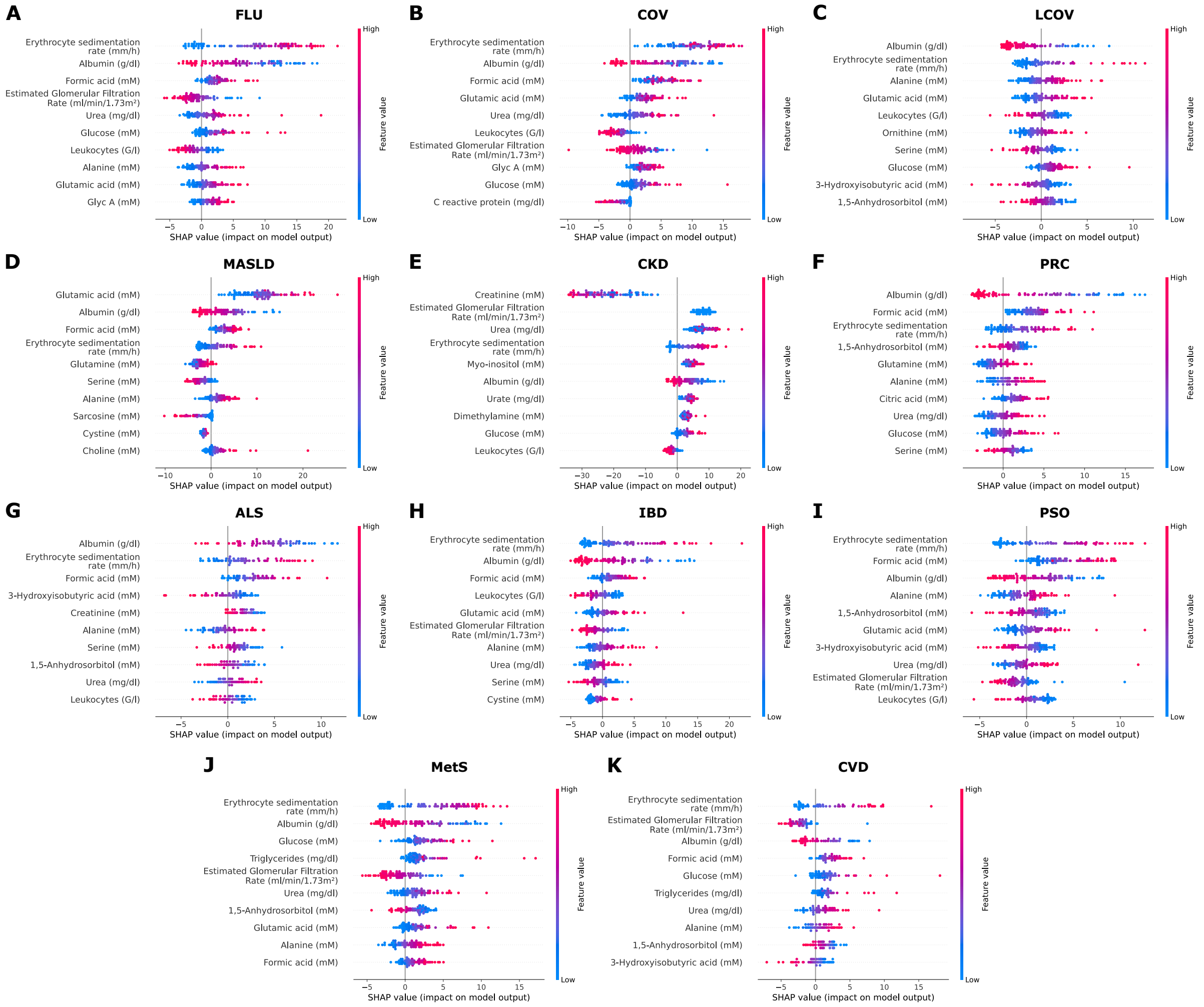

**Figure S14 | SHAP summary plots for the disease cohorts. A-K,** Contribution of the top 10 features to the predicted metabolic age for each specific disease cohort. Features are ordered by mean absolute SHAP value, and point color represents relative feature magnitude (low to high).

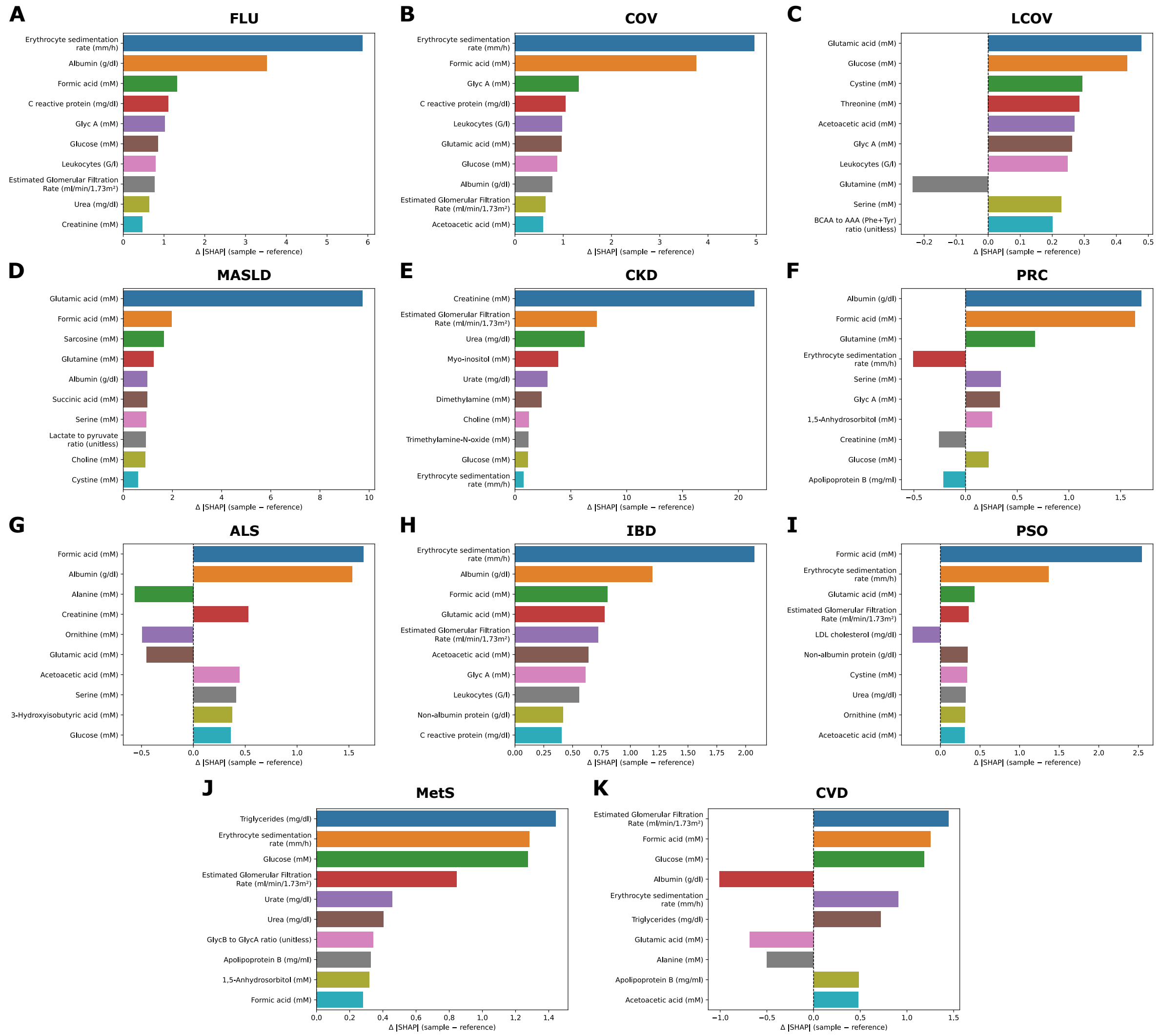

**Figure S15 | SHAP difference plots for the disease cohorts. A-K,** Top 10 features with the largest deviation relative to the reference group for each specific disease. Features are ordered by the mean absolute magnitude of the SHAP difference (Δ|SHAP|), illustrating the specific shifts in metabolic age drivers for each condition compared to their matched controls.

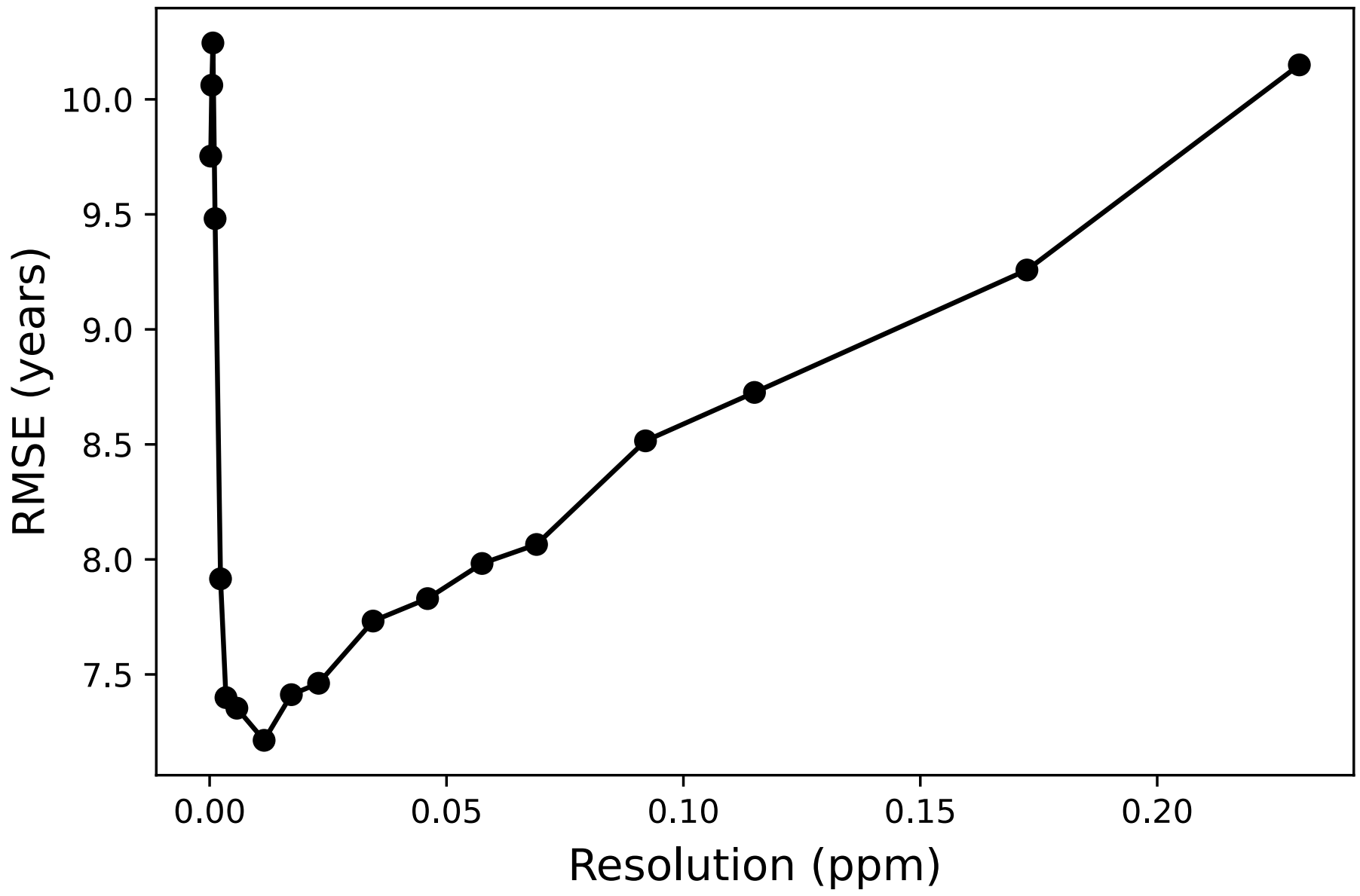

**Figure S16 | Optimization of NMR spectral bin width for machine-learning–based age prediction.** The x-axis represents bin size in parts per million (ppm), and the y-axis reports the root-mean-square error (RMSE, in years) obtained on the validation set. The selected bin width of 0.0115 ppm corresponds to a 50-fold reduction in spectral resolution and lies within the optimal range (15–100×), providing minimal prediction error.

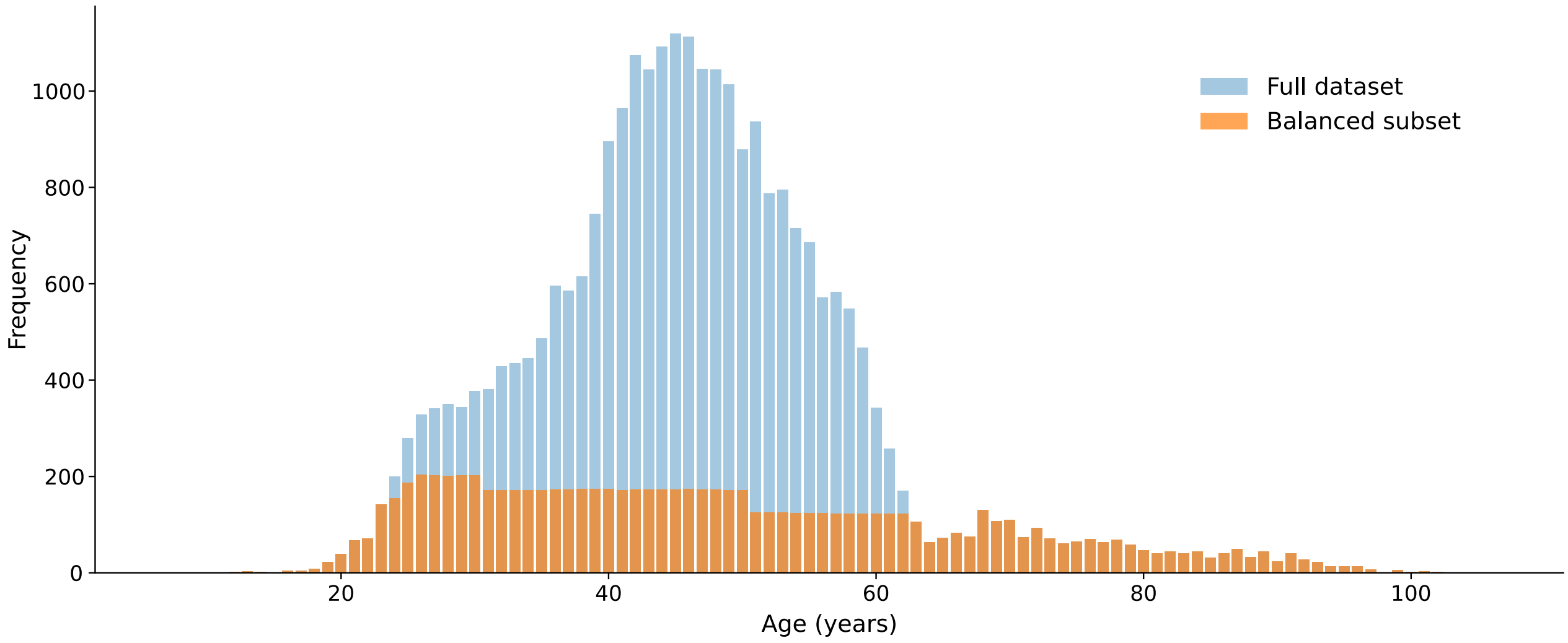

**Figure S17 | Age distribution of the full and balanced datasets.** Age distribution of the full dataset (blue) compared with the stratified and balanced subset used for model training (orange). Bars represent the frequency of samples within each age year. The balancing procedure reduces the overrepresentation of middle-aged individuals and yields a uniformly distributed age profile across the selected cohort.

**Table S1 | General characteristics for the Mondragón Corporation cohort.** Values are shown as mean ± standard deviation for continuous variables, and as counts (percentages) for categorical variables.

| Variable | All (N=23909) | Women (N=9341, 39.1%) | Men (N=14566, 60.9%) | N |
|---|---|---|---|---|
| Age | 44.07 ± 9.42 | 44.97 ± 9.09 | 43.49 ± 9.58 | 23909 |
| BMI | 25.26 ± 3.92 | 24.26 ± 4.22 | 25.90 ± 3.57 | 23647 |
| Weight (kg) | 74.82 ± 14.33 | 65.04 ± 11.82 | 81.10 ± 12.12 | 23720 |
| Height (cm) | 171.76 ± 9.05 | 163.74 ± 6.11 | 176.91 ± 6.53 | 23689 |
| Ethnic group | | | | 15203 |
| Asian | 13 (0.1%) | 4 (0.1%) | 9 (0.1%) | |
| Black | 19 (0.1%) | 8 (0.1%) | 11 (0.1%) | |
| Caucasian | 12221 (80.4%) | 5000 (81.1%) | 7219 (79.9%) | |
| Hispanic | 2926 (19.2%) | 1147 (18.6%) | 1779 (19.7%) | |
| Maghrebi | 24 (0.2%) | 5 (0.1%) | 19 (0.2%) | |
| Smoker | | | | 23778 |
| No | 19266 (81.0%) | 7656 (82.4%) | 11608 (80.2%) | |
| Yes | 4512 (19.0%) | 1638 (17.6%) | 2874 (19.8%) | |
| Cigarettes per day | | | | 4340 |
| 31 or more | 3 (0.1%) | 0 (0.0%) | 3 (0.1%) | |
| Between 11 and 20 | 1560 (35.9%) | 569 (36.1%) | 991 (35.9%) | |
| Between 21 and 30 | 142 (3.3%) | 56 (3.6%) | 86 (3.1%) | |
| Less than 10 | 2635 (60.7%) | 952 (60.4%) | 1683 (60.9%) | |
| Physical exercise | | | | 23733 |
| No | 5458 (23.0%) | 2708 (29.2%) | 2750 (19.0%) | |
| Yes | 18275 (77.0%) | 6562 (70.8%) | 11711 (81.0%) | |
| Physical exercise frequency | | | | 16666 |
| Once per week | 5300 (31.8%) | 1980 (33.3%) | 3320 (31.0%) | |
| Several days per week | 11366 (68.2%) | 3963 (66.7%) | 7401 (69.0%) | |
| Drink alcohol | | | | 23710 |
| Daily intake of alcohol | 589 (2.5%) | 58 (0.6%) | 531 (3.7%) | |
| Never | 3712 (15.7%) | 2086 (22.5%) | 1626 (11.3%) | |
| Only during meals | 1569 (6.6%) | 358 (3.9%) | 1211 (8.4%) | |
| Social drinker | 17840 (75.2%) | 6767 (73.0%) | 11071 (76.7%) | |
| Hypertension | | | | 23720 |
| No | 21919 (92.4%) | 8815 (94.8%) | 13102 (90.8%) | |
| Yes | 1801 (7.6%) | 481 (5.2%) | 1320 (9.2%) | |
| ESR (mm/h) | 7.66 ± 5.21 | 10.25 ± 6.24 | 6.00 ± 3.55 | 23597 |
| Leukocytes (n/L) | 6.70 ± 1.71 | 6.59 ± 1.69 | 6.77 ± 1.72 | 23901 |
| Neutrophils (%) | 51.38 ± 8.57 | 52.71 ± 8.84 | 50.52 ± 8.28 | 23901 |
| Neutrophils (n/L) | 3.49 ± 1.28 | 3.53 ± 1.31 | 3.46 ± 1.27 | 23901 |
| Lymphocytes (%) | 35.37 ± 7.84 | 34.81 ± 7.98 | 35.73 ± 7.73 | 23901 |
| Lymphocytes (n/L) | 2.33 ± 0.67 | 2.25 ± 0.64 | 2.39 ± 0.69 | 23901 |
| Monocytes (%) | 9.17 ± 2.06 | 8.63 ± 1.95 | 9.52 ± 2.06 | 23901 |
| Monocytes (n/L) | 0.61 ± 0.20 | 0.56 ± 0.16 | 0.64 ± 0.21 | 23901 |
| Eosinophils (%) | 3.42 ± 2.28 | 3.17 ± 2.22 | 3.58 ± 2.30 | 23901 |
| Eosinophils (n/L) | 0.23 ± 0.17 | 0.21 ± 0.16 | 0.24 ± 0.17 | 23901 |
| Basophils (%) | 0.66 ± 0.35 | 0.67 ± 0.35 | 0.66 ± 0.34 | 23901 |
| Basophils (n/L) | 0.04 ± 0.02 | 0.04 ± 0.02 | 0.04 ± 0.02 | 23901 |
| Erythrocytes (n/L) | 4.84 ± 0.43 | 4.51 ± 0.32 | 5.05 ± 0.35 | 23901 |
| Hemo (g/dL) | 14.65 ± 1.25 | 13.58 ± 0.97 | 15.33 ± 0.88 | 23901 |
| Hematocrit (%) | 43.58 ± 3.32 | 40.87 ± 2.62 | 45.31 ± 2.45 | 23901 |
| Corpuscular volume (fL) | 90.26 ± 4.21 | 90.74 ± 4.56 | 89.96 ± 3.94 | 23901 |

| Variable | All (N=23909) | Women (N=9341, 39.1%) | Men (N=14566, 60.9%) | N |
|---|---|---|---|---|
| Corpuscular hemoglobin (pg) | 30.33 ± 1.69 | 30.15 ± 1.86 | 30.44 ± 1.56 | 23901 |
| Corpuscular hemoglobin (g/dL) | 33.59 ± 1.01 | 33.21 ± 1.01 | 33.84 ± 0.93 | 23901 |
| Red cell distribution width | 13.17 ± 0.87 | 13.27 ± 0.99 | 13.11 ± 0.77 | 23851 |
| Platelets (n/L) | 242.61 ± 53.44 | 254.95 ± 56.34 | 234.71 ± 49.92 | 23901 |
| Platelet volume (fL) | 8.48 ± 0.71 | 8.55 ± 0.73 | 8.43 ± 0.69 | 23897 |
| Cholesterol (mg/dL) | 196.69 ± 35.15 | 195.29 ± 34.97 | 197.59 ± 35.25 | 23706 |
| Cholesterol HDL (mg/dL) | 61.31 ± 15.57 | 68.55 ± 15.09 | 56.67 ± 14.02 | 23705 |
| Cholesterol no HDL (mg/dL) | 135.38 ± 36.13 | 126.73 ± 34.03 | 140.93 ± 36.35 | 23719 |
| Cholesterol LDL (mg/dL) | 115.48 ± 31.65 | 110.46 ± 30.72 | 118.90 ± 31.81 | 22491 |
| Triglycerides (mg/dL) | 96.76 ± 65.74 | 79.76 ± 47.92 | 107.68 ± 72.92 | 23720 |
| Glucose (mg/dL) | 87.25 ± 12.76 | 85.89 ± 10.57 | 88.12 ± 13.91 | 23722 |
| Urate (mg/dL) | 5.10 ± 1.29 | 4.16 ± 0.96 | 5.70 ± 1.10 | 23718 |
| Creatinine (mg/dL) | 0.87 ± 0.16 | 0.74 ± 0.11 | 0.95 ± 0.12 | 23720 |
| Alanine aminotransferase (U/L) | 21.92 ± 13.30 | 16.55 ± 11.27 | 25.37 ± 13.36 | 23719 |
| Gamma-glutamyl transferase (U/L) | 21.38 ± 22.59 | 15.36 ± 13.09 | 25.25 ± 26.27 | 23718 |
| Urine pH | 5.93 ± 0.62 | 5.92 ± 0.67 | 5.94 ± 0.59 | 12689 |
| Urine proteins (mg/dL) | 24.21 ± 22.24 | 23.43 ± 20.12 | 24.68 ± 23.45 | 295 |
| Urine glucose (mg/dL) | 1562.58 ± 2170.87 | 1036.42 ± 1638.11 | 1697.49 ± 2272.43 | 196 |
| Urine ketone bodies | | | | 22538 |
| + (15 mg/dL) | 183 (0.8%) | 66 (0.8%) | 117 (0.8%) | |
| ++ (40 mg/dL) | 37 (0.2%) | 15 (0.2%) | 22 (0.2%) | |
| +++ (80-160 mg/dL) | 15 (0.1%) | 6 (0.1%) | 9 (0.1%) | |
| Not detected | 21998 (97.6%) | 8306 (97.8%) | 13690 (97.5%) | |
| Traces (5 mg/dL) | 305 (1.4%) | 104 (1.2%) | 201 (1.4%) | |
| Urine urobilinogen | | | | 22538 |
| + (2 mg/dL) | 17 (0.1%) | 1 (0.0%) | 16 (0.1%) | |
| ++ (4 mg/dL) | 31 (0.1%) | 5 (0.1%) | 26 (0.2%) | |
| Not detected | 22169 (98.4%) | 8400 (98.9%) | 13767 (98.1%) | |
| Traces (1 mg/dL) | 321 (1.4%) | 91 (1.1%) | 230 (1.6%) | |
| Urine bilirubin | | | | 22538 |
| Not detected | 22316 (99.0%) | 8407 (98.9%) | 13908 (99.1%) | |
| Positive (+) | 220 (1.0%) | 88 (1.0%) | 131 (0.9%) | |
| Positive (++) | 2 (0.0%) | 2 (0.0%) | 0 (0.0%) | |
| Urine red blood cells | | | | 22516 |
| Not detected | 21432 (95.2%) | 7698 (90.7%) | 13732 (97.9%) | |
| Positive (+) | 831 (3.7%) | 598 (7.0%) | 233 (1.7%) | |
| Positive (++) | 150 (0.7%) | 117 (1.4%) | 33 (0.2%) | |
| Positive (+++) | 103 (0.5%) | 76 (0.9%) | 27 (0.2%) | |
| Urine leukocytes | | | | 22516 |
| Not detected | 21235 (94.3%) | 7323 (86.3%) | 13910 (99.2%) | |
| Positive (+) | 525 (2.3%) | 474 (5.6%) | 51 (0.4%) | |
| Positive (++) | 603 (2.7%) | 549 (6.5%) | 54 (0.4%) | |
| Positive (+++) | 153 (0.7%) | 143 (1.7%) | 10 (0.1%) | |
| Urine nitrites | | | | 22514 |
| Not detected | 22270 (98.9%) | 8253 (97.2%) | 14015 (99.9%) | |
| Positive (+) | 217 (1.0%) | 208 (2.5%) | 9 (0.1%) | |

| Variable | All (N=23909) | Women (N=9341, 39.1%) | Men (N=14566, 60.9%) | N |
|---|---|---|---|---|
| Positive (++) | 27 (0.1%) | 26 (0.3%) | 1 (0.0%) | |
| Urine density (g/mL) | 1.02 ± 0.01 | 1.02 ± 0.01 | 1.02 ± 0.01 | 14060 |
| Aspartate aminotransferase (U/L) | 37.35 ± 45.24 | 32.42 ± 24.70 | 38.84 ± 49.75 | 533 |
| Bilirubin (mg/dL) | 0.61 ± 0.38 | 0.51 ± 0.41 | 0.63 ± 0.37 | 245 |
| Ferritin (ng/mL) | 74.50 ± 109.88 | 23.17 ± 37.59 | 149.07 ± 135.07 | 260 |
| Alkaline phosphatase (U/L) | 73.40 ± 25.57 | 84.60 ± 40.09 | 70.37 ± 18.96 | 225 |
| Fructosamine (µmol/L) | 255.38 ± 46.85 | 240.68 ± 38.43 | 259.28 ± 48.19 | 210 |
| Iron (µg/dL) | 72.62 ± 42.35 | 59.83 ± 39.39 | 85.40 ± 41.53 | 178 |
| Transferrin saturation (%) | 22.28 ± 15.98 | 16.76 ± 14.20 | 28.11 ± 15.74 | 177 |
| Transferrin (mg/dL) | 282.68 ± 59.92 | 308.56 ± 55.93 | 256.50 ± 52.16 | 177 |

**Table S2 | General characteristics for the DDM-Madrid cohort.** Values are shown as mean ± standard deviation for continuous variables, and as counts (percentages) for categorical variables.

| Variable | Women (N=933) | N |
|---|---|---|
| Age | 44.17 ± 2.79 | 933 |
| BMI | 23.96 ± 3.96 | 931 |
| Weight (kg) | 63.00 ± 10.94 | 932 |
| Height (cm) | 162.13 ± 5.87 | 931 |
| Diabetes | | 929 |
| No | 928 (99.9%) | |
| Yes | 1 (0.1%) | |
| Medication for diabetes | | 929 |
| No | 928 (99.9%) | |
| Yes | 1 (0.1%) | |
| Hyperglycemia | | 925 |
| No | 761 (82.3%) | |
| Yes | 164 (17.7%) | |
| Hypertension | | 907 |
| No | 870 (95.9%) | |
| Yes | 37 (4.1%) | |
| Medication for hypertension | | 907 |
| No | 870 (95.9%) | |
| Yes | 37 (4.1%) | |
| Hypercholesterolemia | | 926 |
| No | 807 (87.1%) | |
| Yes | 119 (12.9%) | |
| Medication for hypercholesterolemia | | 926 |
| No | 807 (87.1%) | |
| Yes | 119 (12.9%) | |
| Active cancer | | 789 |
| No | 773 (98.0%) | |
| Yes | 16 (2.0%) | |
| Menopause stage | | 915 |
| Perimenopause | 29 (3.2%) | |
| Premenopause | 886 (96.8%) | |
| Smoker | | 595 |
| No | 369 (62.0%) | |
| Yes | 226 (38.0%) | |
| Former smoker | | 595 |
| No | 251 (42.2%) | |
| Yes | 344 (57.8%) | |
| Physical exercise | | 933 |
| No | 135 (14.5%) | |
| Yes | 798 (85.5%) | |
| Physical exercise frequency | | 798 |
| Once per week | 617 (77.3%) | |
| Several days per week | 181 (22.7%) | |

**Table S3 | General characteristics for the Biosilver cohort.** Values are shown as mean ± standard deviation for continuous variables, and as counts (percentages) for categorical variables.

| Variable | All (N=686) | Women (N=500, 72.9%) | Men (N=186, 27.1%) | N |
|---|---|---|---|---|
| Age | 79.36 ± 9.36 | 79.01 ± 9.52 | 80.32 ± 8.85 | 686 |
| Menopause age | 48.87 ± 5.32 | 48.90 ± 5.33 | 45.00 ± nan | 151 |
| Ethnic group | | | | 235 |
| Caucasian | 235 (100.0%) | 170 (100.0%) | 65 (100.0%) | |
| Smoker | | | | 232 |
| No | 220 (94.8%) | 163 (96.4%) | 57 (90.5%) | |
| Yes | 12 (5.2%) | 6 (3.6%) | 6 (9.5%) | |
| Drink alcohol | | | | 224 |
| Daily intake | 58 (25.9%) | 33 (19.9%) | 25 (43.1%) | |
| Never | 53 (23.7%) | 44 (26.5%) | 9 (15.5%) | |
| Social drinker | 113 (50.4%) | 89 (53.6%) | 24 (41.4%) | |
| Active cancer | | | | 235 |
| No | 221 (94.0%) | 162 (95.3%) | 59 (90.8%) | |
| Yes | 14 (6.0%) | 8 (4.7%) | 6 (9.2%) | |
| Chronic respiratory disease | | | | 235 |
| No | 178 (75.7%) | 133 (78.2%) | 45 (69.2%) | |
| Yes | 57 (24.3%) | 37 (21.8%) | 20 (30.8%) | |
| Chronic heart disease | | | | 235 |
| No | 165 (70.2%) | 127 (74.7%) | 38 (58.5%) | |
| Yes | 70 (29.8%) | 43 (25.3%) | 27 (41.5%) | |
| Neuro disease or stroke | | | | 235 |
| No | 191 (81.3%) | 144 (84.7%) | 47 (72.3%) | |
| Yes | 44 (18.7%) | 26 (15.3%) | 18 (27.7%) | |
| Chronic kidney failure | | | | 235 |
| No | 202 (86.0%) | 148 (87.1%) | 54 (83.1%) | |
| Yes | 33 (14.0%) | 22 (12.9%) | 11 (16.9%) | |
| Chronic digestive disease | | | | 235 |
| No | 160 (68.1%) | 120 (70.6%) | 40 (61.5%) | |
| Yes | 75 (31.9%) | 50 (29.4%) | 25 (38.5%) | |
| Height (cm) | 158.66 ± 8.51 | 155.49 ± 6.37 | 168.30 ± 6.78 | 320 |
| Weight (kg) | 65.60 ± 13.58 | 62.78 ± 12.35 | 74.31 ± 13.57 | 319 |
| Systolic blood pressure (mmHg) | 136.93 ± 22.22 | 136.90 ± 22.13 | 137.03 ± 22.63 | 322 |
| Diastolic blood pressure (mmHg) | 75.58 ± 10.64 | 75.92 ± 10.44 | 74.56 ± 11.21 | 322 |
| BMI | 26.02 ± 4.67 | 25.96 ± 4.85 | 26.21 ± 4.11 | 318 |
| Former smoker | | | | 232 |
| No | 123 (53.0%) | 97 (57.4%) | 26 (41.3%) | |
| Yes | 109 (47.0%) | 72 (42.6%) | 37 (58.7%) | |
| Red blood cell count (millions/µL) | 4.31 ± 0.53 | 4.31 ± 0.47 | 4.32 ± 0.66 | 334 |
| Hemo (g/dL) | 13.29 ± 1.55 | 13.24 ± 1.33 | 13.46 ± 2.04 | 333 |
| Hematocrit (%) | 40.90 ± 4.54 | 40.76 ± 3.92 | 41.29 ± 5.95 | 334 |
| Corpuscular volume (fL) | 95.10 ± 6.00 | 94.79 ± 5.46 | 95.97 ± 7.25 | 334 |
| Platelets (n/L) | 229.62 ± 68.77 | 237.00 ± 69.13 | 208.99 ± 63.73 | 334 |
| Leukocytes (n/L) | 6.68 ± 3.99 | 6.48 ± 3.58 | 7.24 ± 4.96 | 334 |
| ESR (mm/h) | 30.88 ± 28.11 | 30.19 ± 26.63 | 32.85 ± 32.08 | 325 |
| Glucose (mg/dL) | 96.96 ± 33.06 | 94.96 ± 31.06 | 102.53 ± 37.74 | 334 |

| Variable | All (N=686) | Women (N=500, 72.9%) | Men (N=186, 27.1%) | N |
|---|---|---|---|---|
| Urea (mg/dL) | 50.14 ± 33.61 | 48.08 ± 32.42 | 55.86 ± 36.29 | 333 |
| Creatinine (mg/dL) | 1.06 ± 3.32 | 1.06 ± 3.86 | 1.05 ± 0.49 | 334 |
| Urate (mg/dL) | 5.35 ± 1.92 | 5.17 ± 1.95 | 5.87 ± 1.74 | 333 |
| Cholesterol (mg/dL) | 186.62 ± 41.01 | 193.58 ± 39.32 | 167.18 ± 39.56 | 334 |
| Cholesterol HDL (mg/dL) | 55.09 ± 18.39 | 57.97 ± 18.45 | 47.03 ± 15.72 | 334 |
| Cholesterol LDL (mg/dL) | 111.28 ± 31.44 | 115.06 ± 30.68 | 100.75 ± 31.31 | 322 |
| Triglycerides (mg/dL) | 102.21 ± 42.80 | 103.92 ± 43.85 | 97.44 ± 39.55 | 334 |
| Bilirubin (mg/dL) | 0.72 ± 0.39 | 0.69 ± 0.33 | 0.81 ± 0.52 | 334 |
| Albumin (g/dL) | 4.06 ± 0.44 | 4.09 ± 0.43 | 4.00 ± 0.46 | 333 |
| Total protein (g/dL) | 6.56 ± 0.56 | 6.57 ± 0.55 | 6.52 ± 0.60 | 293 |
| Estimated glomerular filtration rate (CKD-EPI) | 73.01 ± 18.40 | 73.38 ± 18.10 | 71.97 ± 19.27 | 334 |
| Creatine kinase (U/L) | 86.70 ± 71.83 | 83.07 ± 64.24 | 96.85 ± 89.34 | 334 |
| Intact parathyroid hormone (ng/L) | 78.90 ± 70.54 | 78.86 ± 66.39 | 79.01 ± 81.47 | 334 |
| Alanine aminotransferase (U/L) | 22.41 ± 18.51 | 21.00 ± 15.05 | 26.35 ± 25.56 | 334 |
| Aspartate aminotransferase (U/L) | 24.40 ± 15.12 | 23.38 ± 13.99 | 27.25 ± 17.67 | 333 |
| Gamma-glutamyl transferase (U/L) | 34.70 ± 48.40 | 32.22 ± 52.86 | 41.65 ± 32.14 | 334 |
| Alkaline phosphatase (U/L) | 80.25 ± 53.01 | 81.22 ± 58.73 | 77.54 ± 32.11 | 334 |
| Calcium (mg/dL) | 9.11 ± 0.54 | 9.15 ± 0.51 | 9.02 ± 0.59 | 332 |
| Phosphorus (mg/dL) | 3.37 ± 0.63 | 3.45 ± 0.62 | 3.13 ± 0.58 | 334 |
| Magnesium (mg/dL) | 2.55 ± 10.78 | 1.98 ± 0.25 | 4.15 ± 21.01 | 334 |
| Sodium (mEq/L) | 140.44 ± 2.97 | 140.48 ± 2.73 | 140.33 ± 3.56 | 334 |
| Potassium (mEq/L) | 4.65 ± 4.01 | 4.60 ± 3.70 | 4.81 ± 4.79 | 334 |
| Chloride (mEq/L) | 105.21 ± 6.44 | 105.60 ± 3.77 | 104.11 ± 10.82 | 334 |
| C-terminal telopeptide (pg/mL) | 547.00 ± 313.00 | 568.60 ± 317.99 | 486.61 ± 291.94 | 334 |
| Procollagen type 1 N-terminal propeptide (µg/dL) | 68.35 ± 214.04 | 58.89 ± 28.85 | 94.43 ± 412.68 | 323 |
| Iron (µg/dL) | 79.98 ± 35.52 | 80.08 ± 33.80 | 79.68 ± 40.13 | 333 |
| Transferrin (mg/dL) | 220.40 ± 48.96 | 222.61 ± 47.62 | 214.22 ± 52.30 | 334 |
| Transferrin saturation (%) | 26.33 ± 13.24 | 26.03 ± 13.02 | 27.16 ± 13.86 | 333 |
| Ferritin (ng/mL) | 147.42 ± 210.68 | 124.64 ± 178.18 | 211.09 ± 273.71 | 334 |
| Folate (ng/mL) | 13.18 ± 25.83 | 14.15 ± 29.93 | 10.46 ± 4.55 | 323 |
| Vitamin B12 (pg/mL) | 454.45 ± 218.11 | 468.99 ± 234.67 | 413.21 ± 156.13 | 330 |
| Vitamin B1 (µg/L) | 51.80 ± 20.22 | 52.81 ± 22.17 | 48.99 ± 13.13 | 331 |
| Vitamin 25 hydroxy D (ng/mL) | 27.76 ± 13.22 | 27.75 ± 13.56 | 27.78 ± 12.31 | 333 |
| Thyroid-stimulating hormone (mIU/L) | 2.47 ± 3.49 | 2.71 ± 3.99 | 1.81 ± 1.16 | 334 |
| CRP (mg/dL) | 1.45 ± 3.21 | 1.31 ± 3.12 | 1.86 ± 3.46 | 334 |

**Table S4 | General characteristics for the AGEPORTUGAL cohort.** Values are shown as mean ± standard deviation for continuous variables, and as counts (percentages) for categorical variables.

| Variable | All (N=246) | Women (N=161, 65.4%) | Men (N=85, 34.6%) | N |
|---|---|---|---|---|
| Age | 83.70 ± 7.81 | 85.17 ± 7.25 | 80.92 ± 8.11 | 246 |
| Weight (kg) | 65.12 ± 12.99 | 63.18 ± 11.38 | 68.62 ± 14.97 | 163 |
| Height (cm) | 155.77 ± 8.87 | 151.27 ± 6.61 | 163.34 ± 6.81 | 188 |
| BMI | 26.87 ± 4.65 | 27.53 ± 4.53 | 25.72 ± 4.66 | 159 |
| Systolic blood pressure (mmHg) | 124.54 ± 20.64 | 124.09 ± 20.65 | 125.36 ± 20.71 | 233 |
| Systolic blood pressure (mmHg) | 69.70 ± 10.71 | 68.91 ± 10.47 | 71.12 ± 11.06 | 233 |
| Hypertension | | | | 235 |
| No | 54 (23.0%) | 33 (21.4%) | 21 (25.9%) | |
| Yes | 181 (77.0%) | 121 (78.6%) | 60 (74.1%) | |
| Atrial fibrillation | | | | 230 |
| No | 203 (88.3%) | 128 (85.3%) | 75 (93.8%) | |
| Yes | 27 (11.7%) | 22 (14.7%) | 5 (6.2%) | |
| Atherosclerosis | | | | 225 |
| No | 217 (96.4%) | 144 (97.3%) | 73 (94.8%) | |
| Yes | 8 (3.6%) | 4 (2.7%) | 4 (5.2%) | |
| Diabetes | | | | 225 |
| No | 157 (69.8%) | 97 (65.1%) | 60 (78.9%) | |
| Yes | 68 (30.2%) | 52 (34.9%) | 16 (21.1%) | |
| Dyslipidemia | | | | 229 |
| No | 117 (51.1%) | 69 (45.7%) | 48 (61.5%) | |
| Yes | 112 (48.9%) | 82 (54.3%) | 30 (38.5%) | |
| Acute myocardial infarction | | | | 225 |
| No | 212 (94.2%) | 140 (94.6%) | 72 (93.5%) | |
| Yes | 13 (5.8%) | 8 (5.4%) | 5 (6.5%) | |
| Chronic congestive heart failure | | | | 225 |
| No | 179 (79.6%) | 111 (75.0%) | 68 (88.3%) | |
| Yes | 46 (20.4%) | 37 (25.0%) | 9 (11.7%) | |
| Angina pectoris | | | | 225 |
| No | 214 (95.1%) | 140 (94.6%) | 74 (96.1%) | |
| Yes | 11 (4.9%) | 8 (5.4%) | 3 (3.9%) | |
| Stroke | | | | 225 |
| No | 191 (84.9%) | 130 (87.8%) | 61 (79.2%) | |
| Yes | 34 (15.1%) | 18 (12.2%) | 16 (20.8%) | |
| Cardiac arrhythmias | | | | 227 |
| No | 227 (100.0%) | 150 (100.0%) | 77 (100.0%) | |
| Peripheral vascular disease | | | | 227 |
| No | 211 (93.0%) | 138 (92.0%) | 73 (94.8%) | |
| Yes | 16 (7.0%) | 12 (8.0%) | 4 (5.2%) | |
| Dementia | | | | 228 |
| No | 169 (74.1%) | 107 (70.9%) | 62 (80.5%) | |
| Yes | 59 (25.9%) | 44 (29.1%) | 15 (19.5%) | |
| Addenbrooke's Cognitive Examination-Revised | 45.59 ± 19.66 | 44.45 ± 19.58 | 47.75 ± 19.77 | 205 |
| Mini-Mental State Examination | 16.85 ± 5.88 | 16.54 ± 5.53 | 17.39 ± 6.47 | 208 |
| Frontal Assessment Battery (FAB) total score | 7.35 ± 3.60 | 7.01 ± 3.50 | 7.96 ± 3.70 | 209 |

| **Variable** | **All (N=246)** | **Women (N=161, 65.4%)** | **Men (N=85, 34.6%)** | **N** |
|---|---|---|---|---|
| Medication for hypertension | | | | 238 |
| No | 47 (19.7%) | 27 (17.2%) | 20 (24.7%) | |
| Yes | 191 (80.3%) | 130 (82.8%) | 61 (75.3%) | |
| Medication for hypercholesterolemia | | | | 238 |
| No | 138 (58.0%) | 84 (53.5%) | 54 (66.7%) | |
| Yes | 100 (42.0%) | 73 (46.5%) | 27 (33.3%) | |
| EUROHIS-QOL 8-item index | 56.69 ± 14.76 | 55.81 ± 15.27 | 58.21 ± 13.78 | 213 |
| Glucose (mg/dL) | 98.29 ± 25.40 | 98.21 ± 26.05 | 98.44 ± 24.28 | 239 |
| Triglycerides (mg/dL) | 113.11 ± 52.30 | 121.80 ± 54.08 | 96.77 ± 44.69 | 239 |
| Cholesterol (mg/dL) | 162.94 ± 42.38 | 167.64 ± 41.84 | 154.00 ± 42.20 | 238 |
| Cholesterol HDL (mg/dL) | 54.84 ± 12.79 | 55.73 ± 13.51 | 53.14 ± 11.17 | 238 |
| Cholesterol LDL (mg/dL) | 85.45 ± 36.20 | 87.55 ± 35.69 | 81.47 ± 37.05 | 238 |
| Smoker | | | | 243 |
| No | 232 (95.5%) | 158 (99.4%) | 74 (88.1%) | |
| Yes | 11 (4.5%) | 1 (0.6%) | 10 (11.9%) | |
| Cigarettes/day | | | | 9 |
| Between 11 and 20 | 2 (22.2%) | 0 (0.0%) | 2 (25.0%) | |
| Less than 10 | 7 (77.8%) | 1 (100.0%) | 6 (75.0%) | |
| Former smoker | | | | 218 |
| No | 186 (85.3%) | 144 (98.0%) | 42 (59.2%) | |
| Yes | 32 (14.7%) | 3 (2.0%) | 29 (40.8%) | |
| Drink alcohol | | | | 237 |
| Daily intake | 1 (0.4%) | 0 (0.0%) | 1 (1.2%) | |
| Social drinker | 22 (9.3%) | 7 (4.6%) | 15 (17.9%) | |
| No | 214 (90.3%) | 146 (95.4%) | 68 (81.0%) | |
| Medication for diabetes | | | | 238 |
| No | 189 (79.4%) | 117 (74.5%) | 72 (88.9%) | |
| Yes | 49 (20.6%) | 40 (25.5%) | 9 (11.1%) | |

**Table S5 | General characteristics for the SPBB cohort (healthy).** Values are shown as mean ± standard deviation for continuous variables, and as counts (percentages) for categorical variables.

| **Variable** | **All (N=3032)** | **Women (N=1540, 50.8%)** | **Men (N=1492, 49.2%)** | **N** |
|---|---|---|---|---|
| Age | 54.91 ± 16.65 | 56.74 ± 16.64 | 53.01 ± 16.45 | 3032 |
| Weight (kg) | 74.42 ± 16.70 | 65.29 ± 12.65 | 83.05 ± 15.43 | 2664 |
| Height (cm) | 166.27 ± 9.96 | 159.32 ± 6.98 | 172.87 ± 7.60 | 2651 |
| BMI | 26.79 ± 4.85 | 25.97 ± 4.95 | 27.66 ± 4.57 | 2943 |
| Waist (cm) | 109.67 ± 14.13 | 107.67 ± 13.97 | 110.74 ± 14.17 | 136 |
| Systolic blood pressure (mmHg) | 139.26 ± 18.48 | 136.19 ± 17.41 | 146.29 ± 19.02 | 303 |
| Diastolic blood pressure (mmHg) | 77.20 ± 9.82 | 75.86 ± 9.79 | 80.27 ± 9.22 | 303 |
| Cholesterol (mg/dL) | 238.92 ± 72.91 | 259.94 ± 68.22 | 219.43 ± 71.72 | 2717 |
| Triglycerides (mg/dL) | 157.94 ± 208.66 | 131.60 ± 97.82 | 183.97 ± 275.28 | 2913 |
| Glucose (mg/dL) | 97.14 ± 27.03 | 94.08 ± 21.43 | 100.17 ± 31.32 | 2908 |
| Diabetes | | | | 2702 |
| No | 2317 (85.8%) | 1217 (89.8%) | 1100 (81.7%) | |
| Yes | 385 (14.2%) | 138 (10.2%) | 247 (18.3%) | |
| Dyslipidemia | | | | 2267 |
| No | 158 (7.0%) | 104 (9.0%) | 54 (4.9%) | |
| Yes | 2109 (93.0%) | 1050 (91.0%) | 1059 (95.1%) | |
| Hypertension | | | | 2887 |
| No | 1812 (62.8%) | 979 (66.5%) | 833 (58.9%) | |
| Yes | 1075 (37.2%) | 494 (33.5%) | 581 (41.1%) | |
| Cholesterol HDL (mg/dL) | 57.08 ± 17.79 | 64.91 ± 17.78 | 49.19 ± 13.89 | 2859 |

**Table S6 | General characteristics for the SPBB cohort (MetS).** Values are shown as mean ± standard deviation for continuous variables, and as counts (percentages) for categorical variables.

| **Variable** | **All (N=458)** | **Women (N=261, 57.0%)** | **Men (N=197, 43.0%)** | **N** |
|---|---|---|---|---|
| Age | 63.08 ± 10.95 | 65.83 ± 9.75 | 59.45 ± 11.40 | 458 |
| Weight (kg) | 84.47 ± 14.55 | 80.02 ± 13.86 | 89.23 ± 13.78 | 387 |
| Height (cm) | 163.51 ± 9.86 | 156.73 ± 6.54 | 170.81 ± 7.27 | 386 |
| BMI | 32.09 ± 5.02 | 33.12 ± 5.45 | 30.68 ± 3.98 | 451 |
| Waist (cm) | 110.09 ± 12.60 | 110.09 ± 12.60 | - | 11 |
| Systolic blood pressure (mmHg) | 147.08 ± 19.37 | 147.07 ± 20.07 | 147.17 ± 9.26 | 73 |
| Diastolic blood pressure (mmHg) | 81.41 ± 10.22 | 81.88 ± 10.30 | 76.17 ± 8.18 | 73 |
| Cholesterol (mg/dL) | 219.95 ± 72.27 | 223.02 ± 67.72 | 216.75 ± 76.78 | 392 |
| Triglycerides (mg/dL) | 256.12 ± 316.84 | 204.45 ± 162.31 | 323.53 ± 435.35 | 454 |
| Glucose (mg/dL) | 130.47 ± 39.69 | 131.92 ± 38.54 | 128.56 ± 41.17 | 454 |
| Diabetes | | | | 358 |
| No | 83 (23.2%) | 44 (26.0%) | 39 (20.6%) | |
| Yes | 275 (76.8%) | 125 (74.0%) | 150 (79.4%) | |
| Dyslipidemia | | | | 380 |
| No | 17 (4.5%) | 17 (8.7%) | 0 (0.0%) | |
| Yes | 363 (95.5%) | 179 (91.3%) | 184 (100.0%) | |
| Hypertension | | | | 451 |
| No | 97 (21.5%) | 77 (29.8%) | 20 (10.4%) | |
| Yes | 354 (78.5%) | 181 (70.2%) | 173 (89.6%) | |
| Cholesterol HDL (mg/dL) | 47.10 ± 12.94 | 51.09 ± 12.56 | 41.98 ± 11.55 | 450 |

**Table S7 | General characteristics for the external control cohort.** Values are shown as mean ± standard deviation for continuous variables, and as counts (percentages) for categorical variables.

| **Variable** | **All (N=121)** | **Women (N=63, 52.1%)** | **Men (N=58, 47.9%)** | **N** |
|---|---|---|---|---|
| Age | 44.60 ± 20.79 | 44.71 ± 22.49 | 44.47 ± 18.96 | 121 |

**Table S8 | General characteristics for the LA2 cohort (MetS).** Values are shown as mean ± standard deviation for continuous variables, and as counts (percentages) for categorical variables.

| Variable | All (N=89) | Women (N=44, 49.4%) | Men (N=45, 50.6%) | N |
|---|---|---|---|---|
| Age | 79.46 ± 7.41 | 82.18 ± 5.82 | 76.80 ± 7.87 | 89 |
| Height (cm) | 160.07 ± 9.57 | 153.82 ± 7.04 | 166.96 ± 6.89 | 82 |
| Weight (kg) | 82.09 ± 15.48 | 77.60 ± 15.56 | 86.80 ± 14.10 | 84 |
| BMI | 31.96 ± 5.12 | 32.72 ± 5.84 | 31.14 ± 4.09 | 82 |
| Hip (cm) | 112.07 ± 18.25 | 115.80 ± 17.54 | 107.93 ± 18.37 | 78 |
| Waist (cm) | 112.88 ± 18.34 | 113.88 ± 15.84 | 111.77 ± 20.94 | 78 |
| Diabetes | | | | 89 |
| Yes | 62 (69.7%) | 30 (68.2%) | 32 (71.1%) | |
| No | 27 (30.3%) | 14 (31.8%) | 13 (28.9%) | |
| Hypertension | | | | 89 |
| Yes | 87 (97.8%) | 43 (97.7%) | 44 (97.8%) | |
| No | 2 (2.2%) | 1 (2.3%) | 1 (2.2%) | |
| Glucose (mg/dL) | 138.29 ± 56.58 | 147.98 ± 72.87 | 128.82 ± 32.08 | 89 |
| Cholesterol HDL (mg/dL) | 40.67 ± 12.92 | 43.77 ± 12.77 | 37.64 ± 12.46 | 89 |
| Triglycerides (mg/dL) | 132.15 ± 55.21 | 141.41 ± 58.92 | 123.09 ± 50.35 | 89 |

**Table S9 | General characteristics for the LiverBible cohort (MetS).** Values are shown as mean ± standard deviation for continuous variables, and as counts (percentages) for categorical variables.

| Variable | All (N=113) | Women (N=21, 18.6%) | Men (N=92, 81.4%) | N |
|---|---|---|---|---|
| Age | 55.94 ± 6.02 | 55.00 ± 6.29 | 56.15 ± 5.97 | 113 |
| Ethnic group | | | | 113 |
| Caucasian | 101 (89.4%) | 18 (85.7%) | 83 (90.2%) | |
| Asian | 10 (8.8%) | 3 (14.3%) | 7 (7.6%) | |
| North african | 2 (1.8%) | 0 (0.0%) | 2 (2.2%) | |
| BMI | 29.98 ± 4.18 | 29.62 ± 4.73 | 30.07 ± 4.07 | 113 |
| Weight (kg) | 88.99 ± 15.19 | 77.71 ± 12.71 | 91.56 ± 14.58 | 113 |
| Height (cm) | 172.03 ± 8.75 | 161.76 ± 5.62 | 174.37 ± 7.57 | 113 |
| Systolic blood pressure (mmHg) | 140.12 ± 17.12 | 142.48 ± 13.22 | 139.58 ± 17.91 | 113 |
| Diastolic blood pressure (mmHg) | 86.97 ± 8.63 | 83.71 ± 10.92 | 87.72 ± 7.91 | 113 |
| Hypertension | | | | 113 |
| Yes | 106 (93.8%) | 18 (85.7%) | 88 (95.7%) | |
| No | 7 (6.2%) | 3 (14.3%) | 4 (4.3%) | |
| Waist (cm) | 105.86 ± 11.14 | 100.86 ± 8.47 | 107.00 ± 11.39 | 113 |
| Aspartate aminotransferase (U/L) | 24.54 ± 7.84 | 21.52 ± 5.18 | 25.23 ± 8.19 | 113 |
| Alanine transaminase (U/L) | 30.51 ± 13.57 | 22.62 ± 8.18 | 32.32 ± 13.93 | 113 |
| Gamma-glutamyl transferase (U/L) | 30.35 ± 18.76 | 22.90 ± 10.20 | 32.05 ± 19.86 | 113 |
| Creatinine (mg/dL) | 0.97 ± 0.16 | 0.82 ± 0.14 | 1.00 ± 0.14 | 113 |
| Estimated glomerular filtration rate (CKD-EPI) | 83.78 ± 13.00 | 81.62 ± 15.14 | 84.28 ± 12.51 | 113 |
| Total protein (g/dL) | 7.08 ± 0.33 | 7.12 ± 0.33 | 7.07 ± 0.33 | 105 |
| Platelets (n/L) | 238.30 ± 55.79 | 284.14 ± 62.39 | 227.84 ± 48.78 | 113 |
| Hemo (g/dL) | 14.79 ± 1.08 | 13.41 ± 1.06 | 15.11 ± 0.81 | 113 |
| Ferritin (ng/mL) | 125.94 ± 120.18 | 71.43 ± 59.44 | 138.38 ± 127.11 | 113 |
| Glucose (mg/dL) | 116.32 ± 15.74 | 112.81 ± 15.72 | 117.12 ± 15.72 | 113 |
| Cholesterol (mg/dL) | 201.07 ± 33.50 | 205.29 ± 36.74 | 200.11 ± 32.85 | 113 |
| Cholesterol HDL (mg/dL) | 47.36 ± 11.48 | 53.00 ± 11.87 | 46.08 ± 11.06 | 113 |
| Triglycerides (mg/dL) | 175.45 ± 91.44 | 162.62 ± 52.56 | 178.38 ± 98.17 | 113 |
| Cholesterol LDL (mg/dL) | 118.62 ± 30.42 | 119.76 ± 30.88 | 118.36 ± 30.49 | 113 |
| Proprotein convertase subtilisin/kexin type 9 (ng/mL) | 303.10 ± 81.01 | 315.84 ± 63.25 | 301.73 ± 83.52 | 31 |
| Hypercholesterolemia | | | | 113 |
| No | 55 (48.7%) | 10 (47.6%) | 45 (48.9%) | |
| Yes | 58 (51.3%) | 11 (52.4%) | 47 (51.1%) | |
| HbA1c (mmol/mol) | 40.96 ± 7.96 | 43.90 ± 13.10 | 40.29 ± 6.15 | 113 |
| Insulin (µIU/mL) | 18.12 ± 10.14 | 15.46 ± 8.24 | 18.73 ± 10.47 | 113 |
| Homeostatic model assessment for insulin resistance | 5.22 ± 3.11 | 4.30 ± 2.66 | 5.43 ± 3.18 | 113 |
| Urine albumin to creatinine ratio (mg/g) | 18.27 ± 46.37 | 32.08 ± 95.53 | 14.98 ± 22.86 | 104 |
| Albuminuria | | | | 104 |
| No | 92 (88.5%) | 18 (90.0%) | 74 (88.1%) | |
| Yes | 12 (11.5%) | 2 (10.0%) | 10 (11.9%) | |
| Hyperglycemia | | | | 113 |
| No | 62 (54.9%) | 10 (47.6%) | 52 (56.5%) | |
| Yes | 51 (45.1%) | 11 (52.4%) | 40 (43.5%) | |

| Variable | All (N=113) | Women (N=21, 18.6%) | Men (N=92, 81.4%) | N |
|---|---|---|---|---|
| Diabetes type 2 | | | | 113 |
| No | 104 (92.0%) | 18 (85.7%) | 86 (93.5%) | |
| Yes | 9 (8.0%) | 3 (14.3%) | 6 (6.5%) | |
| Basophils (n/L) | 0.05 ± 0.02 | 0.05 ± 0.02 | 0.05 ± 0.02 | 113 |
| Hematocrit (%) | 42.82 ± 2.82 | 39.45 ± 2.90 | 43.59 ± 2.17 | 113 |
| Eosinophils (n/L) | 0.18 ± 0.11 | 0.16 ± 0.10 | 0.18 ± 0.12 | 113 |
| Red blood cell count (millions/µL) | 4.98 ± 0.38 | 4.57 ± 0.32 | 5.07 ± 0.33 | 113 |
| Leukocytes (n/L) | 6.46 ± 1.38 | 6.64 ± 1.35 | 6.42 ± 1.38 | 113 |
| Lymphocytes (n/L) | 2.03 ± 0.52 | 2.26 ± 0.47 | 1.98 ± 0.52 | 113 |
| Corpuscular hemoglobin (pg) | 29.78 ± 1.73 | 29.36 ± 1.78 | 29.87 ± 1.72 | 113 |
| Corpuscular hemoglobin (g/dL) | 34.54 ± 1.11 | 33.99 ± 0.87 | 34.67 ± 1.12 | 113 |
| Monocytes (n/L) | 0.53 ± 0.13 | 0.46 ± 0.11 | 0.54 ± 0.13 | 113 |
| Red cell distribution width | 12.99 ± 0.84 | 13.07 ± 1.38 | 12.97 ± 0.67 | 113 |
| Corpuscular volume (fL) | 86.27 ± 3.88 | 86.38 ± 3.87 | 86.25 ± 3.90 | 113 |
| Smoker | | | | 112 |
| No | 105 (93.8%) | 20 (95.2%) | 85 (93.4%) | |
| Yes | 7 (6.2%) | 1 (4.8%) | 6 (6.6%) | |
| Atherogenic risk index | 0.07 ± 0.04 | 0.02 ± 0.01 | 0.08 ± 0.04 | 113 |
| CRP (mg/dL) | 0.28 ± 0.31 | 0.29 ± 0.26 | 0.28 ± 0.32 | 109 |
| Apolipoprotein B (mg/mL) | 0.97 ± 0.28 | 0.84 ± 0.31 | 1.01 ± 0.27 | 62 |

**Table S10 | General characteristics for the Covid-19 cohort.** Values are shown as mean ± standard deviation for continuous variables, and as counts (percentages) for categorical variables.

| Variable | All (N=470) | Women (N=244, 51.9%) | Men (N=226, 48.1%) | N |
|---|---|---|---|---|
| Age | 68.42 ± 17.50 | 68.67 ± 20.01 | 68.15 ± 14.35 | 470 |
| Smoker | | | | 388 |
| Former | 98 (25.3%) | 27 (13.3%) | 71 (38.4%) | |
| No | 269 (69.3%) | 166 (81.8%) | 103 (55.7%) | |
| Yes | 21 (5.4%) | 10 (4.9%) | 11 (5.9%) | |
| Chronic obstructive pulmonary disease | | | | 389 |
| No | 355 (91.3%) | 194 (95.6%) | 161 (86.6%) | |
| Yes | 34 (8.7%) | 9 (4.4%) | 25 (13.4%) | |
| Diabetes | | | | 389 |
| No | 297 (76.3%) | 161 (79.3%) | 136 (73.1%) | |
| Yes | 92 (23.7%) | 42 (20.7%) | 50 (26.9%) | |
| Hypertension | | | | 389 |
| No | 211 (54.2%) | 115 (56.7%) | 96 (51.6%) | |
| Yes | 178 (45.8%) | 88 (43.3%) | 90 (48.4%) | |
| Cardiovascular | | | | 389 |
| No | 282 (72.5%) | 154 (75.9%) | 128 (68.8%) | |
| Yes | 107 (27.5%) | 49 (24.1%) | 58 (31.2%) | |
| Cerebrovascular | | | | 389 |
| No | 346 (88.9%) | 179 (88.2%) | 167 (89.8%) | |
| Yes | 43 (11.1%) | 24 (11.8%) | 19 (10.2%) | |
| Hepatitis B virus | | | | 389 |
| No | 388 (99.7%) | 202 (99.5%) | 186 (100.0%) | |
| Yes | 1 (0.3%) | 1 (0.5%) | 0 (0.0%) | |
| Neoplasm | | | | 261 |
| No | 229 (87.7%) | 106 (88.3%) | 123 (87.2%) | |
| Yes | 32 (12.3%) | 14 (11.7%) | 18 (12.8%) | |
| Renal insufficiency | | | | 389 |
| No | 359 (92.3%) | 189 (93.1%) | 170 (91.4%) | |
| Yes | 30 (7.7%) | 14 (6.9%) | 16 (8.6%) | |
| Immunodeficiency | | | | 389 |
| Yes (neoplasm) | 2 (0.5%) | 2 (1.0%) | 0 (0.0%) | |
| No | 374 (96.1%) | 193 (95.1%) | 181 (97.3%) | |
| Yes | 13 (3.3%) | 8 (3.9%) | 5 (2.7%) | |
| Liver failure | | | | 389 |
| No | 382 (98.2%) | 201 (99.0%) | 181 (97.3%) | |
| Yes | 7 (1.8%) | 2 (1.0%) | 5 (2.7%) | |
| Conjunctival congestion | | | | 261 |
| No | 258 (98.9%) | 117 (97.5%) | 141 (100.0%) | |
| Yes | 3 (1.1%) | 3 (2.5%) | 0 (0.0%) | |
| Nasal congestion | | | | 261 |
| No | 250 (95.8%) | 118 (98.3%) | 132 (93.6%) | |
| Yes | 11 (4.2%) | 2 (1.7%) | 9 (6.4%) | |
| Headache | | | | 261 |
| No | 210 (80.5%) | 90 (75.0%) | 120 (85.1%) | |

| Variable | All (N=470) | Women (N=244, 51.9%) | Men (N=226, 48.1%) | N |
|---|---|---|---|---|
| Yes | 51 (19.5%) | 30 (25.0%) | 21 (14.9%) | |
| Dry cough | | | | 261 |
| No | 123 (47.1%) | 47 (39.2%) | 76 (53.9%) | |
| Yes | 138 (52.9%) | 73 (60.8%) | 65 (46.1%) | |
| Odynophagia | | | | 261 |
| No | 230 (88.1%) | 101 (84.2%) | 129 (91.5%) | |
| Yes | 31 (11.9%) | 19 (15.8%) | 12 (8.5%) | |
| Productive cough | | | | 261 |
| No | 203 (77.8%) | 102 (85.0%) | 101 (71.6%) | |
| Yes | 58 (22.2%) | 18 (15.0%) | 40 (28.4%) | |
| Fatigue | | | | 261 |
| No | 111 (42.5%) | 52 (43.3%) | 59 (41.8%) | |
| Yes | 150 (57.5%) | 68 (56.7%) | 82 (58.2%) | |
| Hemoptysis | | | | 261 |
| No | 260 (99.6%) | 120 (100.0%) | 140 (99.3%) | |
| Yes | 1 (0.4%) | 0 (0.0%) | 1 (0.7%) | |
| Diarrhea | | | | 261 |
| No | 184 (70.5%) | 80 (66.7%) | 104 (73.8%) | |
| Yes | 77 (29.5%) | 40 (33.3%) | 37 (26.2%) | |
| Nausea vomiting | | | | 260 |
| No | 221 (85.0%) | 94 (78.3%) | 127 (90.7%) | |
| Yes | 39 (15.0%) | 26 (21.7%) | 13 (9.3%) | |
| Myalgia | | | | 261 |
| No | 185 (70.9%) | 85 (70.8%) | 100 (70.9%) | |
| Yes | 76 (29.1%) | 35 (29.2%) | 41 (29.1%) | |
| Shaking chills | | | | 261 |
| No | 205 (78.5%) | 95 (79.2%) | 110 (78.0%) | |
| Yes | 56 (21.5%) | 25 (20.8%) | 31 (22.0%) | |
| Oropharyngeal congestion | | | | 261 |
| No | 260 (99.6%) | 119 (99.2%) | 141 (100.0%) | |
| Yes | 1 (0.4%) | 1 (0.8%) | 0 (0.0%) | |
| Tonsil inflammation | | | | 261 |
| No | 261 (100.0%) | 120 (100.0%) | 141 (100.0%) | |
| Skin rash | | | | 261 |
| No | 258 (98.9%) | 119 (99.2%) | 139 (98.6%) | |
| Yes | 3 (1.1%) | 1 (0.8%) | 2 (1.4%) | |
| Lymphadenopathy | | | | 261 |
| No | 258 (98.9%) | 119 (99.2%) | 139 (98.6%) | |
| Yes | 3 (1.1%) | 1 (0.8%) | 2 (1.4%) | |
| Clouding of consciousness | | | | 261 |
| No | 241 (92.3%) | 109 (90.8%) | 132 (93.6%) | |
| Yes | 20 (7.7%) | 11 (9.2%) | 9 (6.4%) | |
| Disorientation | | | | 380 |
| No | 351 (92.4%) | 182 (92.4%) | 169 (92.3%) | |
| Yes | 29 (7.6%) | 15 (7.6%) | 14 (7.7%) | |

| Variable | All (N=470) | Women (N=244, 51.9%) | Men (N=226, 48.1%) | N |
|---|---|---|---|---|
| Fever | | | | 261 |
| No | 82 (31.4%) | 48 (40.0%) | 34 (24.1%) | |
| Yes | 179 (68.6%) | 72 (60.0%) | 107 (75.9%) | |
| Temperature (ºC) | 36.48 ± 0.91 | 36.42 ± 0.83 | 36.52 ± 0.98 | 253 |
| Breathing frequency (n/min) | 22.45 ± 7.97 | 20.81 ± 7.50 | 23.55 ± 8.17 | 67 |
| Heart rate (n/min) | 91.24 ± 17.66 | 90.90 ± 16.05 | 91.54 ± 18.98 | 254 |
| Systolic blood pressure (mmHg) | 135.69 ± 23.15 | 133.58 ± 21.88 | 137.52 ± 24.13 | 254 |
| Diastolic blood pressure (mmHg) | 78.20 ± 12.70 | 76.92 ± 13.04 | 79.31 ± 12.34 | 255 |
| Glucose (mg/dL) | 140.94 ± 107.46 | 142.32 ± 131.40 | 139.74 ± 81.51 | 264 |
| Creatinine (mg/dL) | 1.06 ± 0.76 | 0.93 ± 0.62 | 1.18 ± 0.86 | 329 |
| Urea (mg/dL) | 45.06 ± 30.29 | 39.81 ± 26.44 | 49.69 ± 32.70 | 265 |
| Bilirubin (mg/dL) | 0.68 ± 0.43 | 0.61 ± 0.39 | 0.75 ± 0.46 | 195 |
| Alanine aminotransferase (U/L) | 37.03 ± 42.14 | 32.70 ± 48.06 | 41.33 ± 34.92 | 317 |
| Protein (g/dL) | 6.35 ± 0.57 | 6.36 ± 0.51 | 6.34 ± 0.61 | 146 |
| Albumin (g/dL) | 3.79 ± 0.46 | 3.82 ± 0.40 | 3.76 ± 0.51 | 180 |
| Lactate dehydrogenase (U/L) | 307.80 ± 192.05 | 291.56 ± 225.54 | 323.62 ± 151.63 | 304 |
| Ferritin (ng/mL) | 638.31 ± 729.01 | 369.14 ± 364.10 | 894.91 ± 882.81 | 293 |
| Procalcitonin (ng/mL) | 0.46 ± 1.34 | 0.45 ± 1.80 | 0.46 ± 0.91 | 143 |
| CRP (mg/L) | 77.55 ± 71.86 | 64.35 ± 66.60 | 90.76 ± 74.64 | 328 |
| Creatine phosphokinase (U/L) | 171.87 ± 469.07 | 182.51 ± 604.99 | 161.37 ± 277.98 | 296 |
| Interleukin-6 (pg/mL) | 19.50 ± 24.78 | - | 19.50 ± 24.78 | 12 |
| Leukocytes (n/L) | 167.82 ± 1200.79 | 189.88 ± 1338.24 | 145.89 ± 1050.14 | 331 |
| Lymphocytes (n/L) | 23.07 ± 175.54 | 30.10 ± 221.16 | 16.04 ± 113.16 | 330 |
| Neutrophils (n/L) | 137.12 ± 1012.94 | 152.20 ± 1122.40 | 122.39 ± 896.53 | 328 |
| Monocytes (n/L) | 0.42 ± 0.24 | 0.41 ± 0.21 | 0.43 ± 0.26 | 261 |
| Platelets (n/L) | 4522.63 ± 33232.96 | 4395.21 ± 32067.14 | 4650.04 ± 34456.63 | 330 |
| Activated partial thromboplastin time (s) | 24.37 ± 4.95 | 24.16 ± 5.25 | 24.55 ± 4.69 | 256 |
| Prothrombin activity (%) | 88.26 ± 20.64 | 90.11 ± 21.41 | 86.36 ± 19.71 | 297 |
| D-dimer (ng/mL) | 2236.99 ± 8130.65 | 2084.56 ± 7668.10 | 2387.51 ± 8584.47 | 318 |
| Antiviral treatment | | | | 380 |
| No | 176 (46.3%) | 104 (52.3%) | 72 (39.8%) | |
| Yes | 204 (53.7%) | 95 (47.7%) | 109 (60.2%) | |
| Antibiotic treatment | | | | 380 |
| No | 143 (37.6%) | 85 (42.7%) | 58 (32.0%) | |
| Yes | 237 (62.4%) | 114 (57.3%) | 123 (68.0%) | |
| Immunomodulatory treatment | | | | 380 |
| No | 358 (94.2%) | 185 (93.0%) | 173 (95.6%) | |
| Yes | 22 (5.8%) | 14 (7.0%) | 8 (4.4%) | |
| Antioxidant treatment | | | | 380 |
| No | 370 (97.4%) | 195 (98.0%) | 175 (96.7%) | |
| Yes | 10 (2.6%) | 4 (2.0%) | 6 (3.3%) | |
| Antiaggregant treatment | | | | 380 |
| No | 365 (96.1%) | 192 (96.5%) | 173 (95.6%) | |
| Yes | 15 (3.9%) | 7 (3.5%) | 8 (4.4%) | |
| Anticoagulant treatment | | | | 380 |
| No | 159 (41.8%) | 88 (44.2%) | 71 (39.2%) | |
| Yes | 221 (58.2%) | 111 (55.8%) | 110 (60.8%) | |

| Variable | All (N=470) | Women (N=244, 51.9%) | Men (N=226, 48.1%) | N |
|---|---|---|---|---|
| Corticoids treatment | | | | 382 |
| No | 229 (59.9%) | 140 (70.4%) | 89 (48.6%) | |
| Yes | 153 (40.1%) | 59 (29.6%) | 94 (51.4%) | |
| Vasopressor treatment | | | | 377 |
| No | 360 (95.5%) | 192 (97.5%) | 168 (93.3%) | |
| Yes | 17 (4.5%) | 5 (2.5%) | 12 (6.7%) | |
| Antipaludic treatment | | | | 380 |
| No | 114 (30.0%) | 68 (34.2%) | 46 (25.4%) | |
| Yes | 266 (70.0%) | 131 (65.8%) | 135 (74.6%) | |
| Pneumonia | | | | 252 |
| Bilateral | 175 (69.4%) | 71 (60.7%) | 104 (77.0%) | |
| Normal | 44 (17.5%) | 30 (25.6%) | 14 (10.4%) | |
| Unilateral | 33 (13.1%) | 16 (13.7%) | 17 (12.6%) | |
| SOFA score | 2.83 ± 3.58 | 3.38 ± 4.41 | 2.53 ± 3.18 | 23 |
| CURB-65 score | | | | 109 |
| 0 | 46 (42.2%) | 26 (51.0%) | 20 (34.5%) | |
| 1 | 30 (27.5%) | 14 (27.5%) | 16 (27.6%) | |
| 2 | 26 (23.9%) | 10 (19.6%) | 16 (27.6%) | |
| 3 | 7 (6.4%) | 1 (2.0%) | 6 (10.3%) | |
| Intubation days | 24.04 ± 17.57 | 22.83 ± 14.61 | 24.40 ± 18.70 | 26 |
| Nasal cannulas | | | | 364 |
| No | 193 (53.0%) | 110 (57.3%) | 83 (48.3%) | |
| Yes | 171 (47.0%) | 82 (42.7%) | 89 (51.7%) | |
| Venturi mask | | | | 360 |
| No | 294 (81.7%) | 169 (89.4%) | 125 (73.1%) | |
| Yes | 66 (18.3%) | 20 (10.6%) | 46 (26.9%) | |
| Continuous positive airway pressure (CPAP) | | | | 350 |
| No | 333 (95.1%) | 178 (96.7%) | 155 (93.4%) | |
| Yes | 17 (4.9%) | 6 (3.3%) | 11 (6.6%) | |
| High flow nasal cannula | | | | 350 |
| No | 330 (94.3%) | 178 (96.7%) | 152 (91.6%) | |
| Yes | 20 (5.7%) | 6 (3.3%) | 14 (8.4%) | |
| Extracorporeal membrane oxygenation (ECMO) | | | | 224 |
| No | 223 (99.6%) | 102 (100.0%) | 121 (99.2%) | |
| Yes | 1 (0.4%) | 0 (0.0%) | 1 (0.8%) | |
| ECMO days | | | | 1 |
| 5.0 | 1 (100.0%) | 0 (0.0%) | 1 (100.0%) | |
| Hemofiltration | | | | 222 |
| No | 219 (98.6%) | 99 (98.0%) | 120 (99.2%) | |
| Yes | 3 (1.4%) | 2 (2.0%) | 1 (0.8%) | |
| Hemofiltration days | 34 | 34 | - | 1 |
| Is dead | | | | 396 |
| No | 357 (90.2%) | 189 (91.3%) | 168 (88.9%) | |
| Yes | 39 (9.8%) | 18 (8.7%) | 21 (11.1%) | |
| Still in ICU | | | | 396 |
| No | 390 (98.5%) | 203 (98.1%) | 187 (98.9%) | |

| Variable | All (N=470) | Women (N=244, 51.9%) | Men (N=226, 48.1%) | N |
|---|---|---|---|---|
| Yes | 6 (1.5%) | 4 (1.9%) | 2 (1.1%) | |
| Still at hospital | | | | 396 |
| No | 384 (97.0%) | 201 (97.1%) | 183 (96.8%) | |
| Yes | 12 (3.0%) | 6 (2.9%) | 6 (3.2%) | |
| Still at home | | | | 396 |
| No | 395 (99.7%) | 206 (99.5%) | 189 (100.0%) | |
| Yes | 1 (0.3%) | 1 (0.5%) | 0 (0.0%) | |
| Number hospital days | 10.78 ± 16.78 | 9.29 ± 16.74 | 12.41 ± 16.72 | 396 |
| Number ICU days | 0.39 ± 1.98 | 0.21 ± 1.25 | 0.59 ± 2.54 | 396 |
| Number home days | 3.31 ± 13.59 | 2.53 ± 13.35 | 4.16 ± 13.83 | 396 |
| Is dead | 0.77 ± 3.45 | 1.03 ± 4.28 | 0.49 ± 2.17 | 396 |

**Table S11 | General characteristics for the Long Covid-19 cohort.** Values are shown as mean ± standard deviation for continuous variables, and as counts (percentages) for categorical variables.

| Variable | All (N=231) | Women (N=179, 77.5%) | Men (N=52, 22.5%) | N |
|---|---|---|---|---|
| Age | 49.31 ± 8.63 | 48.87 ± 8.64 | 50.83 ± 8.48 | 231 |
| BMI | 26.03 ± 5.50 | 25.82 ± 5.65 | 26.84 ± 4.91 | 205 |
| Weight (kg) | 71.36 ± 16.80 | 68.25 ± 15.70 | 83.07 ± 15.79 | 205 |
| Height (cm) | 165.30 ± 8.45 | 162.51 ± 6.30 | 175.73 ± 7.19 | 208 |
| Days from Covid | 549.39 ± 195.94 | 546.41 ± 195.45 | 560.49 ± 199.69 | 203 |
| Number hospital days | 3.08 ± 6.43 | 2.49 ± 6.47 | 5.27 ± 5.88 | 208 |
| Number ICU days | 0.54 ± 3.20 | 0.31 ± 1.88 | 1.42 ± 5.95 | 206 |
| Has Covid antibodies | | | | 172 |
| Don´t know | 60 (34.9%) | 48 (35.6%) | 12 (32.4%) | |
| No | 31 (18.0%) | 24 (17.8%) | 7 (18.9%) | |
| Yes | 81 (47.1%) | 63 (46.7%) | 18 (48.6%) | |
| Smoker | | | | 208 |
| Former | 47 (22.6%) | 38 (23.2%) | 9 (20.5%) | |
| No | 153 (73.6%) | 120 (73.2%) | 33 (75.0%) | |
| Yes | 8 (3.8%) | 6 (3.7%) | 2 (4.5%) | |
| Diabetes | | | | 208 |
| No | 199 (95.7%) | 157 (95.7%) | 42 (95.5%) | |
| Yes | 9 (4.3%) | 7 (4.3%) | 2 (4.5%) | |
| Diabetes type 1 | | | | 208 |
| No | 206 (99.0%) | 162 (98.8%) | 44 (100.0%) | |
| Yes | 2 (1.0%) | 2 (1.2%) | 0 (0.0%) | |
| Diabetes type 2 | | | | 208 |
| No | 202 (97.1%) | 160 (97.6%) | 42 (95.5%) | |
| Yes | 6 (2.9%) | 4 (2.4%) | 2 (4.5%) | |
| Cardiovascular | | | | 208 |
| No | 190 (91.3%) | 151 (92.1%) | 39 (88.6%) | |
| Yes | 18 (8.7%) | 13 (7.9%) | 5 (11.4%) | |
| Cerebrovascular | | | | 208 |
| No | 206 (99.0%) | 162 (98.8%) | 44 (100.0%) | |
| Yes | 2 (1.0%) | 2 (1.2%) | 0 (0.0%) | |
| Chronic obstructive pulmonary disease | | | | 208 |
| No | 203 (97.6%) | 162 (98.8%) | 41 (93.2%) | |
| Yes | 5 (2.4%) | 2 (1.2%) | 3 (6.8%) | |
| Immunodeficiency | | | | 208 |
| No | 203 (97.6%) | 159 (97.0%) | 44 (100.0%) | |
| Yes | 5 (2.4%) | 5 (3.0%) | 0 (0.0%) | |
| Hepatitis B virus | | | | 208 |
| No | 205 (98.6%) | 162 (98.8%) | 43 (97.7%) | |
| Yes | 3 (1.4%) | 2 (1.2%) | 1 (2.3%) | |
| Liver failure | | | | 208 |
| No | 206 (99.0%) | 164 (100.0%) | 42 (95.5%) | |
| Yes | 2 (1.0%) | 0 (0.0%) | 2 (4.5%) | |
| Renal insufficiency | | | | 208 |
| No | 202 (97.1%) | 159 (97.0%) | 43 (97.7%) | |
| Yes | 6 (2.9%) | 5 (3.0%) | 1 (2.3%) | |

| Variable | All (N=231) | Women (N=179, 77.5%) | Men (N=52, 22.5%) | N |
|---|---|---|---|---|
| IBD | | | | 205 |
| No | 194 (94.6%) | 153 (95.0%) | 41 (93.2%) | |
| Yes | 11 (5.4%) | 8 (5.0%) | 3 (6.8%) | |
| Hypertension | | | | 205 |
| No | 182 (88.8%) | 147 (91.3%) | 35 (79.5%) | |
| Yes | 23 (11.2%) | 14 (8.7%) | 9 (20.5%) | |
| Arthritis | | | | 203 |
| No | 190 (93.6%) | 151 (95.0%) | 39 (88.6%) | |
| Yes | 13 (6.4%) | 8 (5.0%) | 5 (11.4%) | |
| Obesity | | | | 205 |
| No | 174 (84.9%) | 137 (85.1%) | 37 (84.1%) | |
| Yes | 31 (15.1%) | 24 (14.9%) | 7 (15.9%) | |
| Current acute Covid | | | | 204 |
| No | 180 (88.2%) | 139 (86.3%) | 41 (95.3%) | |
| Yes | 24 (11.8%) | 22 (13.7%) | 2 (4.7%) | |
| Asthenia | | | | 207 |
| No | 15 (7.2%) | 11 (6.7%) | 4 (9.1%) | |
| Yes | 192 (92.8%) | 152 (93.3%) | 40 (90.9%) | |
| Asthenia (post Covid) | | | | 205 |
| No | 5 (2.4%) | 4 (2.5%) | 1 (2.3%) | |
| Yes | 200 (97.6%) | 157 (97.5%) | 43 (97.7%) | |
| Malaise | | | | 207 |
| No | 11 (5.3%) | 7 (4.3%) | 4 (9.1%) | |
| Yes | 196 (94.7%) | 156 (95.7%) | 40 (90.9%) | |
| Malaise (post Covid) | | | | 204 |
| No | 34 (16.7%) | 24 (15.0%) | 10 (22.7%) | |
| Yes | 170 (83.3%) | 136 (85.0%) | 34 (77.3%) | |
| Headache | | | | 206 |
| No | 43 (20.9%) | 29 (17.9%) | 14 (31.8%) | |
| Yes | 163 (79.1%) | 133 (82.1%) | 30 (68.2%) | |
| Headache (post Covid) | | | | 225 |
| No | 69 (30.7%) | 46 (26.6%) | 23 (44.2%) | |
| Yes | 156 (69.3%) | 127 (73.4%) | 29 (55.8%) | |
| Ageusia | | | | 205 |
| No | 104 (50.7%) | 78 (48.4%) | 26 (59.1%) | |
| Yes | 101 (49.3%) | 83 (51.6%) | 18 (40.9%) | |
| Ageusia (post Covid) | | | | 202 |
| No | 163 (80.7%) | 122 (77.2%) | 41 (93.2%) | |
| Yes | 39 (19.3%) | 36 (22.8%) | 3 (6.8%) | |
| Anosmia | | | | 204 |
| No | 104 (51.0%) | 77 (48.1%) | 27 (61.4%) | |
| Yes | 100 (49.0%) | 83 (51.9%) | 17 (38.6%) | |
| Anosmia (post Covid) | | | | 201 |
| No | 151 (75.1%) | 112 (71.3%) | 39 (88.6%) | |
| Yes | 50 (24.9%) | 45 (28.7%) | 5 (11.4%) | |

| Variable | All (N=231) | Women (N=179, 77.5%) | Men (N=52, 22.5%) | N |
|---|---|---|---|---|
| Low mood | | | | 205 |
| No | 84 (41.0%) | 72 (44.7%) | 12 (27.3%) | |
| Yes | 121 (59.0%) | 89 (55.3%) | 32 (72.7%) | |
| Low mood (post Covid) | | | | 204 |
| No | 46 (22.5%) | 37 (23.1%) | 9 (20.5%) | |
| Yes | 158 (77.5%) | 123 (76.9%) | 35 (79.5%) | |
| Myalgia | | | | 207 |
| No | 30 (14.5%) | 22 (13.5%) | 8 (18.2%) | |
| Yes | 177 (85.5%) | 141 (86.5%) | 36 (81.8%) | |
| Myalgia (post Covid) | | | | 204 |
| No | 19 (9.3%) | 15 (9.4%) | 4 (9.1%) | |
| Yes | 185 (90.7%) | 145 (90.6%) | 40 (90.9%) | |
| Dyspnea | | | | 207 |
| No | 52 (25.1%) | 44 (27.0%) | 8 (18.2%) | |
| Yes | 155 (74.9%) | 119 (73.0%) | 36 (81.8%) | |
| Dyspnea (post Covid) | | | | 227 |
| No | 56 (24.7%) | 43 (24.6%) | 13 (25.0%) | |
| Yes | 171 (75.3%) | 132 (75.4%) | 39 (75.0%) | |
| Arthralgia | | | | 207 |
| No | 63 (30.4%) | 48 (29.4%) | 15 (34.1%) | |
| Yes | 144 (69.6%) | 115 (70.6%) | 29 (65.9%) | |
| Arthralgia (post Covid) | | | | 205 |
| No | 32 (15.6%) | 23 (14.3%) | 9 (20.5%) | |
| Yes | 173 (84.4%) | 138 (85.7%) | 35 (79.5%) | |
| Absent mindedness | | | | 205 |
| No | 95 (46.3%) | 74 (46.0%) | 21 (47.7%) | |
| Yes | 110 (53.7%) | 87 (54.0%) | 23 (52.3%) | |
| Absent mindedness (post Covid) | | | | 205 |
| No | 21 (10.2%) | 11 (6.8%) | 10 (22.7%) | |
| Yes | 184 (89.8%) | 150 (93.2%) | 34 (77.3%) | |
| Back pain | | | | 206 |
| No | 73 (35.4%) | 52 (31.9%) | 21 (48.8%) | |
| Yes | 133 (64.6%) | 111 (68.1%) | 22 (51.2%) | |
| Back pain (post Covid) | | | | 202 |
| No | 58 (28.7%) | 42 (26.6%) | 16 (36.4%) | |
| Yes | 144 (71.3%) | 116 (73.4%) | 28 (63.6%) | |
| Chest pressure | | | | 206 |
| No | 88 (42.7%) | 67 (41.4%) | 21 (47.7%) | |
| Yes | 118 (57.3%) | 95 (58.6%) | 23 (52.3%) | |
| Chest pressure (post Covid) | | | | 203 |
| No | 86 (42.4%) | 65 (40.9%) | 21 (47.7%) | |
| Yes | 117 (57.6%) | 94 (59.1%) | 23 (52.3%) | |
| Anxiety | | | | 206 |
| No | 128 (62.1%) | 105 (64.8%) | 23 (52.3%) | |
| Yes | 78 (37.9%) | 57 (35.2%) | 21 (47.7%) | |

| Variable | All (N=231) | Women (N=179, 77.5%) | Men (N=52, 22.5%) | N |
|---|---|---|---|---|
| Anxiety (post Covid) | | | | 204 |
| No | 88 (43.1%) | 71 (44.4%) | 17 (38.6%) | |
| Yes | 116 (56.9%) | 89 (55.6%) | 27 (61.4%) | |
| Fever | | | | 207 |
| No | 51 (24.6%) | 41 (25.2%) | 10 (22.7%) | |
| Yes | 156 (75.4%) | 122 (74.8%) | 34 (77.3%) | |
| Fever (post Covid) | | | | 204 |
| No | 144 (70.6%) | 108 (67.1%) | 36 (83.7%) | |
| Yes | 60 (29.4%) | 53 (32.9%) | 7 (16.3%) | |
| Cough | | | | 207 |
| No | 64 (30.9%) | 47 (28.8%) | 17 (38.6%) | |
| Yes | 143 (69.1%) | 116 (71.2%) | 27 (61.4%) | |
| Cough (post Covid) | | | | 204 |
| No | 121 (59.3%) | 89 (55.6%) | 32 (72.7%) | |
| Yes | 83 (40.7%) | 71 (44.4%) | 12 (27.3%) | |
| Memory problems | | | | 206 |
| No | 117 (56.8%) | 90 (55.2%) | 27 (62.8%) | |
| Yes | 89 (43.2%) | 73 (44.8%) | 16 (37.2%) | |
| Memory problems (post Covid) | | | | 203 |
| No | 26 (12.8%) | 15 (9.4%) | 11 (25.6%) | |
| Yes | 177 (87.2%) | 145 (90.6%) | 32 (74.4%) | |
| Neck pain | | | | 208 |
| No | 92 (44.2%) | 68 (41.5%) | 24 (54.5%) | |
| Yes | 116 (55.8%) | 96 (58.5%) | 20 (45.5%) | |
| Neck pain (post Covid) | | | | 202 |
| No | 64 (31.7%) | 44 (27.8%) | 20 (45.5%) | |
| Yes | 138 (68.3%) | 114 (72.2%) | 24 (54.5%) | |
| Diarrhea | | | | 207 |
| No | 104 (50.2%) | 79 (48.5%) | 25 (56.8%) | |
| Yes | 103 (49.8%) | 84 (51.5%) | 19 (43.2%) | |
| Diarrhea (post Covid) | | | | 204 |
| No | 117 (57.4%) | 91 (56.9%) | 26 (59.1%) | |
| Yes | 87 (42.6%) | 69 (43.1%) | 18 (40.9%) | |
| Chest pain | | | | 207 |
| No | 100 (48.3%) | 78 (47.9%) | 22 (50.0%) | |
| Yes | 107 (51.7%) | 85 (52.1%) | 22 (50.0%) | |
| Chest pain (post Covid) | | | | 203 |
| No | 91 (44.8%) | 69 (43.4%) | 22 (50.0%) | |
| Yes | 112 (55.2%) | 90 (56.6%) | 22 (50.0%) | |
| Palpitations | | | | 207 |
| No | 120 (58.0%) | 94 (57.7%) | 26 (59.1%) | |
| Yes | 87 (42.0%) | 69 (42.3%) | 18 (40.9%) | |
| Palpitations (post Covid) | | | | 227 |
| No | 84 (37.0%) | 59 (33.7%) | 25 (48.1%) | |
| Yes | 143 (63.0%) | 116 (66.3%) | 27 (51.9%) | |

| **Variable** | **All (N=231)** | **Women (N=179, 77.5%)** | **Men (N=52, 22.5%)** | **N** |
|---|---|---|---|---|
| Dizziness | | | | 207 |
| No | 110 (53.1%) | 81 (49.7%) | 29 (65.9%) | |
| Yes | 97 (46.9%) | 82 (50.3%) | 15 (34.1%) | |
| Dizziness (post Covid) | | | | 225 |
| No | 91 (40.4%) | 66 (38.2%) | 25 (48.1%) | |
| Yes | 134 (59.6%) | 107 (61.8%) | 27 (51.9%) | |
| Paresthesia | | | | 207 |
| No | 124 (59.9%) | 96 (58.9%) | 28 (63.6%) | |
| Yes | 83 (40.1%) | 67 (41.1%) | 16 (36.4%) | |
| Paresthesia (post Covid) | | | | 227 |
| No | 81 (35.7%) | 60 (34.3%) | 21 (40.4%) | |
| Yes | 146 (64.3%) | 115 (65.7%) | 31 (59.6%) | |
| Pain (post Covid) | | | | 23 |
| No | 7 (30.4%) | 3 (20.0%) | 4 (50.0%) | |
| Yes | 16 (69.6%) | 12 (80.0%) | 4 (50.0%) | |
| Insomnia (post Covid) | | | | 23 |
| No | 8 (34.8%) | 4 (26.7%) | 4 (50.0%) | |
| Yes | 15 (65.2%) | 11 (73.3%) | 4 (50.0%) | |
| Fatigue (post Covid) | | | | 23 |
| No | 4 (17.4%) | 2 (13.3%) | 2 (25.0%) | |
| Yes | 19 (82.6%) | 13 (86.7%) | 6 (75.0%) | |
| Brain fog (post Covid) | | | | 23 |
| No | 6 (26.1%) | 2 (13.3%) | 4 (50.0%) | |
| Yes | 17 (73.9%) | 13 (86.7%) | 4 (50.0%) | |
| Digestive problem (post Covid) | | | | 23 |
| No | 15 (65.2%) | 9 (60.0%) | 6 (75.0%) | |
| Yes | 8 (34.8%) | 6 (40.0%) | 2 (25.0%) | |

**Table S12 | General characteristics for the Influenza cohort.** Values are shown as mean ± standard deviation for continuous variables, and as counts (percentages) for categorical variables.

| Variable | All (N=158) | Women (N=72, 45.6%) | Men (N=86, 54.4%) | N |
|---|---|---|---|---|
| Age | 49.73 ± 19.71 | 48.42 ± 18.30 | 50.84 ± 20.85 | 158 |
| Virus | | | | 56 |
| A | 20 (35.7%) | 6 (33.3%) | 14 (36.8%) | |
| A H1 | 10 (17.9%) | 3 (16.7%) | 7 (18.4%) | |
| A H1 N1 | 5 (8.9%) | 3 (16.7%) | 2 (5.3%) | |
| A H3 | 1 (1.8%) | 0 (0.0%) | 1 (2.6%) | |
| B | 1 (1.8%) | 0 (0.0%) | 1 (2.6%) | |
| Seasonal | 19 (33.9%) | 6 (33.3%) | 13 (34.2%) | |

**Table S13 | General characteristics for the IBD cohort.** Values are shown as mean ± standard deviation for continuous variables, and as counts (percentages) for categorical variables.

| **Variable** | **All (N=310)** | **Women (N=132, 42.6%)** | **Men (N=178, 57.4%)** | **N** |
|---|---|---|---|---|
| Age | 42.42 ± 14.97 | 43.02 ± 15.21 | 41.97 ± 14.81 | 310 |
| Smoker | | | | 309 |
| No | 189 (61.2%) | 85 (64.4%) | 104 (58.8%) | |
| Yes | 52 (16.8%) | 22 (16.7%) | 30 (16.9%) | |
| Former | 68 (22.0%) | 25 (18.9%) | 43 (24.3%) | |
| IBD type | | | | 310 |
| Crohn's disease | 125 (40.3%) | 56 (42.4%) | 69 (38.8%) | |
| Ulcerative colitis | 179 (57.7%) | 76 (57.6%) | 103 (57.9%) | |
| Indeterminate colitis | 6 (1.9%) | 0 (0.0%) | 6 (3.4%) | |
| Weight (kg) | 73.86 ± 15.73 | 66.43 ± 13.28 | 79.38 ± 15.15 | 307 |
| Height (cm) | 169.81 ± 8.82 | 163.07 ± 6.75 | 174.90 ± 6.48 | 307 |
| BMI | 25.54 ± 4.90 | 25.10 ± 5.36 | 25.88 ± 4.51 | 306 |
| Hemo (g/dL) | 13.53 ± 1.96 | 12.55 ± 1.77 | 14.33 ± 1.74 | 188 |
| Creatinine (mg/dL) | 0.82 ± 0.19 | 0.71 ± 0.17 | 0.90 ± 0.15 | 181 |
| Urea (mg/dL) | 29.53 ± 8.77 | 28.94 ± 9.23 | 30.06 ± 8.36 | 167 |
| Glucose (mg/dL) | 91.65 ± 17.60 | 89.15 ± 14.33 | 93.68 ± 19.70 | 181 |
| Cholesterol (mg/dL) | 166.62 ± 41.37 | 170.54 ± 38.23 | 163.31 ± 43.80 | 155 |
| Triglycerides (mg/dL) | 96.89 ± 55.84 | 90.35 ± 43.95 | 102.04 ± 63.47 | 143 |
| Albumin (g/dL) | 4.15 ± 0.57 | 4.12 ± 0.55 | 4.17 ± 0.58 | 160 |
| CRP (mg/dL) | 1.76 ± 3.46 | 1.34 ± 1.63 | 2.08 ± 4.38 | 171 |
| Hematocrit (%) | 40.81 ± 5.33 | 38.00 ± 4.54 | 43.09 ± 4.82 | 188 |
| Platelets (n/L) | 296.56 ± 112.12 | 321.24 ± 112.15 | 276.63 ± 108.59 | 188 |
| Leukocytes (n/L) | 7.77 ± 2.77 | 7.64 ± 3.00 | 7.87 ± 2.58 | 187 |
| Alanine aminotransferase (U/L) | 22.51 ± 18.28 | 21.95 ± 22.66 | 22.96 ± 13.88 | 177 |
| Aspartate aminotransferase (U/L) | 22.37 ± 12.50 | 21.68 ± 13.64 | 22.91 ± 11.58 | 171 |
| Alkaline phosphatase (U/L) | 76.10 ± 31.02 | 78.33 ± 32.29 | 74.32 ± 30.03 | 169 |
| Gamma-glutamyl transferase (U/L) | 26.20 ± 19.13 | 21.20 ± 15.13 | 30.17 ± 21.03 | 172 |
| Bilirubin (mg/dL) | 0.93 ± 2.06 | 0.97 ± 2.10 | 0.90 ± 2.04 | 156 |

**Table S14 | General characteristics for the Psoriasis cohort.** Values are shown as mean ± standard deviation for continuous variables, and as counts (percentages) for categorical variables.

| Variable | All (N=226) | Women (N=109, 48.2%) | Men (N=117, 51.8%) | N |
|---|---|---|---|---|
| Age | 48.19 ± 15.40 | 47.04 ± 16.20 | 49.27 ± 14.59 | 226 |

**Table S15 | General characteristics for the ALS cohort.** Values are shown as mean ± standard deviation for continuous variables, and as counts (percentages) for categorical variables.

| **Variable** | **All (N=64)** | **Women (N=20, 31.2%)** | **Men (N=44, 68.8%)** | **N** |
|---|---|---|---|---|
| Age | 58.83 ± 11.28 | 55.75 ± 11.26 | 60.23 ± 11.14 | 64 |

**Table S16 | General characteristics for the Kidney Disease cohort.** Values are shown as mean ± standard deviation for continuous variables, and as counts (percentages) for categorical variables.

| **Variable** | **All (N=828)** | **Women (N=267, 32.2%)** | **Men (N=561, 67.8%)** | **N** |
|---|---|---|---|---|
| Age | 67.63 ± 14.44 | 67.80 ± 15.14 | 67.56 ± 14.10 | 828 |
| BMI | 26.65 ± 5.32 | 26.33 ± 5.81 | 26.80 ± 5.06 | 815 |
| Weight (kg) | 72.44 ± 15.98 | 64.30 ± 14.45 | 76.27 ± 15.22 | 816 |
| Height (cm) | 164.38 ± 11.42 | 155.41 ± 12.01 | 168.67 ± 8.19 | 816 |
| Dialysate formula | | | | 823 |
| Acetate | 52 (6.3%) | 17 (6.4%) | 35 (6.3%) | |
| Citrate | 771 (93.7%) | 247 (93.6%) | 524 (93.7%) | |
| Charlson comorbidity index | 7.76 ± 2.89 | 7.42 ± 2.73 | 7.92 ± 2.96 | 818 |
| SARC-F score | 2.55 ± 2.61 | 3.13 ± 2.60 | 2.27 ± 2.58 | 740 |
| FRAIL scale | 1.61 ± 1.36 | 1.75 ± 1.41 | 1.55 ± 1.33 | 747 |
| Erythrocytes (n/L) | 3.63 ± 0.50 | 3.61 ± 0.50 | 3.64 ± 0.50 | 760 |
| Hemo (g/dL) | 11.42 ± 1.33 | 11.32 ± 1.29 | 11.46 ± 1.34 | 774 |
| Hematocrit (%) | 35.23 ± 4.24 | 35.25 ± 4.07 | 35.22 ± 4.32 | 773 |
| Mean corpuscular volume | 97.17 ± 7.55 | 98.05 ± 7.89 | 96.76 ± 7.35 | 800 |
| Platelets (n/L) | 193.73 ± 65.99 | 196.04 ± 64.34 | 192.65 ± 66.78 | 800 |
| Leukocytes (n/L) | 20.41 ± 348.23 | 6.79 ± 2.45 | 26.79 ± 421.87 | 800 |
| Glucose (mg/dL) | 141.73 ± 68.19 | 143.11 ± 72.28 | 141.08 ± 66.25 | 800 |
| Urea (mg/dL) | 119.20 ± 39.07 | 110.84 ± 35.86 | 122.95 ± 39.89 | 749 |
| Creatinine (mg/dL) | 7.34 ± 2.28 | 6.43 ± 1.72 | 7.77 ± 2.38 | 800 |
| Urate (mg/dL) | 5.60 ± 1.30 | 5.38 ± 1.15 | 5.70 ± 1.35 | 644 |
| Cholesterol (mg/dL) | 138.32 ± 37.35 | 149.82 ± 36.96 | 132.93 ± 36.34 | 767 |
| Cholesterol HDL (mg/dL) | 45.55 ± 15.21 | 51.32 ± 17.06 | 42.86 ± 13.46 | 670 |
| Cholesterol LDL (mg/dL) | 68.10 ± 30.14 | 74.06 ± 30.47 | 65.30 ± 29.61 | 665 |
| Triglycerides (mg/dL) | 124.52 ± 72.74 | 127.70 ± 62.51 | 123.03 ± 77.08 | 793 |
| Bilirubin (mg/dL) | 0.45 ± 0.27 | 0.43 ± 0.26 | 0.46 ± 0.27 | 546 |
| Albumin (g/dL) | 3.95 ± 0.34 | 3.92 ± 0.33 | 3.97 ± 0.35 | 800 |
| Total protein (g/dL) | 6.64 ± 2.26 | 6.78 ± 3.92 | 6.58 ± 0.57 | 688 |
| Intact parathyroid hormone (ng/L) | 366.19 ± 351.72 | 384.98 ± 392.57 | 357.40 ± 330.89 | 800 |
| Alanine aminotransferase (U/L) | 16.40 ± 31.96 | 16.94 ± 51.20 | 16.15 ± 16.71 | 799 |
| Aspartate aminotransferase (U/L) | 18.37 ± 24.76 | 20.38 ± 37.65 | 17.44 ± 15.52 | 725 |
| Gamma-glutamyl transferase (U/L) | 38.73 ± 51.36 | 39.49 ± 43.37 | 38.39 ± 54.59 | 617 |
| Alkaline phosphatase (U/L) | 113.66 ± 87.85 | 120.07 ± 82.91 | 110.66 ± 89.99 | 799 |
| Calcium (mg/dL) | 8.89 ± 0.67 | 8.94 ± 0.71 | 8.86 ± 0.66 | 800 |
| Phosphorus (mg/dL) | 4.77 ± 1.37 | 4.77 ± 1.32 | 4.77 ± 1.40 | 800 |
| Magnesium (mg/dL) | 1.97 ± 0.58 | 1.94 ± 0.57 | 1.99 ± 0.58 | 595 |
| Sodium (mEq/L) | 137.77 ± 5.52 | 137.12 ± 8.66 | 138.07 ± 3.07 | 800 |
| Potassium (mEq/L) | 4.95 ± 0.71 | 4.89 ± 0.71 | 4.99 ± 0.70 | 800 |
| Iron (µg/dL) | 65.39 ± 26.21 | 63.54 ± 27.36 | 66.25 ± 25.63 | 800 |
| Transferrin (mg/dL) | 163.94 ± 50.97 | 165.49 ± 46.31 | 163.22 ± 53.03 | 792 |
| Transferrin saturation index (%) | 28.97 ± 12.20 | 28.79 ± 12.88 | 29.05 ± 11.87 | 799 |
| Ferritin (ng/mL) | 577.87 ± 602.43 | 591.72 ± 466.82 | 571.39 ± 656.63 | 800 |
| Folate (ng/mL) | 8.50 ± 7.03 | 8.94 ± 7.47 | 8.30 ± 6.81 | 613 |
| Vitamin B12 (pg/mL) | 507.13 ± 233.38 | 520.93 ± 236.17 | 500.72 ± 232.05 | 697 |
| Vitamin 25 hydroxy D (ng/mL) | 25.07 ± 18.43 | 24.05 ± 16.54 | 25.55 ± 19.24 | 738 |
| Thyroid-stimulating hormone (mIU/L) | 2.51 ± 3.97 | 3.06 ± 4.52 | 2.24 ± 3.67 | 341 |
| CRP (mg/dL) | 8.19 ± 19.37 | 7.29 ± 16.49 | 8.61 ± 20.58 | 715 |
| Total carbon dioxide (mmol/L) | 22.28 ± 3.25 | 22.15 ± 3.13 | 22.35 ± 3.31 | 395 |

**Table S17 | General characteristics for the Liver Disease cohort.** Values are shown as mean ± standard deviation for continuous variables, and as counts (percentages) for categorical variables.

| Variable | All (N=160) | Women (N=49, 30.6%) | Men (N=111, 69.4%) | N |
|---|---|---|---|---|
| Age | 48.61 ± 12.44 | 55.04 ± 10.47 | 45.77 ± 12.22 | 160 |
| BMI | 29.34 ± 4.86 | 29.65 ± 5.00 | 29.21 ± 4.81 | 160 |
| NAS steatosis score | | | | 157 |
| 0 | 7 (4.5%) | 3 (6.1%) | 4 (3.7%) | |
| 1 | 76 (48.4%) | 19 (38.8%) | 57 (52.8%) | |
| 2 | 47 (29.9%) | 15 (30.6%) | 32 (29.6%) | |
| 3 | 27 (17.2%) | 12 (24.5%) | 15 (13.9%) | |
| NAS lobular inflammation score | | | | 156 |
| 0 | 36 (23.1%) | 9 (18.8%) | 27 (25.0%) | |
| 1 | 103 (66.0%) | 31 (64.6%) | 72 (66.7%) | |
| 2 | 16 (10.3%) | 7 (14.6%) | 9 (8.3%) | |
| 3 | 1 (0.6%) | 1 (2.1%) | 0 (0.0%) | |
| NAS hepatocellular ballooning score | | | | 156 |
| 0 | 26 (16.7%) | 8 (16.7%) | 18 (16.7%) | |
| 1 | 83 (53.2%) | 18 (37.5%) | 65 (60.2%) | |
| 2 | 47 (30.1%) | 22 (45.8%) | 25 (23.1%) | |
| Total NAFLD activity score (NAS) | 3.62 ± 1.43 | 4.04 ± 1.69 | 3.44 ± 1.26 | 156 |
| Fibrosis stage | | | | 160 |
| 0 | 41 (25.6%) | 12 (24.5%) | 29 (26.1%) | |
| 1 | 32 (20.0%) | 6 (12.2%) | 26 (23.4%) | |
| 1a | 2 (1.2%) | 1 (2.0%) | 1 (0.9%) | |
| 1c | 6 (3.8%) | 2 (4.1%) | 4 (3.6%) | |
| 2 | 30 (18.8%) | 6 (12.2%) | 24 (21.6%) | |
| 3 | 26 (16.2%) | 11 (22.4%) | 15 (13.5%) | |
| 4 | 23 (14.4%) | 11 (22.4%) | 12 (10.8%) | |
| NAFLD diagnosis | | | | 156 |
| NASH | 105 (67.3%) | 35 (72.9%) | 70 (64.8%) | |
| Normal | 3 (1.9%) | 2 (4.2%) | 1 (0.9%) | |
| Other pathology | 4 (2.6%) | 1 (2.1%) | 3 (2.8%) | |
| Steatofibrosis | 17 (10.9%) | 3 (6.2%) | 14 (13.0%) | |
| Steatosis | 27 (17.3%) | 7 (14.6%) | 20 (18.5%) | |
| Ethnic group | | | | 160 |
| White | 160 (100.0%) | 49 (100.0%) | 111 (100.0%) | |
| Waist (cm) | 100.45 ± 10.99 | 100.18 ± 11.18 | 100.55 ± 10.98 | 119 |
| Glucose (mg/dL) | 102.55 ± 25.67 | 113.01 ± 31.97 | 97.93 ± 20.90 | 160 |
| Triglycerides (mg/dL) | 144.54 ± 95.70 | 146.40 ± 77.41 | 143.69 ± 103.26 | 157 |
| Albumin (g/L) | 43.97 ± 3.37 | 42.80 ± 3.23 | 44.47 ± 3.31 | 139 |
| Alanine aminotransferase (U/L) | 60.69 ± 43.06 | 54.57 ± 42.40 | 63.40 ± 43.26 | 160 |
| Aspartate aminotransferase (U/L) | 39.59 ± 32.83 | 42.78 ± 39.40 | 38.18 ± 29.56 | 160 |
| Diabetes | | | | 160 |
| No | 95 (59.4%) | 20 (40.8%) | 75 (67.6%) | |
| Yes | 65 (40.6%) | 29 (59.2%) | 36 (32.4%) | |
| Medication for hypertension | | | | 154 |
| No | 88 (57.1%) | 21 (43.8%) | 67 (63.2%) | |
| Yes | 66 (42.9%) | 27 (56.2%) | 39 (36.8%) | |

| **Variable** | **All (N=160)** | **Women (N=49, 30.6%)** | **Men (N=111, 69.4%)** | **N** |
|---|---|---|---|---|
| Medication for hypercholesterolemia | | | | 154 |
| No | 108 (70.1%) | 28 (58.3%) | 80 (75.5%) | |
| Yes | 46 (29.9%) | 20 (41.7%) | 26 (24.5%) | |
| Medication for diabetes | | | | 158 |
| No | 118 (74.7%) | 33 (67.3%) | 85 (78.0%) | |
| Yes | 40 (25.3%) | 16 (32.7%) | 24 (22.0%) | |

**Table S18 | General characteristics for the Prostate Cancer cohort.** Values are shown as mean ± standard deviation for continuous variables, and as counts (percentages) for categorical variables.

| **Variable** | **Men (N=717)** | **N** |
|---|---|---|
| Age | 64.49 ± 6.45 | 717 |
| BMI | 27.02 ± 3.23 | 710 |
| Waist (cm) | 101.03 ± 10.36 | 701 |
| Hip (cm) | 103.54 ± 9.05 | 701 |
| Waist/hip ratio | 0.98 ± 0.10 | 701 |
| Gleason score (prostate cancer) (urine) | | 383 |
| 3+3 | 54 (14.1%) | |
| 3+4 | 234 (61.1%) | |
| 3+5 | 8 (2.1%) | |
| 4+3 | 44 (11.5%) | |
| 4+4 | 19 (5.0%) | |
| 4+4 // 3+4 | 1 (0.3%) | |
| 4+5 | 22 (5.7%) | |
| 5+5 | 1 (0.3%) | |
| Gleason score (prostate cancer) (blood) | | 689 |
| 3+3 | 103 (14.9%) | |
| 3+3 // 3+4 | 2 (0.3%) | |
| 3+3 // 4+5 | 1 (0.1%) | |
| 3+4 | 410 (59.5%) | |
| 3+4 // 3+3 | 1 (0.1%) | |
| 3+4 // 3+5 | 2 (0.3%) | |
| 3+4 // 4+3 | 2 (0.3%) | |
| 3+4 // 4+5 | 1 (0.1%) | |
| 3+5 | 13 (1.9%) | |
| 4+3 | 75 (10.9%) | |
| 4+4 | 27 (3.9%) | |
| 4+4 // 3+3 | 1 (0.1%) | |
| 4+4 // 3+4 | 1 (0.1%) | |
| 4+5 | 48 (7.0%) | |
| 4+5 // 3+4 | 1 (0.1%) | |
| 5+4 | 1 (0.1%) | |
| Cancer stage (urine) | | 375 |
| pT1c | 2 (0.5%) | |
| pT2 | 1 (0.3%) | |
| pT2a | 7 (1.9%) | |
| pT2b | 1 (0.3%) | |
| pT2c | 277 (73.9%) | |
| pT3a | 65 (17.3%) | |
| pT3b | 22 (5.9%) | |
| Cancer stage (blood) | | 682 |
| pT1c | 5 (0.7%) | |
| pT2 | 78 (11.4%) | |
| pT2a | 13 (1.9%) | |
| pT2b | 2 (0.3%) | |
| pT2c | 372 (54.5%) | |
| pT3a | 152 (22.3%) | |
| pT3b | 59 (8.7%) | |
| pT4 | 1 (0.1%) | |
| Perineural invasion (urine) | | 354 |
| Pn0 | 47 (13.3%) | |
| Pn1 | 306 (86.4%) | |
| PnX | 1 (0.3%) | |
| Perineural invasion (blood) | | 650 |
| Pn0 | 57 (8.8%) | |
| Pn1 | 593 (91.2%) | |
| Lymph node status (urine) | | 359 |
| N0 | 21 (5.8%) | |
| N1 | 11 (3.1%) | |
| NX | 327 (91.1%) | |
| Lymph node status (blood) | | 655 |
| N0 | 68 (10.4%) | |
| N1 | 19 (2.9%) | |
| NX | 568 (86.7%) | |
| Total prostate-specific antigen (ng/mL) (urine) | 7.87 ± 4.12 | 349 |
| Total prostate-specific antigen (ng/mL) (blood) | 8.03 ± 4.62 | 683 |
| Diabetes | | 705 |
| No | 629 (89.2%) | |
| Yes | 76 (10.8%) | |
| Medication for diabetes | | 97 |
| No | 36 (37.1%) | |
| Yes | 61 (62.9%) | |
| Hypercholesterolemia | | 699 |
| No | 415 (59.4%) | |
| Yes | 284 (40.6%) | |
| Medication for hypercholesterolemia | | 245 |
| No | 37 (15.1%) | |
| Yes | 208 (84.9%) | |
| Hypertension | | 605 |
| No | 366 (60.5%) | |
| Yes | 239 (39.5%) | |
| Medication for hypertension | | 282 |
| No | 76 (27.0%) | |
| Yes | 206 (73.0%) | |

**Table S19 | General characteristics for the Prostate Cancer cohort (BHP).** Values are shown as mean ± standard deviation for continuous variables, and as counts (percentages) for categorical variables.

| Variable | Men (N=315) | N |
|---|---|---|
| Age | 70.65 ± 8.54 | 315 |
| BMI | 27.24 ± 3.74 | 312 |
| Waist (cm) | 103.64 ± 10.51 | 307 |
| Hip (cm) | 105.57 ± 7.17 | 308 |
| Waist/hip ratio | 0.98 ± 0.06 | 307 |
| Diabetes | | 312 |
| No | 244 (78.2%) | |
| Yes | 68 (21.8%) | |
| Medication for diabetes | | 65 |
| No | 9 (13.8%) | |
| Yes | 56 (86.2%) | |
| Hypercholesterolemia | | 308 |
| No | 194 (63.0%) | |
| Yes | 114 (37.0%) | |
| Medication for hypercholesterolemia | | 108 |
| No | 8 (7.4%) | |
| Yes | 100 (92.6%) | |
| Hypertension | | 267 |
| No | 131 (49.1%) | |
| Yes | 136 (50.9%) | |
| Medication for hypertension | | 153 |
| No | 33 (21.6%) | |
| Yes | 120 (78.4%) | |

**Table S20 | General characteristics for the CVD cohort.** Values are shown as mean ± standard deviation for continuous variables, and as counts (percentages) for categorical variables.

| **Variable** | **All (N=59)** | **Women (N=21, 35.6%)** | **Men (N=38, 64.4%)** | **N** |
|---|---|---|---|---|
| Age | 60.76 ± 10.15 | 61.90 ± 11.18 | 60.13 ± 9.63 | 59 |
| Weight (kg) | 81.48 ± 16.27 | 74.39 ± 13.69 | 85.23 ± 16.45 | 52 |
| Height (cm) | 167.13 ± 11.03 | 157.11 ± 8.46 | 172.44 ± 8.20 | 52 |
| BMI | 29.15 ± 5.09 | 30.31 ± 6.07 | 28.53 ± 4.46 | 52 |

**Table S21 | List of metabolites and clinical parameters included in the metabolic age model.** All NMR-derived metabolite concentrations are expressed in mM while the clinical parameters are reported alongside their specific laboratory units.

| Metabolites (mM) | | Clinical parameters | |
|---|---|---|---|
| | | Parameter | Unit |
| 1,5-Anhydrosorbitol | Glutamic acid | Albumin | g/dl |
| 2-Aminobutyric acid | Glutamine | Apolipoprotein B | mg/ml |
| 2-Hydroxybutyric acid | Glycerol | Bilirubin | mg/dl |
| 2-Oxoglutaric acid | Glycine | Calcium | mg/dl |
| 3-Hydroxybutyric acid | Histidine | C-reactive protein | mg/dl |
| 3-Hydroxyisobutyric acid | Isoleucine | Erythrocyte sedimentation rate | mm/h |
| Acetic acid | Lactic acid | Erythrocytes | T/l |
| Acetoacetic acid | Leucine | Estimated glomerular filtration rate | ml/min/1.73m² |
| Acetone | Lysine | Fructosamine | µM |
| Alanine | Methanol | Glyc A | mM |
| Arginine | Methionine | Glyc B | mM |
| Asparagine | Myo-inositol | HDL cholesterol | mg/dl |
| Aspartate | N,N-Dimethylglycine | Hemoglobin | g/dl |
| Betaine | Ornithine | Iron | µg/dl |
| Choline | Phenylalanine | LDL cholesterol | mg/dl |
| Citric acid | Proline | Leukocytes | G/l |
| Creatine | Pyruvic acid | Lipoprotein(a) | mg/dl |
| Creatinine | Sarcosine | Platelets | G/l |
| Cystine | Serine | SPC (Supramolecular Phospholipid Composite) | mM |
| D-Galactose | Succinic acid | Total cholesterol | mg/dl |
| Dimethylamine | Threonine | Total protein | g/dl |
| Dimethylsulfone | Trimethylamine-N-oxide | Transferrin | mg/dl |
| Ethanol | Tyrosine | Triglycerides | mg/dl |
| Formic acid | Valine | Urate | mg/dl |
| Glucose | | Urea | mg/dl |

**Table S22 | Definition and biological rationale of predefined SHAP groups.** The Table Sdetails the assignment of individual quantified biomarkers into 14 distinct biological or clinical domains, alongside the physiological significance of each group in the context of systemic aging and metabolic function.

| SHAP group | Constituent biomarkers | Biological rationale |
|---|---|---|
| **Energy metabolism** | Glucose (mM), Pyruvic acid (mM), Lactic acid (mM), Lactate to pyruvate ratio (unitless), D-Galactose (mM), 1,5-Anhydrosorbitol (mM), Acetic acid (mM), Citric acid (mM), Succinic acid (mM), 2-Oxoglutaric acid (mM) | This group captures the efficiency of central carbon metabolism, including glycolysis and the tricarboxylic acid (TCA) cycle. As biological aging progresses, mitochondrial dysfunction becomes a primary hallmark, leading to altered cellular respiration, shifts in energy substrates, and impaired metabolic flexibility. |
| **Lipid and ketone metabolism** | Acetoacetic acid (mM), Acetone (mM), 3-Hydroxybutyric acid (mM), Glycerol (mM) | These metabolites reflect lipolysis, fatty acid oxidation, and hepatic ketogenesis. Changes in this profile indicate shifts in systemic energy partitioning, often associated with insulin resistance, age-related metabolic inflexibility, or adaptations to energy deprivation. |
| **One-carbon and methylation metabolism** | Betaine (mM), N,N-Dimethylglycine (mM), Sarcosine (mM), Choline (mM), Methionine (mM), Formic acid (mM), Serine (mM), Glycine (mM) | This pathway is fundamental for DNA methylation, nucleotide synthesis, and epigenetic maintenance. Dysregulation in one-carbon metabolism is tightly linked to the "epigenetic clock" of aging, cellular senescence, and age-related neurocognitive decline. |
| **Lipoprotein and cardiovascular risk markers** | LDL cholesterol (mg/dl), Triglycerides (mg/dl), Apolipoprotein B (mg/ml), Lipoprotein(a) (mg/dl), SPC (mM), HDL cholesterol (mg/dl) | This cluster represents systemic lipid transport and atherogenic potential. The gradual deterioration of vascular health, endothelial dysfunction, and increased cardiovascular risk are defining features of physiological aging, heavily influenced by the balance of these lipoproteins. |
| **Kidney function and nitrogen waste** | Creatinine (mM), Urea (mg/dl), Urea to creatinine ratio (unitless), Estimated Glomerular Filtration Rate (ml/min/1.73m$^2$), Urate (mg/dl), Myo-inositol (mM) | These markers evaluate renal filtration capacity and the accumulation of uremic solutes. The gradual decline in the glomerular filtration rate and the reduced efficiency in clearing nitrogenous waste are standard, quantifiable metrics of renal senescence. |
| **Inflammation and immune activity** | Glyc A (mM), Erythrocyte sedimentation rate (mm/h), Albumin (g/dl), GlycB to GlycA ratio (unitless), SPC (mM), C-reactive protein (mg/dl), Leukocytes (G/l) | This group quantifies "inflammaging"—the chronic, low-grade, sterile inflammation characteristic of older age. Sustained systemic inflammation drives tissue degeneration, immune exhaustion, and accelerates the biological aging process across all organ systems. |

| SHAP group | Constituent biomarkers | Biological rationale |
|---|---|---|
| **Microbiome-derived metabolites** | Trimethylamine-N-oxide (mM), Dimethylamine (mM), Methanol (mM) | These compounds reflect the interplay between the host and the gut microbiome. Aging is associated with increased intestinal permeability and dysbiosis, leading to higher systemic absorption of microbial metabolites that can contribute to cardiovascular and metabolic aging. |
| **Exposome and xenobiotics** | Dimethylsulfone (mM), Ethanol (mM), Methanol (mM), Formic acid (mM) | This category encapsulates the "exposome," representing the cumulative impact of environmental exposures, dietary habits, and lifestyle factors. These exogenous compounds interact continuously with intrinsic biological pathways to modulate the trajectory of aging. |
| **Hematological resilience and oxygen transport** | Erythrocytes (T/l), Iron (µg/dl), Transferrin (mg/dl), Platelets (G/l) | These markers assess bone marrow reserve, oxygen-carrying capacity, and coagulation status. Declines in this domain often manifest as the "anemia of aging" and a pro-thrombotic state, reducing physical endurance and increasing vulnerability to physiological stress. |
| **Muscle and BCAA metabolism** | Valine (mM), Leucine (mM), Isoleucine (mM), 3-Hydroxyisobutyric acid (mM), Creatine (mM) | This group monitors skeletal muscle turnover, protein anabolism, and the catabolism of branched-chain amino acids. Alterations in these specific metabolites are direct biochemical correlates of sarcopenia, muscle atrophy, and declining physical function. |
| **Structural integrity and systemic frailty** | Calcium (mg/dl), Albumin (g/dl), Non-albumin protein (g/dl), Proline (mM), Lysine (mM) | This domain acts as a composite indicator of physical frailty. It assesses the degradation of bone mass (calcium), the breakdown of the extracellular matrix and collagen (proline, lysine), and overall systemic nutritional reserves (serum proteins). |
| **Redox homeostasis** | Cystine (mM), Methionine (mM), Serine (mM), Glycine (mM), 2-Aminobutyric acid (mM), 2-Hydroxybutyric acid (mM) | These metabolites reflect the cellular capacity to synthesize core antioxidants, primarily glutathione, and manage oxidative stress. The progressive loss of redox balance leads to accumulated oxidative damage to DNA, lipids, and proteins, driving the aging phenotype. |

| SHAP group | Constituent biomarkers | Biological rationale |
|---|---|---|
| **Nitrogen homeostasis** | Glutamine to glutamate ratio (unitless), Glutamic acid (mM), Glutamine (mM), Aspartate (mM), Asparagine (mM), Alanine (mM), Threonine (mM), Arginine (mM), Ornithine (mM) | This group maps the systemic transport of nitrogen, ammonia buffering mechanisms, and the urea cycle. The efficiency of inter-organ nitrogen shuttling (especially between muscle and liver) declines with age, reflecting early metabolic senescence and reduced homeostatic capacity. |
| **Hepatic clearance of aromatic amino acids** | Bilirubin (mg/dl), BCAA to AAA (Phe+Tyr) ratio (unitless), Phenylalanine (mM), Tyrosine (mM), Histidine (mM) | This cluster measures the detoxification and filtration capacity of the liver. As hepatic function wanes with age, the liver loses its efficiency in clearing aromatic amino acids, leading to their systemic accumulation, which can disrupt blood-brain barrier transport and central nervous system signaling. |

**Table S23 | Number of samples selected for each age bracket and sex to create the balanced dataset.** Values correspond to the subset of 8,640 individuals obtained after random stratified sampling across the five predefined age groups.

| | **female** | **male** |
|---|---|---|
| **0-30** | 864 | 864 |
| **31-40** | 864 | 864 |
| **41-50** | 864 | 864 |
| **51-65** | 735 | 993 |
| **>65** | 993 | 735 |